\pdfoutput=1
\documentclass[a4paper,12pt]{article}

\usepackage{amsfonts}
\usepackage{mathrsfs}
\usepackage{amsmath}
\usepackage{amssymb}
\usepackage{framed}

\usepackage[medium]{titlesec}
\usepackage{bm}
\usepackage{cite}
\usepackage{stmaryrd}

\usepackage[normalem]{ulem}
\usepackage{extarrows}
\usepackage{slashed}
\usepackage{isodateo}
\usepackage{graphicx}
\usepackage[dvipsnames]{xcolor}
\usepackage[bookmarksnumbered=true,bookmarksopen=true]{hyperref}
 \hypersetup{colorlinks,%
             linkcolor=NavyBlue, %
             citecolor=PineGreen, %
             urlcolor=PineGreen}
\usepackage[hmargin=.7in,vmargin=1.1in]{geometry}
\usepackage{indentfirst}
\usepackage{booktabs}

\usepackage{bbm}

\newcommand{\FR}[2]{\displaystyle\frac{\,{#1}\,}{#2}}
\newcommand{\fr}[2]{\mbox{$\frac{\,{#1}\,}{#2}$}}
\newcommand{\n}{\nonumber}

\makeatletter
\newcommand{\subalign}[1]{%
  \vcenter{%
    \Let@ \restore@math@cr \default@tag
    \baselineskip\fontdimen10 \scriptfont\tw@
    \advance\baselineskip\fontdimen12 \scriptfont\tw@
    \lineskip\thr@@\fontdimen8 \scriptfont\thr@@
    \lineskiplimit\lineskip
    \ialign{\hfil$\m@th\scriptstyle##$&$\m@th\scriptstyle{}##$\hfil\crcr
      #1\crcr
    }%
  }%
}
\makeatother

\graphicspath{{fig/}}

\def\bge{\begin{equation}}
\def\ede{\end{equation}}
\def\bga{\begin{aligned}}
\def\eda{\end{aligned}}
\def\bgb{\begin{bmatrix}}
\def\edb{\end{bmatrix}}
\def\bgp{\begin{pmatrix}}
\def\edp{\end{pmatrix}}
\def\bgm{\begin{matrix}}
\def\edm{\end{matrix}}
\def\bgs{\begin{subequations}}
\def\eds{\end{subequations}}
\newcommand{\order}[1]{\mathcal{O}({#1})}
\def\di{{\mathrm{d}}}

\def\mb{\mathbf}

\def\pd{\partial}

\def\la{\langle}\def\ra{\rangle}
\def\sla{\slashed}

\def\to{\rightarrow}
\def\To{\Rightarrow}
\def\ii{\mathrm{i}}

\def\al{\alpha}
\def\be{\beta}
\def\ga{\gamma}
\def\de{\delta}

\def\si{\sigma}

\def\aa{\mathsf{a}}
\def\bb{\mathsf{b}}
\def\cc{\mathsf{c}}

\def\G{{\wt{\mathcal{G}}}}
\def\G{{\mathcal{G}}}

\def\gL{{\mathcal{L}}}
\def\gR{{\mathcal{R}}}

\usepackage{mdframed} 
\usepackage{tikz}

\newmdenv[skipabove=0pt,%
          skipbelow=5pt,%
          leftmargin=0pt,%
          rightmargin=0pt,%
          innertopmargin=-5pt,%
          innerbottommargin=7pt,%
          innerleftmargin=2pt,%
          innerrightmargin=2pt,%
          splittopskip=0pt,%
          splitbottomskip=0pt,%
          linewidth=0pt,%
          nobreak=true]%
          {keyeqn2}

\newmdenv[backgroundcolor=gray!15,%
          skipabove=0pt,%
          skipbelow=5pt,%
          leftmargin=0pt,%
          rightmargin=0pt,%
          innertopmargin=-5pt,%
          innerbottommargin=7pt,%
          innerleftmargin=2pt,%
          innerrightmargin=2pt,%
          splittopskip=0pt,%
          splitbottomskip=0pt,%
          linewidth=0pt,%
          nobreak=true]%
          {keyeqn}
           
\newmdenv[font=\small,
		  linecolor=black,
          skipabove=10pt,%
          skipbelow=10pt,%
          leftmargin=0pt,%
          rightmargin=0pt,%
          innertopmargin=14pt,%
          innerbottommargin=14pt,%
          innerleftmargin=12pt,%
          innerrightmargin=12pt,%
          splittopskip=15pt,%
          splitbottomskip=5pt,%
          linewidth=0.8pt]%
          {boxedtext}

\usepackage{titlesec}          
\titleformat{\section}
{\normalfont\fontsize{15}{20}\bfseries}{\thesection}{1em}{}

\newcommand{\wt}[1]{\mkern 2mu \widetilde{\mkern -2mu #1 \mkern -2mu}\mkern 2mu}
\newcommand{\wh}[1]{\mkern 2mu \widehat{\mkern-2mu#1\mkern-2mu}\mkern 2mu}

\newcommand{\mft}[1]{\big\llbracket{#1}\big\rrbracket}

\newcommand{\fnemail}[1]{\footnote{Email: \href{mailto:#1}{\nolinkurl{#1}}}}

\begin{document}

\title{\Large\textbf{Massive Inflationary Amplitudes:\\ Differential Equations and Complete Solutions for General Trees\\[2mm]}}

\author{Haoyuan Liu$^{\,a\,}$\fnemail{liuhy23@mails.tsinghua.edu.cn}~~~~~and~~~~~Zhong-Zhi Xianyu$^{\,a,b\,}$\fnemail{zxianyu@tsinghua.edu.cn}\\[5mm]
$^a\,$\normalsize{\emph{Department of Physics, Tsinghua University, Beijing 100084, China} }\\
$^b\,$\normalsize{\emph{Peng Huanwu Center for Fundamental Theory, Hefei, Anhui 230026, China }}
}

\date{}
\maketitle

\begin{tikzpicture}[overlay]
\node[minimum width=40mm,minimum height=15mm] (b) at (16,10.6){\small\verb+USTC-ICTS/PCFT-24-56+};
\end{tikzpicture}

\vspace{20mm}

\begin{abstract}
\vspace{10mm}

We construct and solve a complete system of differential equations for general tree-level inflation correlators with an arbitrary number of massive scalar exchanges and time-dependent couplings. Any massive tree correlators can be uniquely fixed by solving this system of equations with appropriate boundary conditions. We take a hybrid approach to solve this system, using the differential equation to get the inhomogeneous solution and the bulk time integrals to determine the homogeneous solution. Altogether, we obtain analytical results for all tree-level massive inflation correlators with generic kinematics, expressed as multivariate hypergeometric series of energy ratios. The result can be neatly organized as a sum of the completely inhomogeneous solution, which we call the massive family tree, and all of its cuts. As simple applications, we provide full analytical expressions for tree correlators with one, two, and three massive exchanges.

\end{abstract}

\newpage
\tableofcontents

\newpage
\section{Introduction}\label{sec_intro}

There have been considerable interests and efforts recently in the study of quantum field theoretic amplitudes in the cosmological background, including the wavefunction coefficients and the equal-time correlation functions. Theory-wise, these amplitudes possess fascinating structures that crystalize many properties of QFTs in cosmological backgrounds. Phenomenologically, they underlie the large-scale correlations of primordial curvature or tensor perturbations, and encode rich physical information about the primordial inflationary universe. 

A major boost of this direction came from a related program of cosmological collider (CC) physics \cite{Chen:2009we,Chen:2009zp,Arkani-Hamed:2015bza,Chen:2016nrs,Chen:2016uwp,Chen:2016hrz,Lee:2016vti,An:2017hlx,An:2017rwo,Iyer:2017qzw,Kumar:2017ecc,Tong:2018tqf,Chen:2018sce,Chen:2018xck,Chua:2018dqh,Wu:2018lmx,Saito:2018omt,Li:2019ves,Lu:2019tjj,Liu:2019fag,Hook:2019zxa,Hook:2019vcn,Kumar:2018jxz,Kumar:2019ebj,Alexander:2019vtb,Wang:2019gbi,Wang:2019gok,Wang:2020uic,Li:2020xwr,Wang:2020ioa,Fan:2020xgh,Aoki:2020zbj,Bodas:2020yho,Maru:2021ezc,Lu:2021gso,Sou:2021juh,Lu:2021wxu,Pinol:2021aun,Cui:2021iie,Tong:2022cdz,Reece:2022soh,Chen:2022vzh,Niu:2022quw,Niu:2022fki,Chen:2023txq,Chakraborty:2023qbp,Tong:2023krn,Jazayeri:2023xcj,Jazayeri:2023kji,Aoki:2023tjm,McCulloch:2024hiz,Craig:2024qgy}, which aims at probing heavy new particles and their interactions at the inflation scale. In CC physics, the central observables are the oscillatory shapes in the scalar/tensor correlation functions generated by resonant interactions between heavy spectator fields and curvature/tensor perturbations. Thus, CC physics necessitates the study of correlators or wavefunction coefficients mediated by massive fields, which we collectively call massive inflationary amplitudes. 

To compute massive correlators appearing in realistic CC models can be a very challenging task. The challenges come in several ways: First, propagators for massive fields in the cosmological background are necessarily special functions (typically Hankel functions or Whittaker functions in inflation) with time orderings, and thus we need to perform time-ordered integrals over products of these special functions. Second, many CC processes involve loop graphs and loop integrals can be nontrivial. Third, CC processes with large signals typically break some spacetime symmetries, including dS boosts (special conformal transformations) and scale invariance (dilatation). 

Despite all these challenges, the study of massive inflation correlators has received great boosts in recent years. Several new techniques have been proposed and many results for massive correlators have been obtained. The program of cosmological bootstrap was proposed to compute massive inflationary correlators by solving differential equations \cite{Baumann:2022jpr}. The original works considered the case with full or weakly broken dS isometries and succeeded in bootstrapping tree graphs with a single massive exchange \cite{Arkani-Hamed:2018kmz,Baumann:2019oyu}. Later studies improved the methods and obtained results with strongly broken dS boosts and/or dilatation \cite{Pimentel:2022fsc,Jazayeri:2022kjy,Qin:2022fbv,Aoki:2023wdc}. In some cases, it is possible to obtain closed-form expressions without infinite sums \cite{Qin:2023ejc}. However, the differential-equation method quickly becomes nontrivial beyond the single massive exchange, as the order and number of differential equations increase fast with the number of massive lines. Currently, the state-of-the-art result from the differential equation approach is the double massive exchanges in a few degenerate configurations \cite{Aoki:2024uyi}. See also \cite{Hillman:2021bnk,Chen:2023iix,Chen:2024glu} for related works.

Incidentally, things get much simplified if one considers the special case of conformal scalar with mass $m^2=2H^2$ ($H$ being the Hubble parameter) instead of scalars of general masses, due to the enhanced symmetry \cite{Arkani-Hamed:2017fdk}. In this case, a system of differential equations has been constructed for arbitrary tree graphs in power-law FRW universes. From the boundary viewpoint, this system emerges from a rather sophisticated kinematic flow \cite{Arkani-Hamed:2023bsv,Arkani-Hamed:2023kig}. However, \cite{Fan:2024iek} showed that the full analytical solutions to this system can be easily written down by computing the bulk time integrals directly using the family-tree decomposition (FTD). Later it was realized that the differential equations for the family tree integrals are also much simpler \cite{He:2024olr}. See also \cite{Hillman:2019wgh,De:2023xue,Benincasa:2024leu,Benincasa:2024lxe,Baumann:2024mvm} for related works.

At the same time, there is an independent line of developments which focus directly on the bulk time integrals for the correlators \cite{Wang:2021qez,Qin:2022lva,Qin:2022fbv,Xianyu:2022jwk,Qin:2023nhv,Qin:2023bjk,Xianyu:2023ytd,Liu:2024xyi,Werth:2024mjg,Qin:2024gtr}. Two useful techniques in this direction are the
 partial Mellin-Barnes representation (PMB) \cite{Qin:2022lva,Qin:2022fbv} and the family tree decomposition (FTD)\cite{Xianyu:2023ytd,Fan:2024iek}. The PMB representation resolves all special functions in the bulk modes into powers, so that the bulk integrals become arbitrarily nested time integrals over powers and exponential functions\footnote{See \cite{Sleight:2019mgd,Sleight:2019hfp,Sleight:2020obc,Sleight:2021plv} for a related full Mellin-space approach.}. The family-tree decomposition is then designed to compute such integrals with the results in a special class of multivariate hypergeometric functions, called family trees. Then, the analytical computation of any tree diagrams is reduced to a routine of collecting poles of the Mellin integrands, which can be trivially done for generic tree graphs. To the best of our knowledge, before this work, the combination of PMB and FTD is the only method that can compute tree correlators with an arbitrary number of massive exchanges. Meanwhile, the numerical computation of tree graphs has also been systematized recently \cite{Werth:2023pfl,Pinol:2023oux,Werth:2024aui}. Beyond the tree level, things are less explored, although progress has also been made at 1-loop and multi-loop orders, including full analytical or numerical results, as well as analytical results for CC signals \cite{Wang:2021qez,Xianyu:2022jwk,Qin:2023bjk,Qin:2023nhv,Liu:2024xyi,Qin:2024gtr}. 
 
Given the methods of PMB and FTD, the problem of analytically computing any massive tree correlators has been solved. However, the situation is not fully satisfactory. First, results obtained with PMB and FTD are not in an optimal form, in the sense that their series representations typically involve too many layers of summations: For a tree diagram with $I$ internal lines, this method yields a series with $3I$ summation variables, among which $2I$ layers come from PMB and $I$ from FTD \cite{Xianyu:2023ytd}. On the other hand, as we shall see, a diagram is fully specified by $2I$ independent kinematic variables. So, if the final answer is to be expressed as a hypergeometric function of $2I$ variables, we would reasonably conjecture that this function involves only $2I$ layers of summations. To find these more economic expressions for general trees was one of the original motivations for this work. 

Second, going through the procedure of PMB and FTD means that we still need to take Mellin representation for special functions, do the FTD, and then collect all relevant poles in the Mellin integrand. This is a trivial procedure, but still involves quite an amount of calculations when the graph becomes large. In this regard, we were further motivated by the  analytical results for conformal scalar amplitudes in \cite{Fan:2024iek}. There, we have a very simple set of rules to write down the full analytical results for arbitrary conformal scalar tree amplitudes, in terms of a family tree integral and its cuts. We were led to suspect that similar rules exist for arbitrary masses, not only the conformal value $m^2=2H^2$. Were this true, we would be able to directly write down the full analytical results for arbitrary massive tree correlators without doing any real computations. 

We achieve this goal in the current work, eliminating the need to do tedious computations, and providing full analytical results for an arbitrary number of massive exchanges for generic kinematics. (By generic kinematics, we mean that all $2I$ kinematic variables are unconstrained by equalities and can vary independently. Constrained cases will be called degenerate.) In particular, we show that the concept of family trees can be properly generalized to arbitrary masses with a compact formula, and that a generic tree diagram can be expressed in terms of a massive family tree and its cuts. The result is in an optimal form, in the sense that the number of summations is reduced to $2I$, matching the number of independent variables. Since the whole procedure only requires us to take a finite number of products and summations of massive family trees, we could say that computing tree correlators in inflation has been made as simple as in flat space, where we take products and sums of propagators and couplings. 

\paragraph{Summary of main results} The final answer, albeit simple, is obtained by going through a rather long journey, along which we also get a number of intermediate results, including a complete system of differential equations for arbitrary trees, recursion relations for building trees from smaller graphs, the cuts of trees, and the relations between the cuts and the CC signals. These results are useful and interesting in their own rights. Thus, before delving into the details in subsequent sections, we outline the basic strategy and summarize the main results of this paper.

\begin{figure}[t]
\centering
\includegraphics[width=0.9\textwidth]{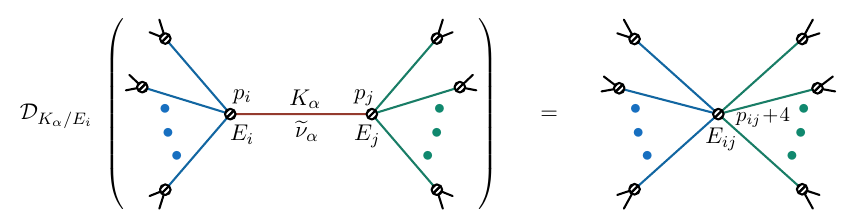}
\caption{Graphic representation of the differential equation (\ref{eq_DEtG}) for the graph $\G$ with respect to Line $\al$ and Vertex $i$.}
\label{fig_DE}
\end{figure}

After defining massive tree amplitudes and briefly reviewing the standard diagrammatic methods for their computations in Sec.\;\ref{sec_review}, we derive a complete set of differential equations for a general tree graph with an arbitrary number of massive exchanges, with couplings carrying arbitrary power-law time dependence in Sec.\;\ref{sec_DE}. For a tree graph of $V$ vertices and thus $I=V-1$ internal lines, there are $2I$ independent kinematic variables, and the system consists of $2I$ coupled second-order partial differential equations, given in (\ref{eq_DEtG}) and (\ref{eq_Dri}) and illustrated in Fig.\;\ref{fig_DE}. We show that these equations can be conveniently derived with the standard method of inserting a Klein-Gordon operator and doing integration by parts. The final system is a coupled Lauricella system \cite{Matsumoto_2020}. 

Then, from Sec.\;\ref{sec_recursion} onward, we set out to solve the differential equations. The strategy is to build the completely inhomogeneous solution (CIS) first, and then obtain the homogeneous solutions as its cuts. Following this strategy, in Sec.\;\ref{sec_recursion} we derive two recursion formulae, (\ref{eq_IngoingInhom}) and (\ref{eq_outgoingRecursion}), to generate inhomogeneous solutions for adding a new line to any massive tree graph. (See Fig.\;\ref{fig_recursion}.) Then, in Sec.\;\ref{sec_CIS}, we use the recursion formulae to build the CIS to an arbitrary massive tree graph. It is somewhat striking that the result can be expressed as a very simple formula (\ref{eq_MFT}), which can be thought of as a generalization of family trees to massive cases. Thus we call it a massive family tree.  Then, in Sec.\;\ref{sec_hom}, we construct the homogeneous solutions by cutting the CIS, and the rule for taking a cut is summarized in (\ref{eq_SingleCut}). (See also Fig.\;\ref{fig_singlecut}.) The full answer for any tree diagram is then presented in Sec.\;\ref{sec_examples} together with a few explicit examples, including tree diagrams with single, double, and triple exchanges. Further discussions and outlooks are left to Sec.\;\ref{sec_conclusion}.  

\paragraph{Notations and conventions} We work with $(3+1)$-dimensional dS spacetime with metric $\di s^2=(-\di\tau^2+\di\bm x^2)/(H\tau)^2$ with $\tau\in(-\infty,0)$ the conformal time and $\bm x\in \mathbb{R}^3$ the comoving coordiates. The Hubble parameter $H$ is set to unity, $H=1$, throughout the work. We draw heavy use of shorthand notations such as $E_{ij}=E_i+E_j$ and $q_{1\cdots N}\equiv q_1+\cdots+q_N$. We use the diagrammatic notation of \cite{Chen:2017ryl} for Feynman diagrams in Schwinger-Keldysh formalism. We also use the massive family tree notation $\mft{\cdots}$, similar to \cite{Fan:2024iek}, which is reviewed in App.\;\ref{app_MFT}. We list some frequently used symbols in App.\;\ref{app_notation} and useful mathematical functions in App.\;\ref{app_functions}.

\section{Brief Review of Massive Inflationary Amplitudes}
\label{sec_review}

In this section, we set the stage by briefly reviewing the perturbative construction of general massive amplitudes, and fixing our notations and conventions. As in \cite{Fan:2024iek}, we use inflationary amplitudes to mean two different types of objects: the wavefunction coefficients and the (equal-time) correlators. They are conceptually different objects but share a lot of technical similarities. Therefore, it is very convenient to consider the both at the same time. A slightly more detailed introduction to these objects can be found in \cite{Fan:2024iek}. More comprehensive reviews on this subject can be found in \cite{Chen:2017ryl} for correlation functions and \cite{Goodhew:2020hob} for wavefunction coefficients. Below we only list the essential equations to be self-contained. 

\paragraph{Massive correlators} In cosmology, the most relevant correlators are those of a massless scalar or tensor field during inflation. Also, in more theoretical studies, one often considers a conformal scalar field (with mass $m^2=2$) since it has a simple mode function closely related to a massless scalar mode. Below, we will use the conformal scalar as an example for illustration, and the generalization to the massless scalar is trivial. 

We assume that the conformal scalar $\phi_c$ and a bunch of massive scalar fields $\si_\al$ are directly and weakly coupled in the bulk via polynomial interactions. (Derivative couplings or spinning exchanges introduce additional tensor structures that can be factored out and are irrelevant to our current study.) The scalar $\si_\al$ has mass $m^2_\al>9/4$ so that the corresponding \emph{mass parameter} $\wt\nu_\al\equiv\sqrt{m_\al^2-9/4}$ is real. The assumption of $m^2_\al>9/4$ is only for easier presentation; Generalization to arbitrary mass is trivial as long as their couplings are IR-finite. Also, we will simply call $\wt\nu_\al$ the mass in the following when no confusion could arise.

Then, we consider a general tree-graph contribution to the $N$-point correlator of $\phi_c$, where all internal lines are from exchanges of massive scalars $\si_\al$. (The case of exchanging an arbitrary number of conformal scalars is much easier and has been completely solved in \cite{Fan:2024iek}). We can conveniently derive the diagrammatic expansion and the corresponding Feynman rules in Schwinger-Keldysh (SK) formalism. See \cite{Chen:2017ryl} for a pedagogical introduction. As a result, we can directly compute a tree graph as a bulk time integral over products of couplings and propagators:
\begin{align}
\label{eq_GraphInt}
  \wh{\mathcal{G}}(\bm k_1,\cdots,\bm k_N)=\sum_{\aa_1,\cdots,\aa_V=\pm} \int_{-\infty}^0\prod_{i=1}^V\Big[\di\tau_i\,\ii\aa_i(-\tau_i)^{P_{i}}\Big]\prod_{n=1}^{N}C_{\aa_n}(k_n,\tau_n)\prod_{\al=1}^{I}D_{\aa_\al\aa_\al'}^{(\wt\nu_\al)}(K_\al;\tau_\al,\tau_\al '),
\end{align}	
where we are considering a tree graph with $N$ external lines, $I$ internal lines, and $V$ vertices. Clearly, for a tree diagram, we have $V=I+1$. It is understood that the time and SK indices in the propagators should be properly identified with the corresponding variables of the vertices to which they attach. Below let us explain this expression more explicitly.

 According to the diagrammatic rule, we assign a time integral $\int\di\tau_i$ to each interaction vertex, together with a factor of $+\ii$ (a consequence of real and direct couplings), an SK index $\aa_i=\pm$ due to the closed path integral, and a monomial factor $(-\tau_i)^{P_i}$ accounting for the background evolution and possible time-dependent couplings. We omit trivial multiplicative coupling constants at all vertices throughout this work. 

Then, for each external line, we assign a bulk-to-boundary propagator $C_{\aa_n}(k_n,\tau_n)$ of a conformal scalar field, whose explicit form is:
\bge
\label{eq_CSProp}
  C_\aa(k;\tau)=\FR{\tau\tau_f}{2k}e^{\aa\ii k\tau},
\ede
where $\tau_f<0$ is a late-time cutoff introduced to take care of the late-time scaling behavior of a conformal scalar, and has no real significance for our calculation below.  
For each internal line, on the other hand, we assign a bulk massive propagator $D_{\aa_\al\aa_\al'}^{(\wt\nu_\al)}(K_\al;\tau_\al,\tau_\al ')$ with arbitrary mass $\wt\nu_\al>0$. There are four types of bulk propagators. The two with opposite SK indices $D_{\mp\pm}^{(\wt\nu)}$ (Wightman functions) are factorized and solve the homogeneous Klein-Gordon equation for the massive scalar in dS, and the two with the same SK indices $D_{\pm\pm}^{(\wt\nu)}$ (Feynman and anti-Feynman propagators) are nested in time and solve the Klein-Gordon equation with a $\delta$-function source. Explicitly, they can be expressed in terms of Hankel function $\text{H}^{(i)}_{\nu}(z)$ $(i=1,2)$ as: 
\begin{align}
\label{eq_Dmp}
  D_{-+}^{(\wt\nu)} (K;\tau_1,\tau_2)
  =&~\FR{\pi}{4}e^{-\pi\wt\nu}(\tau_1\tau_2)^{3/2}\mathrm{H}_{\ii\wt\nu}^{(1)}(-K\tau_1)\mathrm{H}_{-\ii\wt\nu}^{(2)}(-K\tau_2),\\
\label{eq_Dpm}
  D_{+-}^{(\wt\nu)} (K;\tau_1,\tau_2)
  =&~\FR{\pi}{4}e^{-\pi\wt\nu}(\tau_1\tau_2)^{3/2}\mathrm{H}_{-\ii\wt\nu}^{(2)}(-K\tau_1)\mathrm{H}_{\ii\wt\nu}^{(1)}(-K\tau_2),\\
\label{eq_Dpmpm}
  D_{\pm\pm}^{(\wt\nu)} (K;\tau_1,\tau_2)=&~D_{\mp\pm}^{(\wt\nu)}(K;\tau_1,\tau_2)\theta(\tau_1-\tau_2)+D_{\pm\mp}^{(\wt\nu)}(K;\tau_1,\tau_2)\theta(\tau_2-\tau_1),
\end{align}

It turns out convenient to introduce a dimensionless version of the graph, which is related to the original graph $\wh{\mathcal{G}}(\bm k_1,\cdots,\bm k_N)$ in (\ref{eq_GraphInt}) by trivial rescaling. The rescaling is based on two useful properties of propagators. The first property is about the bulk-to-boundary propagators (\ref{eq_CSProp}): Suppose there are $A$ external lines attached to the vertex $\tau_i$. Then,
\begin{align}\label{eq_Ctau}
  \prod_{n=1}^A C_{\aa_i}(k_n;\tau_i)(-\tau_i)^{P_i}=\bigg(\prod_{n=1}^A\FR{-\tau_f}{2k_n}\bigg)e^{+\ii\aa_i E_i\tau_i}(-\tau_i)^{P_i+A},
\end{align}
where $E_i\equiv k_1+\cdots+k_A$ will be called the \emph{vertex energy} in the following. Of course, the parameter $P_i$ on each vertex gets shifted after using the explicit expressions for external lines. We denote the shifted exponent by $p_i\equiv P_i+A$, and call it the \emph{twist} of the vertex. Thus, we see that, if we remove the trivial factors $-\tau_f/(2k_n)$, the graph $\wh{\G}$ will depend only on $V$ vertex energies $\{E_1,\cdots,E_V\}$, together with the energy on all internal lines $\{K_1,\cdots,K_I\}$, which we call \emph{line energies}.  

The second property is about the bulk propagators (\ref{eq_Dmp})-(\ref{eq_Dpmpm}):
\begin{align}\label{eq_wtD}
  \wt D_{\aa\bb}^{(\wt\nu)}(K_\al\tau_i,K_\al\tau_j)\equiv K_\al^3 D_{\aa\bb}^{(\wt\nu)}(K_\al;\tau_i,\tau_j) .
\end{align} 
That is, the dimensionless bulk propagator $\wt{D}_{\aa\bb}^{(\wt\nu)}$ depends on the line energy and the two time variables only through the two dimensionless combinations $K_\al\tau_i$ and $K_\al\tau_j$.  

The above two properties motivate us to define the \emph{dimensionless graph} $\G(\{E\},\{K\})$ in the following way:
\bge
\label{eq_DimlessGraph}
  \wh{\mathcal{G}}(\bm k_1,\cdots,\bm k_N)=\bigg(\prod_{n=1}^N\FR{-\tau_f}{2k_n}\bigg)\bigg(\prod_{i=1}^V\FR{1}{E_{i}^{1+p_i}}\bigg)\bigg(\prod_{\al=1}^I \FR{1}{K_\al^3}\bigg) \mathcal{G}(\{E\},\{K\}).
\ede
The notation $\mathcal{G}(\{E\},\{K\})$ highlights that the dimensionless graph $\G$ is fully determined by the $I$ line energies and $V=I+1$ vertex energies. However, since $\G$ is dimensionless, it depends on these $2I+1$ variables only through their ratios. Therefore, we have the freedom to remove one more variable so that there are only $2I$ independent energy ratios\footnote{In (\ref{eq_DimlessGraph}), all energy variables are assumed nonvanishing. When an energy variable is set to zero, we should slightly modify the definition of a dimensionless graph. We consider an example with a zero vertex energy in Sec.\;\ref{sec_4site}.}.

A convenient choice for the $2I$ independent energy ratios can be obtained as follows. For any internal line with energy $K_\al$ with two endpoints with vertex energies $E_i$ and $E_j$, we can form two energy ratios:
\begin{align}\label{eq_r_ali}
  &r_{(\al i)}\equiv \FR{K_\al}{E_i},
  &&r_{(\al j)}\equiv \FR{K_\al}{E_j}.
\end{align}
Then, for a total of $I$ internal lines, we can form $2I$ energy ratios $\{r_A\}$ $A=1,\cdots,2I$. It is clear that these $2I$ energy ratios are independent variables in a generic graph. However, there are special cases in which some of these ratios are constrained. For instance, in the case of a two-point mixing between an external line and an internal line with line energy $K$, the vertex energy of the two-point vertex is also $K$, resulting in an energy ratio constrained to be 1. We call situations like this degenerate kinematics. In this work, we only consider nondegenerate kinematics, where all $2I$ energy ratios can vary independently. We call such cases generic. 

It is easy to see that, in the generic case, the $2I$ energy ratios $\{r_A\}$ are not only independent but also complete, in the sense that they fully specify the kinematics of a dimensionless graph $\G$. To make this point clear, we introduce the dimensionless time $z_i$ at each vertex:
\bge
  z_i\equiv E_i\tau_i.
\ede
Then, it is straightforward to see that the dimensionless graph $\G$ can be computed from the following integral:
\begin{align}
\label{eq_dimlessIntG}
  \G (\{E\},\{K\})=&\sum_{\aa_1,\cdots,\aa_V=\pm}\int_{-\infty}^0\prod_{i=1}^V\Big[\di z_i\, \ii\aa_i (-z_i)^{p_i}e^{\ii \aa_i z_i}\Big]\prod_{\al=1}^I \wt{D}_{\aa_{i}\aa_{j}}^{(\wt\nu_\al)}(r_{(\al i)}z_{i},r_{(\al j)}z_{j}).
\end{align}
Again, it is understood that the $i,j$ indices appearing in the rescaled bulk propagator $\wt{D}$ should be identified with the labels of $\wt{D}$'s two endpoints. 

To summarize, for a general tree-graphic contribution $\wh{\G}$ to a massive inflation correlator, we can construct a dimensionless graph $\G$, which is related to $\wh{\G}$ by trivial rescaling, and is fully specified by a vertex energy $E_i$ and a twist $p_i$ at each vertex, together with a line energy $K_\al$ and a mass $\wt\nu_\al$ for each internal line. Below, we will work exclusively with the dimensionless graph $\G$, and we will simply call it a graph for short. No confusion could arise since the original graph $\wh{\G}$ is always hatted.

\paragraph{Wavefunction coefficients} The wavefunction coefficients of a theory are defined by expanding the Schrödinger-picture wavefunctional around the Gaussian ansatz. These coefficients have diagrammatic expansions quite similar to correlators, and are also computed by bulk time integrals. Explicitly, for the same diagram as specified above, the corresponding wavefunction coefficient ${\Psi}$ is computed by:
\begin{align}\label{eq_wavefunction}
  {\Psi}(\bm k_1,\cdots,\bm k_N)=\int_{-\infty}^0 \prod_{i=1}^V\Big[\di \tau_i (+\ii)(-\tau_i)^{P_i}e^{\ii E_i\tau_i}\Big]\prod_{\al=1}^I G^{(\wt\nu)}(K_\al;\tau_\al,\tau_\al').
\end{align}
For wavefunction coefficients, the bulk-to-boundary propagator of a conformal scalar is simply $B(k;\tau)=e^{\ii k\tau}$. So, unlike correlators, we don't have to amputate the wavefunction coefficients by removing factors $-\tau_f/(2k_n)$. It is automatically a function of vertex energies $E_i$ and line energies $K_\al$. The bulk propagator $G^{(\wt\nu)}$ for a scalar of mass $\wt\nu$, is given by:
\begin{align}\label{eq_G_wavefunc}
  &G^{(\wt\nu)}(K;\tau_1,\tau_2)= \FR{\pi}{4}e^{-\pi\wt\nu}(\tau_1\tau_2)^{3/2}\bigg[\text{H}_{\ii\wt\nu}^{(1)}(-K\tau_1)\text{H}_{-\ii\wt\nu}^{(2)}(-K\tau_2)\theta(\tau_1-\tau_2)\n\\
  &+\text{H}_{-\ii\wt\nu}^{(2)}(-K\tau_1)\text{H}_{\ii\wt\nu}^{(1)}(-K\tau_2)\theta(\tau_2-\tau_1)-\FR{\text{H}_{\ii\wt\nu}^{(1)}(-K\tau_f)}{\text{H}_{-\ii\wt\nu}^{(2)}(-K\tau_f)}\text{H}_{-\ii\wt\nu}^{(2)}(-K\tau_1)\text{H}_{-\ii\wt\nu}^{(2)}(-K\tau_2) \bigg].
\end{align}
Importantly, the bulk propagator satisfies the same Klein-Gordon equation with $\de$-function source as the Feynman propagator $D_{++}^{(\wt\nu)}$. They are distinguished by the boundary conditions: The Feynman propagator is specified by the Bunch-Davies initial condition with canonical normalization, while the bulk propagator for the wavefunction coefficients is determined by the vanishing boundary condition at the future boundary $G^{(\wt\nu)}(K;\tau_1\to 0,\tau_2)=G^{(\wt\nu)}(K;\tau_1,\tau_2\to 0)=0$. As a result, the system of differential equations satisfied by a tree graph of a wavefunction coefficient is identical to the one for the correlator of the same graph, although the solutions are different in two cases due to different boundary conditions. Therefore, from now on, we will focus on correlators, and the wavefunction coefficients can be treated in a similar fashion.

\section{Differential Equations for Massive Tree Correlators}
\label{sec_DE}
  
In this section, we first present the system of differential equations for arbitrary tree-level massive inflation correlators, and then provide the details of the derivation. From now on, we only focus on correlators, and it will become clear that the same set of equations also applies to wavefunction coefficients.

\subsection{The differential equations}
 
In this subsection, we present the differential equations satisfied by an arbitrary tree graph $\G$ in (\ref{eq_dimlessIntG}). As a preparation, we first introduce the concept of graph contraction.

For an arbitrary graph $\G$, we define its \emph{contraction} with respect to its internal line with label $\al$ to be a new dimensionless graph, denoted as $\mathsf{C}_\al[\G]$. Let the two endpoints of Line $\al$ be Vertices $i$ and $j$, with vertex energies $E_i$, $E_j$, and twists $p_i$ and $p_j$, respectively. Then, the contracted graph $\mathsf{C}_\al[\G]$ is obtained by the following two steps: 
\begin{enumerate}
\item
Remove Line $\al$ in $\G$ and pinch Vertices $i$ and $j$ together as a new effective vertex; 
\item
Assign the vertex energy $E_{ij}$ and twist $p_{ij}+4$ to the new effective vertex.
\end{enumerate}
The operation of contraction is quite intuitive: When pinching the two vertices together, we are not changing the total energies flowing into these two vertices from \emph{external} lines. So the vertex energy of the new vertex is simply the sum of the original two. On the other hand, in a twist parameter $p_i$, there is always a contribution of $-4$ coming from the metric determinant $\sqrt{-g}=a^4(\tau)=\tau^{-4}$. When pinching the two vertices into one, we need to remove one metric determinant and collect the remaining time powers, and this is why we have $p_{ij}+4$ for the new vertex, instead of $p_{ij}$. 

With the above preparation, we can now present the system of differential equations for $\G$. There are $2I$ independent equations, where $I$ is the number of internal lines. (See (\ref{eq_dimlessIntG}).) Each of these $2I$ equations can be specified as follows. First, we pick up an internal line, say Line $\al$ (with line energy $K_\al$ and mass $\wt\nu_\al$), which connects Vertex $i$ (with vertex energy $E_i$ and twist $p_i$) and Vertex $j$ (with $E_j$ and $p_j$). Second, we pick up one vertex (say Vertex $i$) from the two endpoints of Line $\al$. Then, there is an energy ratio $r_{(\al i)}=K_\al/E_i$ as introduced before, and there is a corresponding differential operator $\mathcal{D}_{(\al i)}$ that contracts the graph $\G$ to $\mathsf{C}_\al[\G]$. Explicitly, we have:
\begin{keyeqn}
\begin{align}
\label{eq_DEtG}
  &\mathcal{D}_{(\al i)}\G=\FR{r_{(\al i)}^{p_j+4}r_{(\al j)}^{p_i+4}}{\big[r_{(\al i)}+r_{(\al j)}\big]^{p_{ij}+5}}\mathsf{C}_{\al}[\G],\\
\label{eq_Dri}
  &\mathcal{D}_{(\al i)}\equiv \Big(\vartheta_{(\al i)}-\FR{3}{2}\Big)^2+\wt\nu_\al^2-r_{(\al i)}^2\big(\vartheta_{\{i\}}+p_i+2\big)\big(\vartheta_{\{i\}}+p_i+1\big).
\end{align}
\end{keyeqn}
where $p_{ij}\equiv p_i+p_j$, $\vartheta_{(\al i)}\equiv r_{(\al i)}(\pd/\pd {r_{(\al i)}})$ is the Euler operator associated with the energy ratio $r_{(\al i)}$, and $\vartheta_{\{i\}}$ is the sum of Euler operators for all line energies attached to Vertex $i$:
\bge
\label{eq_varthetai}
  \vartheta_{\{i\}}\equiv \sum_{\be\in\mathcal{N}(i)}\vartheta_{(\be i)}.
\ede
Here we use $\mathcal{N}(i)$ to denote the neighbor of Vertex $i$, namely, the set of all lines attached to $i$.  

The $2I$ differential equations in (\ref{eq_DEtG}) form a complete set of differential equations and can fully determine the graph with appropriate boundary conditions, which we will show in subsequent sections. Incidentally, it is straightforward to show that all $2I$ operators  $\mathcal{D}_{(\al i)}$ defined in (\ref{eq_Dri}) commute with each other:
\begin{align}
  \big[\mathcal{D}_{(\al i)},\mathcal{D}_{(\be j)}\big]=0.
\end{align}

\begin{figure}[t]
\centering
\includegraphics[width=0.95\textwidth]{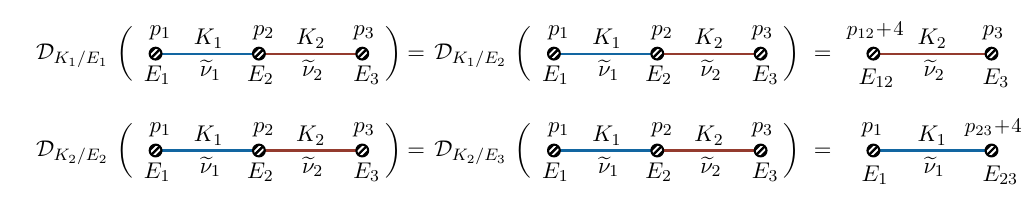}
\caption{The system of differential equations for the 3-site chain graph.}
\label{eq_fd_3chain_DE}
\end{figure}

\paragraph{Example: 3-site chain} Let us spell out all equations with a concrete example. Consider a 3-site chain as shown in Fig.\;\ref{eq_fd_3chain_DE}, which is a tree graph with 3 vertices, 2 bulk massive lines with masses $\wt\nu_1$ and $\wt\nu_2$, and an arbitrary number of external lines. The dimensionless graph is fully specified by 3 twists $p_i$, 2 masses $\wt\nu_\al$ $(\al=1,2)$, and 4 energy ratios $r_1\equiv K_1/E_1$, $r_2\equiv K_1/E_2$, $r_3\equiv K_2/E_2$, and $r_4\equiv K_2/E_3$. We assume that all 4 energy ratios are independent variables. Then, a complete set of differential equations can be written as:
\begin{align}
&\Big[\big(\vartheta_{r_1}-\fr32\big)^2+\wt\nu_1^2-r_1^2 (\vartheta_{r_1}+p_1+2)(\vartheta_{r_1}+p_1+1)\Big]\G_{\wt\nu_1\wt\nu_2}^{p_1p_2p_3}(r_1,\cdots,r_4)\n\\ 
=&\Big[\big(\vartheta_{r_2}-\fr32\big)^2+\wt\nu_1^2-r_2^2(\vartheta_{r_2}+\vartheta_{r_3}+p_2+2)(\vartheta_{r_2}+\vartheta_{r_3}+p_2+1)\Big]\G_{\wt\nu_1\wt\nu_2}^{p_1p_2p_3}(r_1,\cdots,r_4)\n\\
=&~\FR{r_1^{p_2+4}r_2^{p_1+4}}{r_{12}^{p_{12}+5}} \G_{\wt\nu_2}^{(p_{12}+4)p_3}(r_1r_3/r_{12},r_4),\\
&\Big[\big(\vartheta_{r_3}-\fr32\big)^2+\wt\nu_2^2-r_3^2(\vartheta_{r_2}+\vartheta_{r_3}+p_2+2)(\vartheta_{r_2}+\vartheta_{r_3}+p_2+1)\Big]\G_{\wt\nu_1\wt\nu_2}^{p_1p_2p_3}(r_1,\cdots,r_4)\n\\
=&\Big[\big(\vartheta_{r_4}-\fr32\big)^2+\wt\nu_2^2-r_4^2 (\vartheta_{r_4}+p_3+2)(\vartheta_{r_4}+p_3+1)\Big]\G_{\wt\nu_1\wt\nu_2}^{p_1p_2p_3}(r_1,\cdots,r_4)\n\\ 
=&~\FR{r_3^{p_3+4}r_4^{p_2+4}}{r_{34}^{p_{23}+5}} \G_{\wt\nu_1}^{p_1(p_{23}+4)}(r_1,r_4r_2/r_{34}). 
\end{align}
These equations can be graphically expressed as in Fig.\;\ref{eq_fd_3chain_DE}.

\subsection{Deriving the differential equations}

Now we derive the system of differential equations (\ref{eq_DEtG}) satisfied by the dimensionless graph $\G$ in (\ref{eq_dimlessIntG}). The basic strategy is obvious: We insert a Klein-Gordon operator into the integrand of the graph, to annihilate one bulk propagator or to collapse it into a $\delta$ function, so that the whole integral becomes simpler, being either zero or a smaller graph with one less vertex. This simpler graph will be the ``right-hand side'' of our differential equations. For the ``left-hand side,'' we use integration by parts to switch the order of the Klein-Gordon operator and the integral, so that we get a differential operator that directly acts on the whole graph. Equating both sides gives the desired equation. Below we spell out this procedure.

\paragraph{Dimensionless propagator} By definition, the bulk propagator of a massive scalar field is a solution to the Klein-Gordon equation with or without a localized source. In momentum space, this statement takes the following form:
\begin{align}
  \big(\tau_1^2\pd_{\tau_1}^2-2\tau_1\pd_{\tau_1}+K^2\tau_1^2+m^2\big)D_{\aa_1\aa_2}^{(\wt\nu)}(K;\tau_1,\tau_2)=-\ii\aa_1(\tau_1\tau_2)^2 \de(\tau_1-\tau_2)\de_{\aa_1\aa_2},
\end{align}
and there is another similar equation with all $\tau_1\to \tau_2$ in the above Klein-Gordon operator. One can check this equation directly from the explicit expressions in (\ref{eq_Dmp})-(\ref{eq_Dpmpm}). 

Now, in terms of the dimensionless propagator $\wt D_{\aa_1\aa_2}^{(\wt\nu)}(r_1z_1,r_2z_2)$, the above Klein-Gordon equation can be rewritten as:
\begin{align}
\label{eq_eom}
  \big[\vartheta_{z_1}^2-3\vartheta_{z_1}+(r_1^2z_1^2+m^2)\big]\wt D_{\aa_1\aa_2}^{(\wt\nu)}(r_1z_1,r_2z_2)=-\ii\aa_1(r_1z_1r_2z_2)^2 \de(r_1z_1-r_2z_2)\de_{\aa_1\aa_2},
\end{align}
where $\vartheta_{z_1}\equiv z_1\pd_{z_1}$. This will be a key equation in our following derivation of the differential equations for the graph. Incidentally, it is worth noting that the dimensionless propagator $\wt D_{\aa_1\aa_2}^{(\wt\nu)}(r_1z_1,r_2z_2)$ satisfies a trivial yet useful relation:
\begin{align}
\label{eq_zdDidentity}
   \vartheta_{z_1}^n\wt D_{\aa_1\aa_2}^{(\wt\nu)}(r_1z_1,r_2z_2)=&~\vartheta_{r_1}^n\wt D_{\aa_1\aa_2}^{(\wt\nu)}(r_1z_1,r_2z_2) ,
\end{align}
with $n\in\mathbb{N}$. Note in particular that this identity does not depend on the choices of SK indices. 

\begin{figure}[t]
\centering
\includegraphics[width=0.9\textwidth]{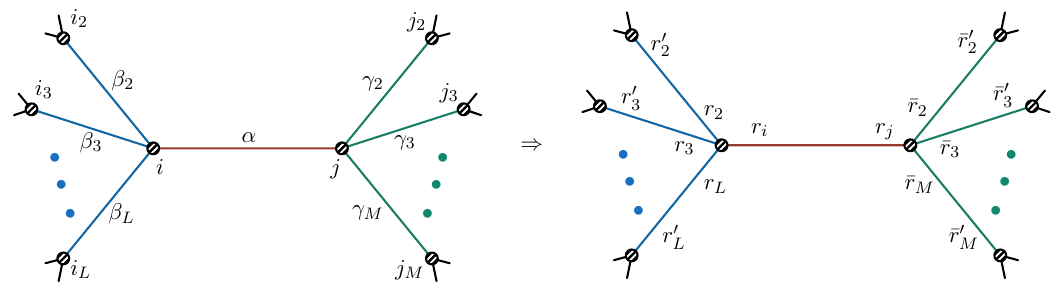}
\caption{Labelling of graph $\G$ for the derivation of its differential equations.}
\label{fig_derivation}
\end{figure}

\paragraph{Differential Operator}
To be specific, we consider an arbitrary internal line, say Line $\al$, which connects two vertices, called Vertices $i$ and $j$, and we want to derive a differential equation associated with the ratio $r_{(\al i)}=K_\al/E_i$. It will be useful to further specify that there are $L\geq 1$ internal lines attached to Vertex $i$ (including Line $\al$), and we label them by $(\al,\be_2,\cdots,\be_{L})$. The other ends of these $L$ internal lines will be labeled by $(j,i_2,\cdots,i_{L})$, respectively. Similarly, we assume that there are $M\geq 1$ internal lines attached to Vertex $j$ (including Line $\al$), and we label them by $(\al,\ga_2,\cdots,\ga_M)$. The other ends of these $M$ internal lines will be labeled by $(i,j_2,\cdots,j_M)$, respectively. To avoid clutter, we use a temporary notation for energy ratios, defined as follows, instead of using $r_{(\al i)}$:
\begin{align}
  &r_i\equiv\FR{K_\al}{E_i},
  &&r_j\equiv\FR{K_\al}{E_j},
  &&r_n\equiv\FR{K_{\be_n}}{E_i},
  &&r_{n}'\equiv\FR{K_{\be_{n}}}{E_{i_{n}}},
  &&\bar r_m=\FR{K_{\ga_m}}{E_j},
  &&\bar r_m'=\FR{K_{\ga_m}}{E_{j_{m}}},
\end{align}
where $n=2,\cdots,L$ and $m=2,\cdots,M$. We show our labeling of the graph in Fig.\;\ref{fig_derivation}.

Now, in the integral (\ref{eq_dimlessIntG}), we can isolate the $z_i$-dependent part as follows:
\begin{align}
\label{eq_GintIsoI}
\G
=&\sum_{\aa_1,\cdots,\aa_V=\pm}\int_{-\infty}^0\prod_{\ell\neq i}\Big[\di z_\ell\, \ii\aa_\ell (-z_\ell)^{p_\ell}e^{\ii\aa_\ell z_\ell}\Big]\prod_{\be \notin\mathcal{N}(i)} \wt{D}_{\aa_{\ell} \aa_{n}}^{(\wt\nu_\be)}(r_{\ell}z_{\ell}, r_{n} z_{n})\int_{-\infty}^0\di z_i\,\mb A_i,
\end{align}
For brevity, we suppress all indices and arguments of $\G$ here and below. In (\ref{eq_GintIsoI}), we introduced $\mb A_i$ as the integrand of the $z_i$-integral:
\bge
 \mb{A}_i\equiv  \ii \aa_i (-z_i)^{p_i}e^{\ii\aa_i z_i}{\color{BrickRed} \wt D_{\aa_i\aa_{j}}^{(\wt\nu_{\al})}\big(r_{i}z_i,r_{j}z_{j}\big)}{\color{RoyalBlue}\prod_{n=2}^{L}\wt D_{\aa_i\aa_{i_n}}^{(\wt\nu_{\be_n})}\big(r_{n}z_i,r_{n}'z_{i_n}\big)}.
\ede
Here and following, we use the same color code as lines in Fig.\;\ref{fig_derivation} for the bulk propagators. To derive the differential equation of $\G$ with respect to $r_{(\al i)}=r_i$, we consider the following expression:
\bge
\label{eq_intDA}
   \int_{-\infty}^0\di z_i\,{\mb D}_\al\mb{A}_i,
\ede
where $\mb D_{\al}\mb{A}_i$ is the integrand of $z_i$ with an additional Klein-Gordon operator acting on Line $\al$ from the side of Vertex $i$:
\begin{align}
\label{eq_DA}
  {\mb D}_\al\mb{A}_i
  \equiv&~  \ii \aa_i (-z_i)^{p_i}e^{\ii\aa_i z_i}
  {\color{RoyalBlue}\prod_{n=2}^{L}\wt D_{\aa_i\aa_{i_n}}^{(\wt\nu_{\be_n})}\big(r_{n}z_i,r_{n}'z_{i_n}\big)}\n\\
  &\times \Big\{\Big[\vartheta_{z_i}^2-3\vartheta_{z_i}+\big(r_{i}^2z_i^2+m_\al^2\big)\Big]{\color{BrickRed}\wt D_{\aa_i\aa_j}^{(\wt\nu_\al)}\big(r_{i}z_i, r_{j}z_j\big)}\Big\}.
\end{align}
To get the left-hand side of the differential equation, we want to pull the Klein-Gordon operator in (\ref{eq_DA}) out of the integral. This is a trivial task for $\vartheta_{z_i}^2-3\vartheta_{z_i}$ thanks to the relations such as (\ref{eq_zdDidentity}). The mass term $\propto m_\al^2$ is trivial as well. Thus, the only trouble is from the term $\propto r_i^2z_i^2$. To deal with it, we use the following relation:
\begin{align}
\label{eq_z2A}
  \int_{-\infty}^0\di z_i\,z_i^2\mb A_i=-\big(\vartheta_{\{i\}}+p_i+2\big)\big(\vartheta_{\{i\}}+p_i+1\big)\int_{-\infty}^0\di z_i\,\mb A_i,
\end{align}
where $\vartheta_{\{i\}}$ is the Euler operator associated with all line energies attached to Vertex $i$, as defined in (\ref{eq_varthetai}). This relation is easily derived using integration by parts, as shown in the box below. 

\begin{boxedtext}
\noindent\textbf{Derivation of (\ref{eq_z2A})~~~} We abbreviate the propagators by:
\begin{align}
  &\wt D_{\al}\equiv \wt D_{\aa_i\aa_j}^{(\wt\nu_\al)}\big(r_{i}z_i, r_{j}z_j\big),
    &&\wt D_{\be_n}\equiv \wt D_{\aa_i\aa_{i_n}}^{(\wt\nu_{\be_n})}\big(r_{n}z_i,r_{n}'z_{i_n}\big).~~~~(n=2,\cdots,L)
\end{align}
Then, we begin with the following equality:
\begin{align}
  0=&\int_{-\infty}^0\di z_i\, \pd_{z_i}\bigg[-(-z_i)^{p_i+1}e^{\ii\aa_iz_i} \wt D_{\al}\prod_{n=2}^L\wt D_{\be_n}\bigg]\n\\
  =&\int_{-\infty}^0\di z_i\,(-z_i)^{p_i}e^{\ii\aa_iz_i}\Big[(p_i+1)+\ii\aa_iz_i+z_i\pd_{z_i} \Big]\wt D_{\al}\prod_{n=2}^L \wt D_{\be_n} \n\\
  =&\int_{-\infty}^0\di z_i\,(-z_i)^{p_i}e^{\ii\aa_iz_i}\bigg[(p_i+1)+\ii\aa_iz_i+\vartheta_{r_i}+\sum_{n=2}^L \vartheta_{r_n} \bigg]\wt D_{\al}\prod_{n=2}^L\wt D_{\be_n}.
\end{align}
In the last line, we have used the relation (\ref{eq_zdDidentity}). Also, the operator $\vartheta_{r_i}+\sum\limits_{n=2}^L \vartheta_{r_{n}}$ is nothing but $\vartheta_{\{i\}}$ as defined in (\ref{eq_varthetai}). Thus we get:
\begin{align}
\label{eq_intzA}
  \int_{-\infty}^0\di z_i\, z_i\mb A_i=-\FR{1}{\ii\aa_i}\big(\vartheta_{\{i\}}+p_i+1\big)\int_{-\infty}^0\di z_i\,\mb A_i.
\end{align}
Likewise, we consider the following the identity:
\begin{align}
  0=&-\int_{-\infty}^0\di z_i\, \pd_{z_i}\bigg[ (-z_i)^{p_i+2}e^{\ii\aa_iz_i} \wt D_{\al}\prod_{n=2}^L\wt D_{\be_n}\bigg]\n\\
  =&\int_{-\infty}^0\di z_i\, (-z_i)^{p_i+1}e^{\ii\aa_iz_i}\Big[(p_i+2)+\ii\aa_iz_i+\vartheta_{\{i\}}\Big]\wt D_{\al}\prod_{n=2}^L\wt D_{\be_n}.
\end{align}
So we get:
\begin{align}
\label{eq_intzzA}
  \int_{-\infty}^0\di z_i\, z_i^2\mb A_i=-\FR{1}{\ii\aa_i}\big(\vartheta_{\{i\}}+p_i+2\big)\int_{-\infty}^0\di z_i\,z_i\mb A_i.
\end{align}
Combining (\ref{eq_intzA}) and (\ref{eq_intzzA}), we get (\ref{eq_z2A}).
\end{boxedtext}

Then, using (\ref{eq_zdDidentity}) and (\ref{eq_z2A}), we can rewrite the expression in (\ref{eq_intDA}) as:
\begin{align}
  \int_{-\infty}^0\di z_i\,{\mb D}_\al\mb{A}_i=&~\Big[\vartheta_{r_i}^2-3\vartheta_{r_i}-r_{i}^2\big(\vartheta_{\{i\}}+p_i+2\big)\big(\vartheta_{\{i\}}+p_i+1\big)+m_\al^2\Big]\int_{-\infty}^0\di z_i\,\mb A_i.
\end{align}
Here $\vartheta_{r_i}\equiv r_{i}\pd_{r_{i}}$. Obviously, all these differential operators trivially penetrate all other integrals and summations in (\ref{eq_GintIsoI}), so that the left-hand side of our differential equation will be:
\bge
  \Big[\Big(\vartheta_{(\al i)}-\FR32\Big)^2+\wt\nu_\al^2-r_{(\al i)}^2\big(\vartheta_{\{i\}}+p_i+2\big)\big(\vartheta_{\{i\}}+p_i+1\big)\Big]\G,
\ede
where we have restored the previous notation $\vartheta_{r_i}=\vartheta_{(\al i)}$ and $r_i=r_{(\al i)}$, completed the square of $\vartheta_{(\al i)}$, and used the relation $\wt\nu_\al^2=m_\al^2-9/4$.

\paragraph{Source term}
For the right-hand side, we use Klein-Gordon equation (\ref{eq_eom}) to evaluate (\ref{eq_intDA}), finish the sum over SK index $\aa_i$, and get:
\begin{align}
\label{eq_sumDAresult}
 \sum_{\aa_i=\pm}\int_{-\infty}^0\di z_i\,{\mb D}_\al \mb{A}_i
  =&\sum_{\aa_i=\pm}\int_{-\infty}^0\di z_i\, \ii \aa_i (-z_i)^{p_i}e^{\ii\aa_i z_i}{\,\color{RoyalBlue}\prod_{n=2}^{L}\wt D_{\aa_i\aa_{i_n}}^{(\wt\nu_{\be_n})}\big(r_{n}z_i,r_{n}'z_{i_n}\big)}\n\\
  &\times (-\ii\aa_i)\big[r_{i}z_i r_{j}z_j\big]^2\de\big(r_{i}z_i- r_{j}z_j\big)\de_{\aa_i\aa_j}\n\\
  =&~ \FR{ r_{j}^{p_i+4}}{r_{i}^{p_i+1}}(-z_j)^{p_i+4}e^{\ii\aa_j (r_{j}/r_{i}) z_j}{\,\color{RoyalBlue} \prod_{n=2}^{L}\wt D_{\aa_j\aa_{i_n}}^{(\wt\nu_{\be_n})}\Big(\FR{r_{j}r_{n}}{r_{i}}z_j,r_{n}'z_{i_n}\Big)}.
\end{align}
To see what it gives us, let us now consider the $z_j$-integral in the original graph (\ref{eq_GintIsoI}):
\begin{align}
\label{eq_zjint}
  &\int_{-\infty}^0\di z_j(\ii\aa_j)(-z_j)^{p_j}e^{\ii\aa_jz_j}{\,\color{PineGreen}\prod_{\ell=2}^M \wt D_{\aa_j\aa_{j_\ell}}^{(\wt\nu_{\ga_\ell})}\big(\bar r_{\ell}z_j,\bar r_{\ell}'z_{j_\ell}\big)}\n\\
  &\times \FR{r_{j}^{p_i+4}}{r_{i}^{p_i+1}}(-z_j)^{p_i+4}e^{\ii\aa_j (r_{j}/r_{i}) z_j}{\,\color{RoyalBlue} \prod_{n=2}^{L}\wt D_{\aa_j\aa_{i_n}}^{(\wt\nu_{\be_n})}\Big(\FR{r_{j}r_{n}}{r_{i}}z_j,r_{n}'z_{i_n}\Big)}.
\end{align}
As shown here, the $z_j$-integrand includes two parts: In the first line, we have factors that come from the original integrand of the graph (\ref{eq_GintIsoI}), and, in the second line, we have contributions from evaluating (\ref{eq_sumDAresult}). 

The $z_j$-integrand in (\ref{eq_zjint}) almost has the form of the integrand for a standard graph, except that the exponent in the exponential factor is not properly ``normalized.'' To normalize this exponent, we change the integral variable $z_j$ to $\tilde z_j\equiv (1+ r_j/r_i)z_j$ (and therefore $z_j=\tilde z_j\times r_i/r_{ij}$, where $r_{ij}=r_i+r_j$). Then, the integral in (\ref{eq_zjint}) becomes:
\begin{align}
\label{eq_zjIntContracted}
  & \FR{r_i^{p_j+4} r_j^{p_{i}+4}}{r_{ij}^{p_{ij}+5}} \int_{-\infty}^0\di \tilde z_j\,(\ii\aa_j)(-\tilde z_j)^{p_{ij}+4}e^{\ii\aa_j\tilde z_j}{\,\color{PineGreen}\prod_{\ell=2}^M \wt D_{\aa_j\aa_{j_\ell}}^{(\wt\nu_{\ga_\ell})}\Big(\FR{r_{i}\bar r_{\ell}}{r_{ij}}\tilde z_j,\bar r_\ell'z_{j_\ell}\Big)} 
   {\,\color{RoyalBlue}\prod_{n=2}^L \wt D_{\aa_j\aa_{i_n}}^{(\wt\nu_{\be_n})}\Big(\FR{r_{j}r_{n}}{r_{ij}}\tilde z_j,r_{n}'z_{i_n}\Big)}.
\end{align}
Now, including all remaining time integrals and summations in (\ref{eq_GintIsoI}), we get the final expression for the right-hand side of our differential equation. These all look rather complicated. Let us spell out their meanings in the following.
\begin{enumerate}
  \item After the $z_i$-integral in (\ref{eq_sumDAresult}), Vertices $i$ and $j$ are effectively pinched together into one vertex (which is still called Vertex $j$). The new Vertex $j$ now has its own time integral (\ref{eq_zjIntContracted}). Clearly, the new Vertex $j$ connects $L+M-2$ bulk lines.  
  
  \item To see the meaning of the complicated $r$-factors in (\ref{eq_zjIntContracted}), we restore the energy variables. First, the coefficients for the first arguments of bulk propagators can be rewritten as:
\begin{align}
   &\FR{r_i}{r_{ij}}\times \bar r_\ell= \FR{E_j}{E_{ij}}\times \FR{K_{\ga_\ell}}{E_j}=\FR{K_{\ga_\ell}}{E_{ij}}, 
   &&\FR{r_j}{r_{ij}}\times r_{n}= \FR{E_i}{E_{ij}}\times \FR{K_{\be_n}}{E_i}=\FR{K_{\be_n}}{E_{ij}}.
\end{align}
The meaning is clear: These factors amount to giving a vertex energy $E_{ij}=E_i+E_j$ to the new Vertex $j$, replacing original vertex energies ($E_j$ at Vertex $j$ and $E_i$ at Vertex $i$).
  
  \item The factor $(-\tilde z_j)^{p_{ij}+4}$ tells us that the twist of the new Vertex $j$ is $p_{ij}+4$. As mentioned before, it simply means that we collect all time powers from the couplings at both ends of the pinched line, and include a term $+4$ to remove one metric determinant $\sqrt{-g}=a^4=\tau^{-4}$. 
  
  \item Finally, the prefactor in front of the integral in (\ref{eq_zjIntContracted}) can be rewritten as:
  \begin{align}
  \label{eq_rfactor}
  \FR{r_{i}^{p_j+4} r_{j}^{p_{i}+4}}{r_{ij}^{p_{ij}+5}}\equiv \Big(\FR{K_\al}{E_{ij}}\Big)^3\Big(\FR{E_i}{E_{ij}}\Big)^{p_i+1}\Big(\FR{E_j}{E_{ij}}\Big)^{p_j+1}.
\end{align}
  It is now easy to recognize that this prefactor comes from our definition of the dimensionless graph (\ref{eq_DimlessGraph}). 
\end{enumerate}

Combining all these points, we see that the right-hand side of the equation is nothing but the factor (\ref{eq_rfactor}) multiplied by $\mathsf{C}_\al[\G]$, namely, the contraction of the original graph $\G$ with respect to Line $\al$. This completes the derivation of the differential equation with respect to $r_{(\al i)}$. Clearly, we can repeat this procedure for all $2I$ independent energy ratios, which leads to $2I$ independent differential equations in (\ref{eq_DEtG}).

\subsection{Strategy of solving the differential equation system} 
\label{sec_strategy}

Before presenting our construction of solutions, let us briefly describe the general structure of the solution. These solutions are multivariate hypergeometric functions that are not very well understood. Therefore, our strategy is to find series expansions in parameter regions of interest. Clearly, one specific series expansion is convergent only in a finite region. It is thus important to know how to expand the solution in different regions.

It is beyond the scope of this work to develop expansions of these hypergeometric functions at all regions, which is likely a nontrivial mathematical problem. Instead, we will focus on regions that are particularly useful and telling for physical applications. Explicitly, we will consider regions where all line energies $K_\al$ can become arbitrarily soft compared to (some of) vertex energies $E_i$, which we call the \emph{signal reigion}. 

It is known that $K_\al\to 0$ limit for any given $\al$ (with all other energies staying in the interior of the physically accessible region) is typically a branch point of the correlator, which we call the \emph{signal branch point}. At this point, the correlator breaks into three pieces, one nonanalytic in $K_\al$, one analytic in $K_\al$ but nonanalytic in some other vertex energies, and one analytic in all energies. Using the terminology of cosmological collider physics, they are respectively called the nonlocal signal, the local signal, and the background with respect to the $K_\al\to 0$ limit.

From the viewpoint of bulk time integrals, the signals (both nonlocal and local) are from the parts of the $K_\al$-propagators that are independent of time-ordering $\theta$-functions, and are thus explicitly factorized. On the contrary, the background comes totally from the nested part of the $K_\al$-propagator. Technically, the signals are relatively easier to compute since they do not involve any nested time integral, and the backgrounds are hard. 

From the viewpoint of differential equations derived above, the signals associated with the $K_\al$-propagator correspond to the pieces of the correlator that are annihilated by the two operators $\mathcal{D}_{(\al i)}$ and $\mathcal{D}_{(\al j)}$ (assuming $K_\al$ line is attached to the two vertices $i$ and $j$), and are thus the homogeneous solutions to the $\mathcal{D}_{(\al i)}$- and $\mathcal{D}_{(\al j)}$-equations. As often happens, the determination of these homogeneous solutions requires appropriate boundary conditions. On the other hand, the background corresponds to a piece that is not annihilated, but contracted by $\mathcal{D}_{(\al i)}$ and $\mathcal{D}_{(\al j)}$. Thus, the background corresponds to the inhomogeneous solution to the $\mathcal{D}_{(\al i)}$- and $\mathcal{D}_{(\al j)}$-equations.

The above descriptions apply to any of the internal lines in a graph. Altogether, when we think of a massive tree graph in the region where all line energies are small, the whole graph exhibits a nice pattern, which can be schematically expressed as:
\bge
\label{eq_Gpattern}
  \G=\sum_{c_1=\,\text{S,B}}\cdots \sum_{c_I=\,\text{S,B}}\G[c_1,\cdots,c_I].
\ede
Here, for each internal line $K_\al$ ($\al=1,\cdots,I$), we introduce a parameter $c_\al$, which takes two possible values S and B, respectively corresponding to the signal/homogeneous and the background/inhomogeneous part with respect to $K_\al$. As mentioned above, every signal can be further decomposed into a nonlocal signal and a local signal. Therefore, a tree graph with $I$ internal lines can be decomposed into $3^I$ distinct pieces according to its analytical behavior around all signal branch points. 

Therefore, to find the complete solution for a massive tree graph in the signal region, we only need to find explicit expressions for all terms $\G[c_1,\cdots,c_I]$. As mentioned, the more lines taking $c_\al=\;$B, the harder to get the solution. The hardest part is thus the piece with all lines nested. We will call it the completely inhomogeneous solution (CIS). At the other extreme, the piece with all lines factorized is pretty straightforward. As we shall see below, they are simply given by products of Lauricella hypergeometric functions. Of course, there are many things in between.

Our strategy is to tackle the hardest part first, namely, to find the CIS of an arbitrary graph with all $c_\al=\;$B in (\ref{eq_Gpattern}). Then, we will show that, changing a $c_\al$ from B to S simply amounts to taking an appropriate cut of CIS. As we will show, taking a single cut to CIS breaks the CIS into a product of two dressed CISs for two subgraphs. The multiple cuts can be executed in a similar way. Thus, all terms in (\ref{eq_Gpattern}) can be expressed as a sum of products of (dressed) CISs of the whole graph and its subgraphs. Therefore, as long as we know how to write down the CIS and how to do the cut (namely, how to dress the CIS of a subgraph), we can immediately get the full solution, in the following form:
\begin{align}
  \G=\sum_{\text{cuts}}\text{CIS}\;[\G],
\end{align}
where we sum over $2^I$ possible cuts of the CIS, including the zero cut. We will achieve this result step by step in the following three sections.  

It will be useful to have an idea of the form of the final solution before we solve for them explicitly. Schematically, the series expansion of the CIS of a graph has the following form:
\begin{align}
\label{eq_CISstructure}
  \text{CIS}\,\big[\G\big]=\sum_{\{\ell,m\}}f\big(\{\ell,m\}\big)\prod_{i=2}^V \Big(\FR{K_i}{E_1}\Big)^{2m_i+3}\Big(\FR{E_i}{E_1}\Big)^{\ell_i+p_i+1}.
\end{align}
That is, we are expanding the CIS as a power series of $1/E_1$ (\footnote{Like the original family tree integrals in \cite{Xianyu:2023ytd,Fan:2024iek}, expanding in a large vertex energy limit is only one of many possible expansion schemes for the CIS. It would be interesting to find more of them. We leave this topic for a future work.}). The series is going to converge if all energy ratios $K_\al/E_1$ and $E_i/E_1$ are small. Therefore, in practice, we need to pick up the largest vertex energy, call it $E_1$, and apply the above expansion. In this series, we have $2(V-1)=2I$ summation variables $\ell_i,m_i$ $(i=2,\cdots,V)$, respectively collecting powers of all $I$ line energies $K_i$ and $I$ remaining vertex energies $E_i$. Notice that we have used the same subscript $i$ to label the line and non-maximal vertex energies, because, after picking up the maximal energy $E_1$, the remaining $V-1$ vertex energies have one-to-one correspondence with the $I=V-1$ line energies. We will see this relation more explicitly below.

The structure of the series (\ref{eq_CISstructure}) tells us that, when writing down the CIS of a graph, all we need to decide is the maximal vertex energy. The relative sizes of other energies are irrelevant. However, this is going to change if we cut the CIS. For one thing, after cutting a diagram open, the subgraph that does not contain the maximal vertex energy $E_1$ is clearly ignorant about $E_1$, and thus we have to pick up the (locally) maximal vertex energy \emph{within} this subgraph to expand its CIS. More importantly, as we shall see, the cut of a graph is \emph{directional}, in the sense that the factorized propagator is not symmetric with respect to the exchange of two endpoint energies. The direction is determined by comparing the two locally maximal vertex energies of the two subgraphs separated by the cut line. Thus, as we execute more cuts, we need more information about the ordering of vertex energies. In the end, when we cut all lines, we need to decide the relative sizes of vertex energies for the two endpoints of every line. 

Therefore, our strategy is to choose a particular ordering for all vertex energies. Then, the structure of the CIS and the directions of all cuts will be determined, as will be detailed below. On the other hand, we do not require any orderings among line energies, but only require that they can be sent to zero so that we can access the signal branch point. For convenience of later presentation, we will make a working assumption that any line energy $K_\al$ is smaller than its two endpoint vertex energies $E_i$ and $E_j$. However, this assumption can be loosened in the final results. We will come back to this point later.

\subsection{Relation to the Lauricella system}
\label{sec_lauricella}

It is curious to observe that the differential equations for arbitrary massive trees (\ref{eq_DEtG}) belong to a coupled Lauricella system \cite{Slater:1966,Matsumoto_2020}. We show this connection explicitly here. This subsection is independent of the rest of the paper, and uninterested readers can safely skip it.

Let us switch off the source term in (\ref{eq_DEtG}) for the moment. As shown above, the solutions to these homogeneous equations correspond to the maximal-cut piece. From the viewpoint of bulk time integrals, this means that the completely homogeneous solutions involve no nested time integrals. As a result, we only need to consider a single time integral for each vertex, and the final result will be a product of these single-vertex integrals. In this sense, the completely homogeneous solution is the simplest part of the whole correlator. 

The single time integral can be performed directly. Let's consider the most general case where a single vertex with dimensionless time $z_i=E_i\tau_i$ is attached to $N$ internal lines with line energy $K_\al$ and mass $\wt\nu_\al$ ($\al=1,\cdots,N$). After cutting all these lines, every line contributes to the $z_i$-integral by a massive mode function: (See (\ref{eq_Dmp})-(\ref{eq_Dpmpm}).) 
\bge
\label{eq_massivemode}
  \si_{\cc_\al}(r_\al z_i)\propto (-r_\al z_i)^{3/2}\text{H}_{\ii \cc_\al \wt\nu_\al}^{(j_\al)}(-r_\al z_i)
\ede
in which we have two choices $j_\al=1,2$ for every line. Once we make a choice, it follows that $\cc_\al=-(-1)^{j_\al}$. Also, $r_\al\equiv K_\al/E_i$. ($\al=1,\cdots,N$). Then, for example, if we choose all $j_\al=+1$, then the corresponding $z_i$-integral has the following form:
\begin{align}
\label{eq_CHS}
  \mathcal{H}(r_1,\cdots,r_N)= \mathcal{C}\int_{-\infty}^0\di z_i\,(-z_i)^{p_i}e^{\ii\aa_i z_i}\prod_{\al=1}^N\Big[(-r_\al z_i)^{3/2}\text{H}_{\ii\wt\nu_\ell}^{(1)}(-r_\al z_i)\Big],
\end{align}
where we have included a constant coefficient $\mathcal{C}$ that is not of interest here. 

The integral (\ref{eq_CHS}) can be directly done by the partial Mellin-Barnes representation \cite{Qin:2022lva,Qin:2022fbv}. That is, we use the MB representation to rewrite all the massive mode functions (\ref{eq_massivemode}). Taking $\cc_\al=+$ in (\ref{eq_massivemode}) as an example, we have: 
\begin{align}
  \si_{+}(r_\al z_i)\propto\int_{-\ii\infty}^{+\ii\infty} \FR{\di s_\al}{2\pi\ii}\FR{e^{(2s_\al-\ii\wt\nu_\al-1)\pi\ii/2}}{\pi} \Gamma\Big[s_\al+\FR{\ii\wt\nu_{\al}}{2},s_\al-\FR{\ii\wt\nu_{\al}}{2}\Big]\Big(\FR{-r_{\al}z_i}{2}\Big)^{-2s_\al}(-r_\al z_i)^{3/2} .
\end{align}
Then the time integral (\ref{eq_CHS}) can be easily performed, and the result is:
\begin{align}
\label{eq_LauricellaMB}
  \mathcal{H}(r_1,\cdots,r_N)= &~\mathcal{C}\int_{-\ii\infty}^{+\ii\infty}\prod_{\al=1}^N\bigg\{\FR{\di s_\al}{2\pi\ii}\FR{e^{(2s_\al-\ii\wt\nu_\al-1)\pi\ii/2}}{\pi} \Gamma\Big[s_\al+\FR{\ii\wt\nu_{\al}}{2},s_\al-\FR{\ii\wt\nu_{\al}}{2}\Big]\Big(\FR{r_\al}{2}\Big)^{-2s_\al}r_\al^{3/2}\bigg\}\n\\
  &\times \Gamma(1+p_i+3N/2-2s_{1\cdots N})(\ii \aa_i)^{-1-p_i-3N/2+2s_{1\cdots N}},
\end{align}
Finally, the Mellin integral over all $s_1,\cdots,s_N$ can be finished by collecting the left poles, namely, setting $s_\al = -n_\al-\ii\cc_\al \wt\nu_\al/2$ with $\cc_\al =\pm$, and we get a multivariate hypergeometric series:
\begin{align}
  \mathcal{H}(r_1,\cdots,r_N)= &~\mathcal{C}\sum_{\cc_1,\cdots,\cc_N=\pm}\sum_{n_1,\cdots, n_N=0}^\infty \Gamma\Big(1+p_i+3N/2+2n_{1\cdots N}+\ii \wt\nu_\text{T}\Big)(\ii\aa_i)^{-1-p_i-3N/2-\ii \nu_\text{T}} \n\\
  &\times \prod_{\al=1}^N \FR{1}{n_\al!}\FR{\ii e^{ (1+\cc_\al)\wt\nu_\al \pi /2}}{\sin(\ii\pi\cc_\al\wt\nu_\al)\Gamma(1+\ii\cc_\al\wt\nu_\al+n_\al)}\Big(\FR{r_\al}2\Big)^{2n_\al+\ii\cc_\al\wt\nu_\al}r_\al^{3/2}. 
\end{align}
Here $\wt\nu_\text{T}\equiv \sum\limits_{\ell=1}^N \cc_\al\wt\nu_\al$. The choices of $\cc_\al=\pm$ ($\al=1,\cdots,N$) correspond to two leading falloffs of the scalar mode $\si_{\cc_\al}$ in the soft limit. We can recognize that all choices lead to Lauricella's $\text{F}_\text{C}$ function. For instance, if we choose all $\cc_\al=+1$. Then the above series sums to the following function up to an overall constant factor:
\begin{align}
  \text{F}_\text{C}\left[\bgm \fr{p_i+1}2+\fr{3N}4+\fr{\ii\wt\nu_T}2,\fr{p_i+2}2+\fr{3N}4+\fr{\ii\wt\nu_T}2 \\ 1+\ii\wt\nu_1,\cdots,1+\ii\wt\nu_N \edm\middle|r_1^2 ,\cdots, r_N^2\right]r_1^{3/2+\ii\wt\nu_1}\cdots r_N^{3/2+\ii\wt\nu_N}. 
\end{align}
From this result, it is straightforward to see that the operators $\mathcal{D}_{(\al i)}$ in our differential equation systems all belong to the Lauricella class \cite{Matsumoto_2020} upon the change of variables $r_i^2\to u_i$. Thus, the differential equation systems can be viewed as a coupled Lauricella system.

\section{Building Inhomogeneous Solution: Recursion Formulae}
\label{sec_recursion}

From this section onwards, we are going to solve the differential equation system for an arbitrary massive tree graph in the ``signal region,'' i.e., when all line energies are small compared to at least some of the vertex energies. As detailed in Sec.\;\ref{sec_strategy}, our strategy will be to solve the ``most difficult'' problem first, by working out the completely inhomogeneous solution (CIS). For this purpose, a very useful tool is a set of recursion formulae that allow us to directly build a series representation for the inhomogeneous solution with respect to a given line, in terms of the solution for a subgraph with the line removed. With these recursion formulae,  we can build solutions for arbitrary trees from a 1-site graph simply by adding lines. In this section, we are going to derive these recursion formulae.

\subsection{Ingoing recursion formula}

In an arbitrary tree graph, we call an internal line a ``leaf,'' if this internal line is attached to a vertex not shared by any other internal lines. Then, it is a trivial observation that any tree diagram can be reduced to a 1-site graph by repetitively removing its ``leaves.''

\begin{figure}[t]
\centering
\includegraphics[width=0.32\textwidth]{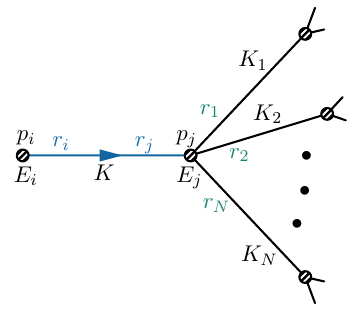}
\hspace{15mm}
\includegraphics[width=0.32\textwidth]{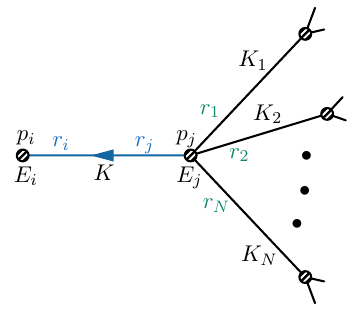}
\caption{Adding a new internal line (blue) with line energy $K$ to an existing graph at Vertex $j$. The ingoing (left plot) and outgoing (right plot) lines correspond to the energy orderings $E_i>E_j$ and $E_i<E_j$, respectively.}
\label{fig_recursion}
\end{figure}
Suppose we have known a series expression for a $V$-site tree graph $\G_{V}$, and we are going to find the series expression for a $(V+1)$-site graph $\G_{V+1}$, which is obtained by adding a leaf to $\G_{V}$. To be specific, we show a general situation in Fig.\;\ref{fig_recursion}, in which we add a new (blue) line, with mass $\wt\nu$ and energy $K$, to Vertex $j$ of an existing graph $\G_V$. The two endpoints of the line are respectively attached to Vertices $i$ and $j$, carrying vertex energies $E_i$ and $E_j$, and twists $p_i$ and $p_j$, respectively. As explained in Sec.\;\ref{sec_strategy}, it is important to distinguish between two cases with $E_i>E_j$ and $E_i<E_j$, corresponding to the ingoing and outgoing lines in Fig.\;\ref{fig_recursion}. We consider the ingoing case in this subsection and the outgoing case in the next. In both cases, we will assume $K<\min\{E_i,E_j\}$. While we only consider small $K_\al$ regions in this work, it is possible to derive recursion formulae for large $K_\al$, which we will consider in a future work.

Another important point is that, before we add the leaf $K$ to $\G_{V}$, we take the twist parameter of Vertex $j$ to be $p_{ij}+4$, so that $\G$ can be treated as a valid source term of the differential equation (\ref{eq_DEtG}).

In the most general situation, Vertex $j$ of $\G_V$ is connected to $N\geq 0$ lines with line energies $K_k$ ($k=1,\cdots,N$). For convenience, we introduce the following notations for the momentum ratios, which are also shown in Fig.\;\ref{fig_recursion}:
\begin{align}
\label{eq_rratios}
  &r_i=\FR{K}{E_i},
  &&r_j=\FR{K}{E_j},
  &&r_k=\FR{K_k}{E_j}.~~~(k=1,\cdots, N)
\end{align}
As assumed, we have an analytical expression for $\G$ as series expansions in $2V-2$ independent energy ratios, including $r_1,\cdots,r_N$. The full series expression for $\G_{V}$ is irrelevant to us. All we need is the dependence on $r_k$ $(k=1,\cdots,N)$ in this series. Without loss of generality, let us parameterize the series expression for $\G_V$ (with the twist of Vertex $j$ being $p_{ij}+4$) as:
\begin{align}
\label{eq_tGseries}
  \G_{V}\big|_{p_j\to p_{ij}+4} = \sum_{\{n\}} c_{\{n\}}r_1^{q_1}\cdots r_N^{q_N}\times \cdots.
\end{align}
Here $\{n\}$ is a collection of summation variables $n_1,\cdots,n_{2V-2}$, taking values from all nonnegative integers, and $c_{\{n\}}$ is a condensed notation for $c_{n_1\cdots n_{2V-2}}$. We spelled out the dependences on $r_1,\cdots,r_N$, and dependences on all other $r$'s are suppressed. In general, the exponents $q_1,\cdots, q_N$ are simple functions of $n_1,\cdots, n_{2V-2}$ (\footnote{\label{fn_qi} More precisely, from the form of the CIS in (\ref{eq_CISstructure}), we see that $q_i$ has the form $q_i=a_{ij}n_j+b_j$, with $a_{ij}\in\{0,\pm 1,\pm 2\}$, and $b_j\in \mathbb{C}$. In addition, for a fixed $i$, there are at most two $a_{ij}$'s nonvanishing. The presence of $b_j$ takes account of possible dependences on the twist or on the mass that arises from the cut. }). Also, the graph $\G_{V}$ is generally a sum of many pieces, each of which has a series expansion in the form of (\ref{eq_tGseries}). Since our recursive algorithm is fully distributive over the summation of these pieces, it suffices to consider one piece at a time.

The strategy for finding $\G_{V+1}$ is to solve its differential equation with respect to $r_i$, which reads:
\begin{align}
\label{eq_Dr1tG}
  &~\Big[\big(\vartheta_i-\fr32\big)^2+\wt\nu^2-r_i^2(\vartheta_i+p_i+2)(\vartheta_i+p_i+1)\Big]\G_{V+1}
  =\FR{r_{i}^{p_j+4}r_{j}^{p_i+4}}{r_{ij}^{p_{ij}+5}}\G_{V}\big|_{E_j\to E_{ij},p_{j}\to p_{ij}+4}\n\\
  =&~\FR{r_{i}^{p_j+4}r_{j}^{p_i+4}}{r_{ij}^{p_{ij}+5}}\sum_{\{n\}} c_{\{n\}}\Big(\FR{r_ir_1}{r_{ij}}\Big)^{q_1}\cdots \Big(\FR{r_ir_N}{r_{ij}}\Big)^{q_N} \times \cdots
\end{align}
Here we have used the fact that shifting $E_j\to E_{ij}$ amounts to multiplying every $r_k$ $(k=1,\cdots,k_N)$ by a factor of $r_i/r_{ij}$.

The equation in (\ref{eq_Dr1tG}) is a standard second-order ordinary differential equation with a source term. So its solution is given by the sum of an inhomogeneous part and a homogeneous part. The inhomogeneous part is solved with the source term, while the homogeneous part is in general a linear combination of two independent solutions, with the coefficients determined by appropriate boundary conditions. We will find the inhomogeneous solution below, and leave the homogeneous solution to Sec.\;\ref{sec_hom}. 

\paragraph{Expansion in small $r_i$} The presence of $r_{ij}$ in (\ref{eq_Dr1tG}) makes the source term troublesome. So, we Taylor expand it. The way to expand it depends on the form of the final series. Since we are interested in the region $E_i>E_j$, namely, $r_i$ being small, we should look for an inhomogeneous solution expanded in $r_i$ and $r_i/r_j$. So, we expand $r_{ij}$ factors in (\ref{eq_Dr1tG}) in the following way:
\begin{align}
  \FR{1}{r_{ij}^{p_{ij}+q_{1\cdots N}+5}}=\FR{1}{r_j^{p_{ij}+q_{1\cdots N}+5}}\sum_{\ell=0}^\infty \FR{(-1)^\ell(\ell+1)_{p_{ij}+q_{1\cdots N}+4}}{\Gamma(p_{ij}+q_{1\cdots N}+5)}\Big(\FR{r_i}{r_j}\Big)^{\ell}.
\end{align}
Then the source term in (\ref{eq_Dr1tG}) can be rewritten as
\begin{align}
\label{eq_source_expanded}
  \FR{r_{i}^{p_j+4}r_{j}^{p_i+4}}{r_{ij}^{p_{ij}+5}}\G_{V}\big|_{E_j\to E_{ij},p_{j}\to p_{ij}+4}=\sum_{\ell=0}^\infty \sum_{\{n\}} \tilde c_{\ell\{n\}}\,r_i^{ 3}\Big(\FR{r_i}{r_j}\Big)^{\ell+q_{1\cdots N}+p_j+1} r_1^{q_1}\cdots r_N^{q_N}\times\cdots,
\end{align}
with the following coefficient:
\begin{align}
\label{eq_ctilde}
  \tilde c_{\ell\{n\}}\equiv \FR{(-1)^\ell(\ell+1)_{p_{ij}+q_{1\cdots N}+4}}{\Gamma(p_{ij}+q_{1\cdots N}+5)}  c_{\{n\}}.
\end{align}
Given the form of the source term (\ref{eq_source_expanded}), we can try the following ansatz for the solution inhomogeneous (Inh) with respect to Line $K$: 
\begin{align}
\label{eq_tGinhsmallr1}
  \mathop{\text{Inh}}_{K}\big[\G_{V+1}\big]= \sum_{\ell,m=0}^\infty \sum_{\{n\}} d_{\ell m\{n\}} \Big(\FR{r_i}2\Big)^{2m+3}\Big(\FR{r_i}{r_j}\Big)^{\ell+q_{1\cdots N}+p_j+1} r_1^{q_1}\cdots r_N^{q_N}\times\cdots .
\end{align}
Here we have included a factor of $1/2$ in $(r_i/2)^{2m+3}$ for later convenience. 
Substituting this ansatz into (\ref{eq_Dr1tG}) and matching the powers of $r_i$ on both sides of the equation, we get a set of recursion relations. In particular, matching the $\order{r_i^{\ell+p_j+q_{1\cdots N}+4}}$ terms, we get an ``initial condition:'' \begin{align}
\label{eq_d0l2n}
  \FR14\Big[\big(\ell+q_{1\cdots N}+p_j+\fr52\big)^2+\wt\nu^2\Big]d_{\ell 0\{n\}}=2\tilde c_{\ell\{n\}}.
\end{align}
Then, matching $\order{r_i^{2m+\ell+p_j+q_{1\cdots N}+4}}$ terms gives the recursion formula:
\begin{align}
\label{eq_dRecursion}
  & \FR14 \Big[\big(2m+\ell+q_{1\cdots N}+p_j+\fr52\big)^2+\wt\nu^2\Big]d_{\ell m\{n\}}\n\\
 =&~(2m+\ell+q_{1\cdots N}+p_{ij}+4)(2m+\ell+q_{1\cdots N}+p_{ij}+3)d_{\ell(m-1)\{n\}}.
\end{align}
It is straightforward to find a general-term formula for the coefficients $d_{\ell m\{n\}}$ by combining (\ref{eq_d0l2n}) and (\ref{eq_dRecursion}):
\begin{align}
  d_{\ell m\{n\}}=\FR{2(\ell+q_{1\cdots N}+p_{ij}+5)_{2m}}{\big(\fr{\ell+q_{1\cdots N}+p_{j}}{2}+\fr54\pm\fr{\ii\wt\nu}2\big)_{m+1} }\tilde c_{\ell \{n\}}.
\end{align}
Here and below, we introduce the shorthand notations for the product of a pair of Pochhammer symbols:
\bge
  \big(\fr{\ell+q_{1\cdots N}+p_{j}}{2}+\fr54\pm\fr{\ii\wt\nu}2\big)_{m+1}\equiv \big(\fr{\ell+q_{1\cdots N}+p_{j}}{2}+\fr54+\fr{\ii\wt\nu}2\big)_{m+1}\big(\fr{\ell+q_{1\cdots N}+p_{j}}{2}+\fr54-\fr{\ii\wt\nu}2\big)_{m+1}.
\ede

Then, using (\ref{eq_ctilde}) to replace $\tilde c_{\ell\{n\}}$ by the original (known) coefficient $c_{\{n\}}$, we finally get a main result of this subsection: 
\begin{keyeqn}
\begin{align}
\label{eq_IngoingInhom} 
\mathop{\text{Inh}}_{K}\big[\G_{V+1}\big]= &\sum_{\ell,m=0}^\infty \sum_{\{n\}} d_{\ell m\{n\}} \,\Big(\FR{r_i}2\Big)^{2m+3}\Big(\FR{r_i}{r_j}\Big)^{\ell+q_{1\cdots N}+p_j+1} r_1^{q_1}\cdots r_N^{q_N}\times\cdots ;\\
\label{eq_IngoingRecursion}
  d_{\ell m\{n\}}=&~\FR{2(-1)^{\ell}(q_{1\cdots N}+p_{ij}+5)_{\ell+2m}}{\ell!\big(\fr{\ell+q_{1\cdots N}+p_{j}}{2}+\fr54\pm\fr{\ii\wt\nu}2\big)_{m+1}} c_{ \{n\}}.
\end{align} 
\end{keyeqn}
That is, we have found an inhomogeneous solution to the $r_i$-equation (\ref{eq_Dr1tG}), expressed as a series (\ref{eq_tGinhsmallr1}), whose coefficients $d_{\ell m\{n\}}$ are obtained by a simple ``dressing'' of the series coefficients of $\G_V$. This formula works for $E_i>E_j$ so that the newly added line is ingoing. So, we call it the \emph{ingoing recursion formula}.

\paragraph{Consistency check} We have found the inhomogeneous solution to the $r_i$-equation alone by considering a series ansatz. This is not quite enough to claim that the series (\ref{eq_IngoingInhom}) \emph{is} the inhomogeneous part of the full diagram $\G_{V+1}$, because, for that purpose, we must check that the expression (\ref{eq_IngoingInhom}) satisfies \emph{all} of $2V$ differential equations for $\G_{V+1}$ besides the $r_i$-equation. 

The $2V-N-1$ differential equations for $r$ variables not explicitly shown in (\ref{eq_rratios}) are trivially satisfied by $[\G_{V+1}]_\text{inh}$ in (\ref{eq_IngoingInhom}), if the ``old'' graph $\G_V$ satisfies the corresponding equations. The only nontrivial equations to be checked are those associated with $r$ variables involving Vertex $j$, namely $r_j$ and all $r_k$ ($k=1,\cdots, N$). 

Let's first check the $r_j$-equation. Acting the operator $\mathcal{D}_{r_j}$ on the ansatz (\ref{eq_tGinhsmallr1}), we have:
\begin{align}
\label{eq_Drjdseries}
  &\Big[(\vartheta_{r_j}-\fr32)+\wt\nu^2-r_j^2(\vartheta_{j1\cdots N}+p_j+2)(\vartheta_{j1\cdots N}+p_j+1)\Big]\mathop{\text{Inh}}_{K}\big[\G_{V+1}\big]\n\\
 =& \sum_{\ell, m=0}^\infty\sum_{\{n\}}d_{\ell m\{n\}}\Big[\big(\ell+q_{1\cdots N}+p_j+\fr52\big)^2+\wt\nu^2-r_j^2\ell(\ell-1)\Big]\n\\
 &\times \Big(\FR{r_i}2\Big)^{2m+3}\Big(\FR{r_i}{r_j}\Big)^{\ell+q_{1\cdots N}+p_j+1} r_1^{q_1}\cdots r_N^{q_N}\times\cdots.
\end{align}
Here $\vartheta_{j1\cdots N}\equiv\vartheta_{j}+\vartheta_{1}+\cdots+\vartheta_{N}$. The result of $\mathcal{D}_{r_j}\mathop{\text{Inh}}_{K}\big[\G_{V+1}\big]$ is again given by (\ref{eq_source_expanded}). Therefore, we can match terms of equal powers of $r_i$ or $r_j$ between (\ref{eq_Drjdseries}) and (\ref{eq_source_expanded}). In particular, matching at $\order{r_i^{4+\ell+q_{1\cdots N}+p_j}}$ yields the same ``initial condition'' (\ref{eq_d0l2n}). On the other hand, matching terms of equal power of $r_j$, we get a new recursion relation for the coefficients $d_{\ell m\{n\}}$:
\begin{align}
   d_{\ell m\{n\}}=\FR{4(\ell+2)(\ell+1)}{(\ell+q_{1\cdots N}+p_j+\fr52)^2+\wt\nu^2}d_{(\ell+2)(m-1)\{n\}}.
\end{align}
One can easily check that this recursion relation is satisfied by our series solution with (\ref{eq_IngoingRecursion}). Thus the series (\ref{eq_IngoingInhom}) is indeed a solution to the $r_j$-equation. 

Next, we check the $r_1$-equation. The procedures for the rest of $r_k$-equations with $k=2,\cdots, N$ are identical. The original $V$-site graph $\G_V$ satisfies a similar set of $r_k$-equations with $k=1,\cdots,N$, but these equations are modified after including the new line. Thus we need to check them.

Now, acting $\mathcal{D}_{r_1}$ operator on $ \mathop{\text{Inh}}_{K}\big[\G_{V+1}\big]$, we get:
\begin{align}
  &\Big[(\vartheta_{r_1}-\fr32)^2+\wt\nu_1^2-r_1^2(\vartheta_{j1\cdots N}+p_j+2)(\vartheta_{j1\cdots N}+p_j+1)\Big]\mathop{\text{Inh}}_{K}\big[\G_{V+1}\big]\n\\
  =&\sum_{\ell,m=0}^\infty\sum_{\{n\}}d_{\ell m\{n\}}
  \Big[\big(q_1-\fr32\big)^2+\wt\nu_1^2-r_1^2\ell(\ell-1)\Big]\n\\
  &\times\Big(\FR{r_i}2\Big)^{2m+3}\Big(\FR{r_i}{r_j}\Big)^{\ell+q_{1\cdots N}+p_j+1} r_1^{q_1}\cdots r_N^{q_N}\times \cdots\n\\
  =&~0~\text{or a $V$-site graph}.
\end{align}
The result of $\mathcal{D}_{r_1}\mathop{\text{Inh}}_{K}\big[\G_{V+1}\big]$ depends on which types of series we are picking up for $\G_V$ in (\ref{eq_tGseries}). In any case, by matching powers of energy ratios, we get a recursion relation:
\begin{align}
\label{eq_drecursion_r1}
  d_{\ell m\{n(q_1)\}}=\FR{(\ell+2)(\ell+1)}{(q_1-\fr32)^2+\wt\nu_1^2}d_{(\ell+2)m \{n(q_1-2)\}}.
\end{align}
Here by $\{n(q_1)\}$ we are highlighting that some of the summation variables $n_1,\cdots,n_{2V-2}$ depend on the power $q_1$, and here we are forming a recursion relation for two $d_{\ell m\{n(q_1)\}}$'s where $q_1$ jumps by 2. Given Footnote \ref{fn_qi}, this normally involves some summation variables in $\{n\}$ jump by 1 or 2. We will see more explicit examples in the next section. 

On the other hand, we note that the coefficient $c_{\{n\}}$ for the series (\ref{eq_tGseries}) of the original subgraph $\G_V$ satisfies another recursion relation coming from the $r_1$-equation for the subgraph $\G_V$:
\begin{align}
  \Big[(\vartheta_1-\fr32)^2+\wt\nu_1^2-r_1^2(\vartheta_{1\cdots N}+p_{ij}+6)(\vartheta_{1\cdots N}+p_{ij}+5)\Big]\G_{V}=0~\text{or a $(V-1)$-site graph}.
\end{align}
Here we have used the facts that the mass for Line $K_1$ (shown in Fig.\;\ref{fig_recursion}) is $\wt\nu_1$, and that the twist for Vertex $j$ in the original graph $\G_{V}$ is $p_{ij}+4$. Again, there are two possibilities for the right-hand side, depending on whether we consider the inhomogeneous or homogeneous piece with respect to Line $K_1$. Importantly, the recursion relation is independent of the right-hand side being zero or not.  In both cases, we have:
\begin{align}
\label{eq_crecursion}
  c_{\{n(q_1)\}}=\FR{(q_{1\cdots N}+p_{ij}+4)(q_{1\cdots N}+p_{ij}+3)}{(q_1-\fr32)^2+\wt\nu_1^2}c_{\{n(q_1-2)\}}.
\end{align}

Now it is trivial to check that the recursion relation (\ref{eq_drecursion_r1}) is satisfied by our solution (\ref{eq_IngoingRecursion}). Explicitly, (\ref{eq_IngoingRecursion}) tells us the following relation:
\begin{align}
  \FR{d_{\ell m\{n(q_1)\}}}{d_{(\ell+2)m\{n(q_1-2)\}}}
  = \FR{(\ell+2)(\ell+1)}{(q_{1\cdots N}+p_{ij}+4)(q_{1\cdots N}+p_{ij}+3)} \FR{c_{\{n(q)\}}}{c_{\{n(q-2)\}}}.
\end{align}
Then, substituting (\ref{eq_crecursion}) into this relation, we immediately get (\ref{eq_drecursion_r1}). This shows that our series for $\mathop{\text{Inh}}_{K}\big[\G_{V+1}\big]$ with coefficients (\ref{eq_IngoingRecursion}) is indeed a valid solution for \emph{all} differential equations for new graph $\G_{V+1}$.  

\subsection{Outgoing recursion formula}
 
Now we repeat the derivation for the inhomogeneous part of the $(V+1)$-site graph $\G_{V+1}$ above, but this time consider the other energy ordering $E_i<E_j$. This corresponds to the right graph of Fig.\;\ref{fig_recursion}. In this region, we need to develop a different expansion, in terms of $r_j$ and $r_j/r_i$. 

As one can anticipate, it is easier to use the $r_j$-equation to generate the solution. After that, we will check the consistency of this solution with other equations. The $r_j$-equation reads:
\bge
\label{eq_rjequation}
  \Big[(\vartheta_{r_j}-\fr32)^2+\wt\nu^2-r_j^2(\vartheta_{j1\cdots N}+p_j+2)(\vartheta_{j1\cdots N}+p_j+1)\Big]\G_{V+1}  =\FR{r_{i}^{p_j+4}r_{j}^{p_i+4}}{r_{ij}^{p_{ij}+5}}\G_{V}\big|_{E_j\to E_{ij},p_j\to p_{ij}+4}.
\ede 
The source term is identical to the $r_i$-equation. However, we need to expand powers of $r_{ij}$ in the small $r_j/r_i$ limit:
\begin{align}
  \FR{1}{r_{ij}^{p_{ij}+q_{1\cdots N}+5}}
  =\FR{1}{r_i^{p_{ij}+q_{1\cdots N}+5}}\sum_{\ell=0}^\infty \FR{(-1)^\ell(\ell+1)_{p_{ij}+q_{1\cdots N}+4}}{\Gamma(p_{ij}+q_{1\cdots N}+5)}\Big(\FR{r_j}{r_i}\Big)^{\ell}.
\end{align}
Correspondingly, the source term can be expanded as:
\begin{align}
\label{eq_sourceExpInrj}
 \FR{r_{i}^{p_j+4}r_{j}^{p_i+4}}{r_{ij}^{p_{ij}+5}}\G_{V}\big|_{E_j\to E_{ij},p_{j}\to p_{ij}+4}
 =\sum_{\ell=0}^\infty \sum_{\{n\}} \tilde c_{\ell\{n\}}\,r_j^{3}\Big(\FR{r_j}{r_i}\Big)^{\ell+p_i+1} r_1^{q_1}\cdots r_N^{q_N}\times\cdots,
\end{align}
where the coefficients $\tilde c_{\ell\{n\}}$ are still given by (\ref{eq_ctilde}). 

The source term (\ref{eq_sourceExpInrj}) suggests to us the following ansatz for the inhomogeneous part of the new graph $\G_{V+1}$: 
\begin{align}
\label{eq_outgoingAnsatz}
  \mathop{\text{Inh}}_{K}\big[\G_{V+1}\big]= \sum_{\ell,m=0}^\infty \sum_{\{n\}} b_{\ell m\{n\}} \Big(\FR{r_j}2\Big)^{2m+3}\Big(\FR{r_j}{r_i}\Big)^{\ell+p_i+1} r_1^{q_1}\cdots r_N^{q_N}\times \cdots .
\end{align}
Again, we included a factor of $1/2$ in $(r_j/2)^{2m+3}$ for later convenience. 
Then, the procedure of determining the coefficients $b_{\ell m\{n\}}$ is identical to the previous case, and so we will be brief. First, we substitute the ansatz (\ref{eq_outgoingAnsatz}) into the $r_j$-equation (\ref{eq_rjequation}), which gives:
\begin{align}
  &\Big[(\vartheta_{r_j}-\fr32)^2+\wt\nu^2-r_j^2(\vartheta_{j1\cdots N}+p_j+2)(\vartheta_{j1\cdots N}+p_j+1)\Big]\mathop{\text{Inh}}_{K}\big[\G_{V+1}\big]\n\\
 =&\sum_{\ell,m,\{n\}} b_{\ell m\{n\}}\Big[(\ell+2m+p_i+\fr52)^2+\wt\nu^2-r_j^2(\ell+2m+q_{1\cdots N}+p_{ij}+6)(\ell+2m+q_{1\cdots N}+p_{ij}+5)\Big]\n\\
 &~\times\Big(\FR{r_j}{2}\Big)^{2m+3}\Big(\FR{r_j}{r_i}\Big)^{\ell+p_i+1}r_1^{q_1}\cdots r_N^{q_N}\times\cdots .
\end{align}
Then, matching the coefficients of $\order{r_j^{\ell+p_i+4}}$ terms on the both sides, we get an ``initial condition:'' 
\bge
  b_{\ell 0\{n\}}=\FR{8\tilde c_{\ell\{n\}}}{(\ell+p_i+\fr52)^2+\wt\nu^2}.
\ede 
Matching coefficients of $\order{r_j^{\ell+2m+p_i+4}}$ terms, we get the recursion relation: 
\bge
  b_{\ell m\{n\}}=\FR{4(\ell+2m+q_{1\cdots N}+p_{ij}+4)(\ell+2m+q_{1\cdots N}+p_{ij}+3)}{(\ell+2m+p_i+\fr52)^2+\wt\nu^2}b_{\ell(m-1)\{n\}}.
\ede
From the above result, together with (\ref{eq_ctilde}), it is again straightforward to find the following \emph{outgoing recursion formula}: 
\begin{keyeqn}
\begin{align} 
\label{eq_outgoingRecursion}
 \mathop{\text{Inh}}_{K}\big[\G_{V+1}\big]=&~ \sum_{\ell,m=0}^\infty \sum_{\{n\}} b_{\ell m\{n\}} \,\Big(\FR{r_j}2\Big)^{2m+3}\Big(\FR{r_j}{r_i}\Big)^{\ell+p_i+1} r_1^{q_1}\cdots r_N^{q_N}\times \cdots ,\\
 \label{eq_bcoef}
  b_{\ell m\{n\}}=&~\FR{2(-1)^{\ell}(q_{1\cdots N}+p_{ij}+5)_{\ell+2m}}{\ell!\big(\fr{\ell+p_i}2+\fr54\pm\fr{\ii\wt\nu}2\big)_{m+1}} c_{\{n\}}.
\end{align}
\end{keyeqn}
This formula gives the inhomogeneous part (with respect to the new line $K$) of the new graph $\G_{V+1}$ in terms of the result of the old graph $\G_V$, and holds when $E_i<E_j$.

Once again, we cannot claim (\ref{eq_outgoingRecursion}) is the right formula unless we have checked that it also solves the $r_i$-equation and all $r_k$-equation with $k=1,\cdots,N$. Going through the same steps as before, it is easy to see that the $r_i$-equation leads to the following recursion relations for the coefficients $b_{\ell m\{n\}}$:
\begin{align}
   b_{\ell m\{n\}}=\FR{4(\ell+1)(\ell+2)}{(\ell+p_i+\fr52)^2+\wt\nu^2} b_{(\ell+2)(m-1)\{n\}} ,
\end{align}
while the $r_1$-equation leads to the following recursion relation:
\begin{align}
  b_{\ell m\{n(q_1)\}}=\FR{(\ell+2m+p_{ij}+q_{1\cdots N}+4)(\ell+2m+p_{ij}+q_{1\cdots N}+3)}{(q_1-\fr32)^2+\wt\nu_1^2}b_{\ell m\{n(q_1-2)\}}.
\end{align}
It is straightforward to verify that our solution (\ref{eq_bcoef}) satisfies both recursion relations. Clearly, the other $r_k$-equations ($k=2,\cdots,N$) can be checked in a way very similar to $r_1$-equation, and we will not elaborate on them.
 
To finish this section, we note that the ingoing recursion formula (\ref{eq_IngoingInhom}) is convergent when both $r_i=K/E_i$ and $r_i/r_j=E_j/E_i$ are small. The outgoing formula (\ref{eq_outgoingRecursion})  works in the other direction, when $r_j=K/E_j$ and $r_j/r_i$ are both small. It may be useful to note that the outgoing formula is not necessarily the analytical continuation of the ingoing formula in the small-$r_j/r_i$ region. This is evident when applying these recursion formulae to build a 2-site graph from a 1-site graph. In that case, if we analytically continue the result from the ingoing formula to the region where the outgoing formula is supposed to work, we will get a piece identical to the outgoing formula. However, we will also get another piece that is nonanalytic in the vertex energy ratios, which is actually a part of the local signal. However, in certain cases, applying the ingoing and outgoing formulae gives identical results. We will see examples in the next section. Indeed, it is this ``happy coincidence'' that allows us to find a simple formula for the CIS of arbitrary graphs with arbitrary vertex energy orderings. 

\section{Completely Inhomogeneous Solution}
\label{sec_CIS}

In this section, we make use of the recursion formulae of the previous section to work out the CIS of general massive tree correlators. We will again work with the dimensionless version (\ref{eq_dimlessIntG}). 

Recall that the recursion formulae derived in the previous section are quite general in the following sense: When we add a new line to a given graph, the recursion formulae give the inhomogeneous piece of the new graph with respect to the newly added line. The given graph consists of many pieces corresponding to the homogeneous or inhomogeneous solutions to all of its differential operators. The recursion formulae work for each of these pieces individually. Thus we can use these formulae to construct a graph recursively. That is, we add a line to a single-vertex graph and use the recursion formulae to build the 2-site inhomogeneous solution, and then use boundary conditions to fix the corresponding homogeneous solution. This gives the full result of a two-vertex graph. Then, we add one more line to the full two-vertex solution to build a three-vertex graph, and repeat this procedure until we reach the final $V$-vertex graph that we want to compute. 

However, it turns out more streamlined to neglect all homogeneous solutions and focus only on the completely inhomogeneous part, which is traditionally considered as the most difficult part of the problem. Then, the homogeneous solutions can be obtained by applying simple cuts as will be detailed in Sec.\ \ref{sec_hom}. Below, we carry out this recursive construction of the CIS. It turns out that the resulting expression has a nice interpretation as a direct generalization of the family tree integrals introduced in \cite{Xianyu:2023ytd,Fan:2024iek}. This massive version of family trees also turns out to be useful when we construct homogeneous solutions below. 

\subsection{Starting point: graph ordering} 
\label{sec_ordering}

Our starting point is a general massive tree correlator of $V$ vertices, described by $V$ vertex energies $E_1,\cdots, E_V$ and $V-1$ line energies $K_2,\cdots, K_{V}$. To construct a series solution, we must specialize in a particular kinematic region where our final series expression is expected to be convergent, as explained in Sec.\;\ref{sec_strategy}. For our purpose of constructing the CIS, we only need a very mild prescription: We only require that the largest energies among all $2V-1$ variables $\{E_i,K_j\}$ is a \emph{vertex energy}. Without loss of generality, we will just call it $E_1$. All other $2V-2$ variables $E_i, K_i$ ($i=2,\cdots,V$) can have arbitrary ordering, as long as they are all suitably smaller than $E_1$. 

\begin{figure}[t]
\centering
\includegraphics[width=\textwidth]{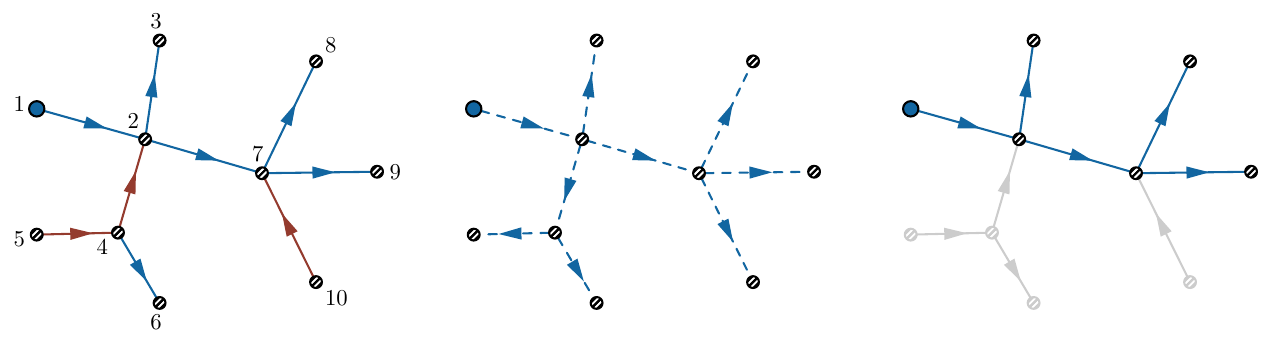}
\caption{An illustration of graph orderings. 
Left: The direction shows the imposed order obtained by comparing the relative sizes of all adjacent vertex energies. The maximal energy vertex (Vertex 1) is marked with a blue dot, the correct lines are colored in blue and the wrong lines in red. Middle: The induced order. Right: The imposed family tree (namely, the maximal connected correct subgraph containing the maximal energy site).}
\label{fig_CIS1}
\end{figure}

Once we impose a particular order to all vertex energies, say $E_1>E_{i_2}>\cdots >E_{i_V}$, the graph acquires a directional structure. (Here $(i_2,\cdots, i_V)$ can be an arbitrary permutation of $(2,\cdots,V)$.) Consequently, to every internal line, we can add a direction which points from the vertex of larger energy towards the vertex of smaller energy. We illustrate this energy direction with an example in the left graph of Fig.\;\ref{fig_CIS1}. We will call this directional structure the \emph{imposed order} since it is imposed from our choice of the kinematic region $E_1>E_{i_2}>\cdots >E_{i_V}$. 

On the other hand, it is crucial to introduce the concept of a partially ordered graph. A partial order means that, to any vertex, we can attach any number of lines with out-going directions, but at most one line with in-going direction. To borrow the convenient nomenclature of \cite{Xianyu:2023ytd,Fan:2024iek}, for any line connecting two vertices, we call the vertex with larger energy the \emph{mother}, and the vertex with smaller energy the \emph{daughter}. Then a partial order means that a mother can have any number of daughters but a daughter can have at most one mother. Consequently, a partially ordered graph naturally acquires an interpretation as a maternal family tree, which explains our use of this terminology. 

With this background knowledge, it is now clear that, once we pick up the largest vertex energy $E_1$, there is a unique partial order of the whole graph that is naturally induced by $E_1$, in which $E_1$ acts as the ancestor of the whole family tree. We call it the \emph{induced partial order}, to distinguish it from the imposed order introduced above. We show the induced order in the middle graph of Fig.\;\ref{fig_CIS1}. 

In general, the imposed order is not a partial order, and is different from the induced partial order. So the two orderings agree on some internal lines and conflict on others. For later convenience, we call internal lines with agreed orders ``correct lines,’’ and lines with conflicting orders ``wrong lines.’’ Correct and wrong lines are respectively colored in blue and red in the left plot of Fig.\;\ref{fig_CIS1}. By construction, all lines attached to the maximal energy $E_1$ must be correct lines. These correct lines can be extended to a maximal partially ordered graph starting from $E_1$, which is in general a subgraph of the whole graph. We call it the \emph{imposed family tree} of $E_1$, as shown in the right plot of Fig.\;\ref{fig_CIS1}.

\begin{figure}[h]
\centering
\includegraphics[width=\textwidth]{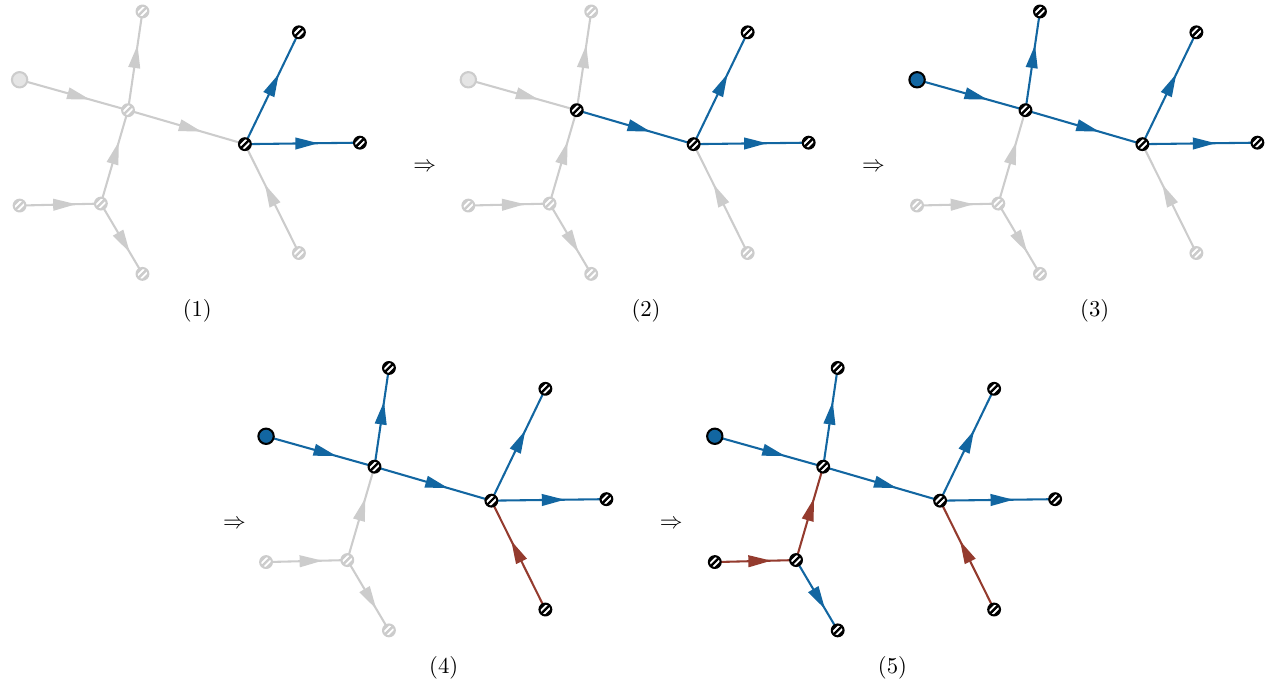}
\caption{The strategy for recursively constructing the CIS of an arbitrary tree with arbitrary vertex energy orderings. (1) By using the outgoing recursion formula, we can find the CIS for a two-generation subfamily.  (2) We then add one ingoing line to enlarge the graph to one earlier generation.  (3) The above two steps allow us to construct the CIS for arbitrary partially ordered graph. Thus we reach the result for the imposed family tree.  (4) We then add one wrong line to the imposed family tree, and show that it is in fact equivalent to a correct line. (5) Using the results of Step 4, we know how to add lines of arbitrary directions, and thus can complete the construction for the CIS of the whole graph.}
\label{fig_CIS2}
\end{figure}

With the above preparation, we can carry out the recursive construction of the CIS. Our procedure is the following:
\begin{enumerate}
  \item We first consider the imposed family tree, and work backward from the last generation to earlier generations. After this step, we will get a formula for CIS for a partially order graph.
  
  \item To make further progress, we need to add a ``wrong line’’ to the imposed family tree. We will show that the resulting expression for the CIS after adding a wrong line is the \emph{same} as adding a ``correct line.’’ Thus, the directions of all wrong lines can be reversed, so that the induced partial order actually works for the whole graph.
  
  \item As a result, the formula we derived for the partially ordered graph actually works when the imposed order is not a partial order, and this finishes our derivation of the completely inhomogeneous solution. 
\end{enumerate}
We show this procedure with an example in Fig.\;\ref{fig_CIS2}.

\subsection{Recursive construction of CIS}

\paragraph{Building one generation} 

\begin{figure}[t]
\centering
\includegraphics[height=0.24\textwidth]{
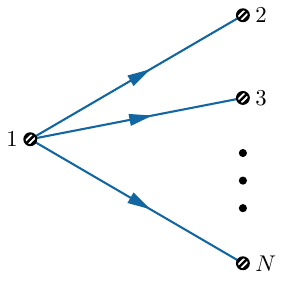}
\caption{The first step of recursively constructing the CIS: the $N$-site star graph $\mft{1(2)\cdots (N)}$.}
\label{fig_1gen}
\end{figure}

As outlined above, we first consider the imposed family tree, starting from the last generation. Thus, in this subsection, we work with a vertex with energy $E_1$ in a second-to-last generation in the imposed family tree, together with all her daughters. Suppose the site $E_1$ has $N-1$ daughters with energies $E_2,\cdots,E_N$, as shown in Fig.\;\ref{fig_1gen}. Also, we assume that the line connecting $E_1$ and $E_i$ carries the line energy $K_i$ and mass $\wt\nu_i$ ($i=2,\cdots,N$), to be in accordance with our general rule of labeling a massive family tree. (See below and App.\;\ref{app_MFT}.) They together form an $N$-site star graph, with all lines pointing outwards.  Below, we construct the CIS of this $N$-site star graph by recursively adding new lines to a single vertex $E_1$. 

First, the one-vertex graph with vertex energy $E_1$ and twist $p_1$ can be trivially computed as:
\begin{align} 
\label{eq_1site}
  \mft{1}=E_{1}^{p_{1}+1}\sum_{\aa=\pm}\int_{-\infty}^0\di\tau\,\ii\aa(-\tau)^{p_{1}}e^{\ii\aa E_{1}\tau}=2\cos( \pi p_{1}/2)\Gamma(p_{1}+1).
\end{align} 
From this point onwards, we introduce our notations for ``massive family trees,'' reviewed in App.\;\ref{app_MFT} (\footnote{The massive family trees $\mft{\cdots}$ in App.\;\ref{app_MFT} are defined via bulk time integrals, while we are using the recursion relations to build the same objects in this subsection. The reason that the two approaches give the same result is discussed below (\ref{eq_CIStoMFT}).}). Here, $\mft{1}$ simply means (the CIS of) the one-site graph. 

Now, we want to add one line connecting $E_1$ with $E_2$, with line energy $K_2$ and mass $\wt\nu_2$. Since $E_2<E_1$ by construction, we should use the outgoing recursion formula (\ref{eq_outgoingRecursion}). The recursion formula requires us to shift the twist $p_1$ of the 1-vertex graph by $p_1\to p_{12}+4$. The shifted 1-vertex graph is:
\begin{align} 
  \mft{1}_{p_1\to p_{12}+4}=2\cos( \pi p_{12}/2)\Gamma(p_{12}+5).
\end{align} 
Then, we get the CIS for the 2-site graph by directly applying the recursion formula (\ref{eq_outgoingRecursion}):
\begin{align}
  \mft{12}=\sum_{\ell_2,m_2=0}^\infty\FR{4(-1)^{\ell_2}\cos( \pi p_{12}/2)
\Gamma(\ell_2+2m_2+p_{12}+5)}{\ell_2!\big(\fr{\ell_2+p_{2}}{2}+\fr54\pm\fr{\ii\wt\nu_2}2\big)_{m_2+1} } \Big(\FR{K_2}{2E_1}\Big)^{2m_2+3}\Big(\FR{E_2}{E_1}\Big)^{\ell_2+p_2+1}.
\end{align}
The summation variables $(\ell_2,m_2)$ are chosen for easier generalization to $N$-site star in the following.

Similarly, we add one more line to $E_1$ of $\mft{12}$ to form the CIS of a 3-site graph $\mft{1(2)(3)}$. Here the notation means that Site 1 has two daughters 2 and 3. (See App.\;\ref{app_MFT}.) Following the recursion method, we first shift the twist $p_1$ in $\mft{12}$ to get $\mft{12}_{p_1\to p_{13}+4}$, and then apply the recursion formula (\ref{eq_outgoingRecursion}). The result is:
\begin{align} 
   \mft{1(2)(3)}
   =&\sum_{\ell_2,\ell_3,m_2,m_3=0}^\infty\FR{4(-1)^{\ell_2}\cos( \pi p_{123}/2)\Gamma(\ell_2+2m_2+p_{123}+9)_{}}{\ell_2!\big(\fr{\ell_2+p_{2}}{2}+\fr54\pm\fr{\ii\wt\nu_2}2\big)_{m_2+1} } \Big(\FR{K_2}{2E_1}\Big)^{\color{RoyalBlue} 2m_2+3}\Big(\FR{E_2}{E_1}\Big)^{\color{RoyalBlue} \ell_2+p_2+1}\n\\
    &\times \FR{2(-1)^{\ell_3}({\color{RoyalBlue}(\ell_2+2m_2+p_2+4)}+p_{13}+5)_{2m_3+\ell_3}}{\ell_3!\big(\fr{\ell_3+p_3}2+\fr54\pm\fr{\ii\wt\nu_3}2\big)_{m_3+1}}
    \Big(\FR{K_3}{2E_1}\Big)^{2m_3+3}\Big(\FR{E_3}{E_1}\Big)^{\ell_3+p_3+1},
\end{align} 
where we have highlighted the powers of $K_2/E_1$ and $E_2/E_1$, which play the roles of $q_1$ and $q_2$ in the recursion formula (\ref{eq_outgoingRecursion}). The above expression can be directly simplified to:
\begin{align}
  \mft{1(2)(3)}
   =&\sum_{\ell_2,\ell_3,m_2,m_3=0}^\infty\FR{8(-1)^{\ell_2}\cos( \pi p_{123}/2)\Gamma(\ell_{23}+2m_{23}+p_{123}+9)}{\ell_2!\big(\fr{\ell_2+p_{2}}{2}+\fr54\pm\fr{\ii\wt\nu_2}2\big)_{m_2+1} } \Big(\FR{K_2}{2E_1}\Big)^{2m_2+3}\Big(\FR{E_2}{E_1}\Big)^{\ell_2+p_2+1}\n\\
    &\times \FR{(-1)^{\ell_3} }{\ell_3!\big(\fr{\ell_3+p_3}2+\fr54\pm\fr{\ii\wt\nu_3}2\big)_{m_3+1}}
    \Big(\FR{K_3}{2E_1}\Big)^{2m_3+3}\Big(\FR{E_3}{E_1}\Big)^{\ell_3+p_3+1}.
\end{align}
We can repeat this procedure $N-1$ times to get the CIS for the $N$-site star graph, as shown in Fig.\;\ref{fig_1gen}. It is not hard to see that the result is:
\begin{align}
\label{eq_N+1siteStar}
  \mft{1(2)\cdots (N)}=&\sum_{\{\ell,m\}} 2^{N}\cos(\pi p_{1\cdots N}/2)\Gamma(\ell_{2\cdots N}+2m_{2\cdots N}+p_{1\cdots N}+4N-3)  \n\\
  &\times\prod_{i=2}^N \FR{(-1)^{\ell_i}}{\ell_i!\big(\fr{\ell_i+p_i}{2}+\fr54\pm\fr{\ii\wt\nu_i}2\big)_{m_i+1} }
  \Big(\FR{K_i}{2E_1}\Big)^{2m_i+3}\Big(\FR{E_i}{E_1}\Big)^{\ell_i+p_i+1},
\end{align}
where we have abbreviated the summation sign:
\bge
  \sum_{\{\ell,m\}}\equiv \sum_{\ell_2,\cdots,\ell_N=0}^\infty \sum_{m_2,\cdots,m_N=0}^\infty .
\ede
We will use similar abbreviated notations below, and the range of the $\{\ell,m\}$ can be easily identified by inspecting the summand. With (\ref{eq_N+1siteStar}), we have completed the first step (Fig.\;\ref{fig_CIS2}(1)) of constructing the CIS.

\begin{figure}[t]
\centering
\includegraphics[height=0.24\textwidth]{
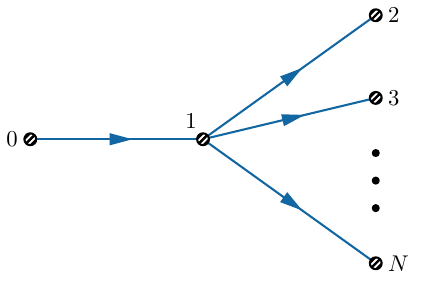}
\caption{The second step of recursively constructing the CIS: the $N$-site star graph with one more generation added: $\mft{01(2)\cdots (N)}$.}
\label{fig_2gen}
\end{figure}

\paragraph{Adding more generations}  
Now we can go one generation earlier, and add the mother site of $E_1$, which we call $E_0$. Thus we need to add an ingoing line to the $E_1$ site, as shown in Fig.\;\ref{fig_2gen}. As we shall see, adding ingoing lines introduces new features to the result, and is a bit subtler than adding outgoing lines. 

We start from the $N$-site star in (\ref{eq_N+1siteStar}), and shift the twist $p_1$ by $p_1\to p_{01}+4$. Then, we apply the ingoing recursion formula (\ref{eq_IngoingInhom}) as:
\begin{align}
\label{eq_012(2N)_1}
  &\mft{01(2)\cdots (N)}=\sum_{\{\ell,m\}} 2^{N+1}\cos(\pi p_{01\cdots N}/2)\Gamma(\ell_{2\cdots N}+2m_{2\cdots N}+p_{01\cdots N}+4N+1)  \n\\
  &\times\prod_{i=2}^N \FR{(-1)^{\ell_i}}{\ell_i!\big(\fr{\ell_i+p_i}{2}+\fr54\pm\fr{\ii\wt\nu_i}2\big)_{m_i+1} }
  \Big(\FR{K_i}{2E_1}\Big)^{2m_i+3}\Big(\FR{E_i}{E_1}\Big)^{\ell_i+p_i+1}\n\\
  &\times\FR{(-1)^{\ell_1}(q_1+p_{01}+5)_{2m_1+\ell_1}}{\ell_1!\big(\fr{\ell_1+{q_1}+p_{1}}{2}+\fr54\pm\fr{\ii\wt\nu}2\big)_{m_1+1}}\Big(\FR{K_1}{2E_0}\Big)^{2m_1+3}\Big(\FR{E_1}{E_0}\Big)^{\ell_1+q_1+p_1+1},
\end{align}
Here we introduced a new parameter: 
\bge
  q_1\equiv\ell_{2\cdots N}+2m_{2\cdots N}+p_{2\cdots N}+4N-4,
\ede
which simply counts the dimension of $1/E_1$ in the series before adding the factors for the new line, as required by the ingoing recursion formula. Note in particular that this $q_1$ parameter appears in the exponent of the newly added $E_1/E_0$ term. As such, it has the effect of replacing all $E_1$ factors by $E_0$ in the original $N$-site star graph, namely, the second line of (\ref{eq_012(2N)_1}). Thus the above expression can be further massaged into the following one:
\begin{align}
\label{eq_01(2N)}
  \mft{01(2)\cdots (N)}=&\sum_{\{\ell,m\}} 2^{N+1}\cos(\pi p_{01\cdots N}/2)\Gamma(q_0+p_0+1)  \n\\
  &\times\prod_{i=1}^N \FR{(-1)^{\ell_i}}{\ell_i!\big(\fr{\ell_i+q_i+p_i}{2}+\fr54\pm\fr{\ii\wt\nu_i}2\big)_{m_i+1} }
  \Big(\FR{K_i}{2E_0}\Big)^{2m_i+3}\Big(\FR{E_i}{E_0}\Big)^{\ell_i+p_i+1}.
\end{align}
Here we have introduced more $q_i$ parameters ($i=0,1,\cdots,N$). Besides $q_1$ defined above, we have:
\begin{align}
  &q_0\equiv q_1+\ell_1+2m_1+p_1+4,
  &&q_i\equiv 0.~~(i=2,\cdots,N)
\end{align}
It is now obvious that we can continue this procedure through all generations of a family tree: Adding a daughter is straightforward, as shown above. On the other hand, when we add a new mother vertex $E_i$, we replace all vertex energies in the denominator of the existing graph with $E_i$. At the same time, we introduce a parameter $q_i$, defined in the following way:
\bge
  q_i\equiv \wt{\ell}_i+2\wt{m}_i+\wt{p}_i+4N_i,
\ede
where $\wt{\ell}_i$ means the sum of all $\ell_j$’s over the descendant vertices of Vertex $i$. $\wt{m}_i$ and $\wt{p}_i$ are defined similarly, and $N_i$ is the number of descendant vertices of Vertex $i$. We call $q_i$ the \emph{family parameter}, since it encodes the family-tree structure of Vertex $i$'s descendants. Then, the expression (\ref{eq_01(2N)}) is equally applicable to all partially ordered graphs, and in particular, to our imposed trees. So, let us put it in a more general way:
\begin{align}
\label{eq_PartOrderTree}
  \mft{\mathscr{P}(\wh{1}2\cdots N)}=&\sum_{\{\ell,m\}} 2^{N}\cos(\pi p_{1\cdots N}/2)\Gamma(q_1+p_1+1)  \n\\
  &\times\prod_{i=2}^N \FR{(-1)^{\ell_i}}{\ell_i!\big(\fr{\ell_i+q_i+p_i}{2}+\fr54\pm\fr{\ii\wt\nu_i}2\big)_{m_i+1} }
  \Big(\FR{K_i}{2E_1}\Big)^{2m_i+3}\Big(\FR{E_i}{E_1}\Big)^{\ell_i+p_i+1},
\end{align}
where $\mathscr{P}(\wt{1}2\cdots N)$ means an arbitrary given partial order of the $N$ sites, with $E_1$ acting as the maximal vertex energy. This completes our recursive construction of the imposed family tree in the whole graph, namely Fig.\;\ref{fig_CIS2}(3).

\paragraph{Adding wrong lines} At this point we have the expression for the CIS of the imposed tree. To enlarge this to the full graph, we need to include the rest of lines, and some of them have ``wrong’’ directions. (Step (4) and (5) in Fig.\;\ref{fig_CIS2}.) So, let us first consider the case of adding one wrong-direction line to any vertex of the imposed tree graph. Since the imposed tree is partially ordered, each of its vertices must have at most one ingoing line but can have any number of outgoing lines. Thus, the most general situation is to add the wrong-direction line to a vertex of the existing imposed tree, which has 1 ingoing line and $N-1$ outgoing lines with $N\geq 1$. (The case of adding an ingoing line to a vertex without ingoing lines is nothing but $\mft{01(2\cdots N)}$ considered above.)

\begin{figure}[t]
\centering
\includegraphics[height=0.24\textwidth]{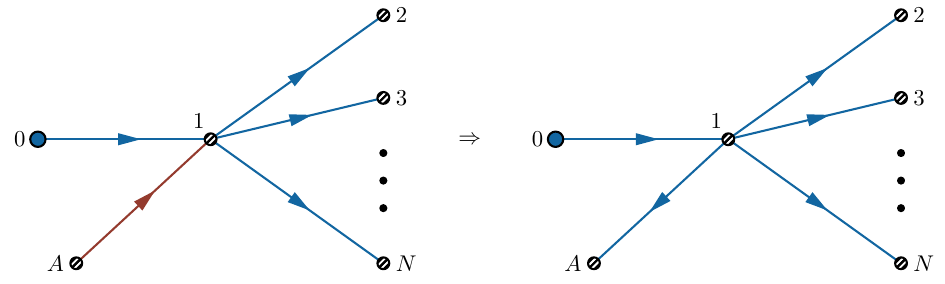}
\caption{The fourth step of recursively constructing the CIS: the $N$-site star graph with one more generation and one wrong line (Left), whose expression is identical to the one with the wrong line reversed (Right). }
\label{fig_wrongline}
\end{figure}

Therefore, we can begin with the three-generation tree $\mft{01(2)\cdots (N)}$ in (\ref{eq_01(2N)}), and add a new line with vertex energy $E_A$ and line energy $K_A$ to the $E_1$ site. The ``wrong direction’’ means $E_A>E_1$, and at this point, we do not assume any ordering between $E_0$ and $E_A$. Then, following the same procedure as above, we first shift the twist of the $E_1$ site in $\mft{01 (2)\cdots (N)}$ by $p_1\to p_{1A}+4$, and apply the ingoing recursion formula (\ref{eq_IngoingInhom})-(\ref{eq_IngoingRecursion}). Then we get:
\begin{align} 
  \mft{01(2)\cdots (N)(\overleftarrow{A})}=&\sum_{\{\ell,m\}} 2^{N+1}\cos(\pi p_{01\cdots NA}/2)\Gamma(\ell_{12\cdots N}+2m_{12\cdots N}+p_{012\cdots NA}+4N+5)  \n\\
  &\times\prod_{i=2}^N \FR{(-1)^{\ell_i}}{\ell_i!\big(\fr{\ell_i+p_i}{2}+\fr54\pm\fr{\ii\wt\nu_i}2\big)_{m_i+1} }
  \Big(\FR{K_i}{2E_0}\Big)^{2m_i+3}\Big(\FR{E_i}{E_0}\Big)^{\ell_i+p_i+1}\n\\
  &\times\FR{(-1)^{\ell_1}}{\ell_1!\big(\fr{\ell_1+q_1+(p_{1A}+4}{2}+\fr54\pm\fr{\ii\wt\nu_1}2\big)_{m_1+1} }
  \Big(\FR{K_1}{2E_0}\Big)^{2m_1+3}\Big(\FR{E_1}{E_0}\Big)^{\ell_1+p_{1A}+5}\n\\
  &\times \FR{2(-1)^{\ell_A}(-\ell_1)_{2m_A+\ell_A}}{\ell_A!\big(\fr{\ell_A-\ell_1-p_{A}}2-\fr54\pm\fr{\ii\wt\nu_A}2\big)_{m_A+1}}\Big(\FR{K_A}{2E_A}\Big)^{2m_A+3}\Big(\FR{E_1}{E_A}\Big)^{\ell_A-\ell_1-p_{A}-4},
\end{align}
in which, for clarity, we have restored $q_0$ and write out the $i=1$ factor explicitly. Here, we encounter a new phenomenon due to the factor $(-\ell_1)_{2m_A+\ell_A}=(-1)^{\ell_A}\ell_1!/\Gamma(1+\ell_1-2m_A-\ell_A)$ in the last line. This factor is zero unless $\ell_1\geq 2m_A+\ell_A$, and it suggests that we should shift the summation variable $\ell_1\to \ell_1+\ell_A+2m_A$. After this shift, we get: 
\begin{align} 
  \mft{01(2)\cdots (N)(\overleftarrow{A})}=&\sum_{\{\ell,m\}} 2^{N+2}\cos(\pi p_{01\cdots NA}/2)\Gamma(\ell_{12\cdots NA}+2m_{12\cdots NA}+p_{012\cdots NA}+4N+5)  \n\\
  &\times\prod_{i=2}^N \FR{(-1)^{\ell_i}}{\ell_i!\big(\fr{\ell_i+p_i}{2}+\fr54\pm\fr{\ii\wt\nu_i}2\big)_{m_i+1} }
  \Big(\FR{K_i}{2E_0}\Big)^{2m_i+3}\Big(\FR{E_i}{E_0}\Big)^{\ell_i+p_i+1}\n\\
  &\times\FR{(-1)^{\ell_{1}}}{\ell_1!\big(\fr{\ell_{1A}+2m_A+q_1+p_{1A}+4}{2}+\fr54\pm\fr{\ii\wt\nu_1}2\big)_{m_1+1} }
  \Big(\FR{K_1}{2E_0}\Big)^{2m_1+3}\Big(\FR{E_1}{E_0}\Big)^{\ell_{1A}+2m_A+p_{1A}+5}\n\\
  &\times \FR{(-1)^{\ell_A}}{\ell_A!\big(\fr{-\ell_1-2m_A-p_{A}}2-\fr54\pm\fr{\ii\wt\nu_A}2\big)_{m_A+1}}\Big(\FR{K_A}{2E_A}\Big)^{2m_A+3}\Big(\FR{E_1}{E_A}\Big)^{\ell_A-\ell_{1A}-2m_A-p_{A}-4}.
\end{align}
Now we can regroup energy-ratio factors, and get:
\begin{align} 
  \mft{01(2)\cdots (N)(\overleftarrow{A})}=&\sum_{\{\ell,m\}} 2^{N+2}\cos(\pi p_{01\cdots NA}/2)\Gamma(\ell_{12\cdots NA}+2m_{12\cdots NA}+p_{012\cdots NA}+4N+5)  \n\\
  &\times\prod_{i=2}^N \FR{(-1)^{\ell_i}}{\ell_i!\big(\fr{\ell_i+p_i}{2}+\fr54\pm\fr{\ii\wt\nu_i}2\big)_{m_i+1} }
  \Big(\FR{K_i}{2E_0}\Big)^{2m_i+3}\Big(\FR{E_i}{E_0}\Big)^{\ell_i+p_i+1}\n\\
  &\times\FR{(-1)^{\ell_{1}}}{\ell_1!\big(\fr{\ell_{1A}+2m_A+q_1+p_{1A}+4}{2}+\fr54\pm\fr{\ii\wt\nu_1}2\big)_{m_1+1} }
  \Big(\FR{K_1}{2E_0}\Big)^{2m_1+3}\Big(\FR{E_1}{E_0}\Big)^{\ell_{A}+p_{1}+1}\n\\
  &\times \FR{(-1)^{\ell_A}}{\ell_A!\big(\fr{-\ell_1-2m_A-p_{A}}2-\fr54\pm\fr{\ii\wt\nu_A}2\big)_{m_A+1}}\Big(\FR{K_A}{2E_0}\Big)^{2m_A+3}\Big(\FR{E_A}{E_0}\Big)^{\ell_{1}+p_{A}+1}.
\end{align}
At this point, it is obvious that we should switch the summation variable $\ell_1\leftrightarrow\ell_A$ to make the expression into a more organized form:
\begin{align} 
  \mft{01(2)\cdots (N)(\overleftarrow{A})}=&\sum_{\{\ell,m\}} 2^{N+2}\cos(\pi p_{01\cdots NA}/2)\Gamma(\ell_{12\cdots NA}+2m_{12\cdots NA}+p_{012\cdots NA}+4N+5)  \n\\
  &\times\prod_{i=2}^N \FR{(-1)^{\ell_i}}{\ell_i!\big(\fr{\ell_i+p_i}{2}+\fr54\pm\fr{\ii\wt\nu_i}2\big)_{m_i+1} }
  \Big(\FR{K_i}{2E_0}\Big)^{2m_i+3}\Big(\FR{E_i}{E_0}\Big)^{\ell_i+p_i+1}\n\\
  &\times\FR{(-1)^{\ell_{1}}}{\ell_1!\big(\fr{\ell_{1A}+2m_A+q_1+p_{1A}+4}{2}+\fr54\pm\fr{\ii\wt\nu_1}2\big)_{m_1+1} }
  \Big(\FR{K_1}{2E_0}\Big)^{2m_1+3}\Big(\FR{E_1}{E_0}\Big)^{\ell_{1}+p_{1}+1}\n\\
  &\times \FR{(-1)^{\ell_A}}{\ell_A!\big(\fr{-\ell_A-2m_A-p_{A}}2-\fr54\pm\fr{\ii\wt\nu_A}2\big)_{m_A+1}}\Big(\FR{K_A}{2E_0}\Big)^{2m_A+3}\Big(\FR{E_A}{E_0}\Big)^{\ell_{A}+p_{A}+1}.
\end{align}
The structure of the series is now quite clear: All energy ratios have $E_0$ in the denominator, meaning that $E_0$ still plays the role of the largest energy, even after we add the new wrong line. This series automatically gives an order $E_A<E_0$. At this point, it seems no further progress can be made, and the coefficients look quite messy. However, let us use the following identity to rewrite the Pochhammer factor in the last line:
\begin{align}
  (-a+b-m)_{m+1}(-a-b-m)_{m+1}=(a+b)_{m+1}(a-b)_{m+1}.~~~~(m\in\mathbb{Z})
\end{align}
Then we get:
\begin{align} 
  \mft{01(2\cdots N\overleftarrow{A})}=&\sum_{\{\ell,m\}} 2^{N+2}\cos(\pi p_{01\cdots NA}/2)\Gamma(\ell_{12\cdots NA}+2m_{12\cdots NA}+p_{012\cdots NA}+4N+5)  \n\\
  &\times\prod_{i=2}^N \FR{(-1)^{\ell_i}}{\ell_i!\big(\fr{\ell_i+p_i}{2}+\fr54\pm\fr{\ii\wt\nu_i}2\big)_{m_i+1} }
  \Big(\FR{K_i}{2E_0}\Big)^{2m_i+3}\Big(\FR{E_i}{E_0}\Big)^{\ell_i+p_i+1}\n\\
  &\times\FR{(-1)^{\ell_{1}}}{\ell_1!\big(\fr{\ell_{1A}+2m_A+q_1+p_{1A}+4}{2}+\fr54\pm\fr{\ii\wt\nu_1}2\big)_{m_1+1} }
  \Big(\FR{K_1}{2E_0}\Big)^{2m_1+3}\Big(\FR{E_1}{E_0}\Big)^{\ell_{1}+p_{1}+1}\n\\
  &\times \FR{(-1)^{\ell_A}}{\ell_A!\big(\fr{\ell_A+p_{A}}2+\fr54\pm\fr{\ii\wt\nu_A}2\big)_{m_A+1}}\Big(\FR{K_A}{2E_0}\Big)^{2m_A+3}\Big(\FR{E_A}{E_0}\Big)^{\ell_{A}+p_{A}+1}.
\end{align}
We see that the new terms generated by adding the wrong line are exactly the same as adding a correct line. Thus, so far as the CIS of the whole graph is concerned, we can flip the directions of all wrong lines to make the graph into a partially ordered tree graph, which is exactly the induced partial order. This result can also be understood directly from the series expression for a partially ordered tree (\ref{eq_PartOrderTree}), where the energy dependence takes the form of $K_i/E_1$ and $E_i/E_1$ for all $i\neq 1$. Thus, the convergence of this series only requires these ratios to be small, and it says nothing about the relative sizes of all $K_i$ and $E_i$ for $i\neq 1$. Thus, we have completed the construction of the CIS for a general tree graph with a general vertex energy ordering.

\subsection{Summary: CISs as massive family trees}

Let us summarize the main result of this section. We have used the inhomogeneous recursion formulae (\ref{eq_IngoingInhom}) and (\ref{eq_outgoingRecursion}) to recursively construct the completely inhomogeneous solution (CIS). It turns out possible to find a general formula for the CIS of arbitrary massive tree graphs $\G_V$ with $V\geq 1$ vertices.

As it appears, the result for the CIS is a multivariate hypergeometric function, and thus a useful way to express this function is a series expansion. To write down a series, we pick up the largest vertex energy, which we call $E_1$. Note that the choice of the largest vertex energy $E_1$ is arbitrary: We can pick up any vertex energy to be the largest one, and call it $E_1$. Once this choice is made, the graph immediately acquires a partial order structure, called the induced order. That is, every line becomes directional, flowing from the $E_1$ site outwards throughout all lines and vertices in the graph, as shown in the middle graph of Fig.\;\ref{fig_CIS1}. This is a natural massive generalization of the family trees introduced in \cite{Xianyu:2023ytd,Fan:2024iek} for conformal scalar amplitudes. Thus, we call it a \emph{massive family tree}. 

Once we have a massive family tree structure for a graph of $V$ vertices, we can express the CIS as a power series in $E_i/E_1$ and $K_i/E_1$, with $i=2,\cdots, N$. Here, we use the convention that a line energy $K_i$ shares the same subscript $i$ with the vertex energy $E_i$ it flows into. There are a total of $2(V-1)$ variables, and correspondingly, there are $2(V-1)$ fold summations. Thus, we can assign a summation variable $\ell_i\in\mathbb{N}$ to $E_i/E_1$, and a summation variable $m_i\in\mathbb{N}$ to $K_i/E_1$. As a result, with any Vertex $i\neq 1$, we associate a set of variables $(E_i,K_i,p_i,\wt\nu_i,\ell_i,m_i)$. 

With these preparations, we can write down the result for the CIS of $\G_V$ as the following series:
\begin{keyeqn}
\begin{align}
\label{eq_MFT}
  \text{CIS}\,\big[\G_V\big]=&~\sum_{\{\ell,m\}} 2^{V}\cos(\pi p_{1\cdots V}/2)\Gamma(q_1+p_1+1)  \n\\
  &\times\prod_{i=2}^V \FR{(-1)^{\ell_i}}{\ell_i!\big(\fr{\ell_i+q_i+p_i}{2}+\fr54\pm\fr{\ii\wt\nu_i}2\big)_{m_i+1} }
  \Big(\FR{K_i}{2E_1}\Big)^{2m_i+3}\Big(\FR{E_i}{E_1}\Big)^{\ell_i+p_i+1}.
\end{align}
\end{keyeqn}
There is one more parameter $q_i$ introduced in this expression, called the family parameter, which encodes the family tree structure of a graph. It is defined by:
\bge
  q_i\equiv \wt{\ell}_i+2\wt{m}_i+\wt{p}_i+4N_i,
\ede
where $\wt{\ell}_i$ means the sum of all $\ell_j$’s over the descendant vertices of Vertex $i$. $\wt{m}_i$ and $\wt{p}_i$ are defined similarly, and $N_i$ is the number of descendant vertices of Vertex $i$. Thus, we see that there is a one-to-one correspondence between a general tree graph with its CIS: Given the graph, we can directly write down the CIS, and given the CIS, we can easily recover the graph by decoding the tree structure from the explicit forms of all $q_i$. Thus, from now on, we will use our notation for massive family trees to denote the CIS of a graph:
\bge
\label{eq_CIStoMFT}
  \text{CIS}\,\big[\G_V\big]=\mft{\mathscr{P}(1\cdots V)},
\ede
where the partial order $\mathscr{P}$ is the induced order from the maximal vertex energy site which can be arbitrarily chosen.

From the viewpoint of bulk time integrals, the CIS corresponds to the part of the integral where all time variables are nested by $\theta$ functions. In a bulk propagator $\wt{D}_{\aa\bb}$, only the same-sign propagators $\wt{D}_{\pm\pm}$ contains $\theta$ function (see (\ref{eq_Dpmpm})). Therefore, in the integral (\ref{eq_dimlessIntG}), the CIS is contributed only by terms with all SK indices being equal, namely $\aa_1=\cdots=\aa_V=\pm$. On the other hand, the $\theta$ function has the freedom to change direction via $\theta(\tau_i-\tau_j)=1-\theta(\tau_j-\tau_i)$. So the property of being fully nested only specifies the nested integral up to additive factorized terms. We need further conditions to identify the CIS from the all-plus and all-minus time integrals, and the condition is that the CIS is analytic for each of $E_i\to 0$ and $K_i\to 0$ ($i=2,\cdots,V$). With this condition, a direct analysis of integral shows that the CIS is contributed by the part of the integral where all $\theta$-functions follow the \emph{induced order} of the graph (the middle graph of Fig.\;\ref{fig_CIS1}). Thus, we see that there is a correspondence between the energy flow and the time flow. This observation will be useful when we cut the CIS below.

The massive family tree (\ref{eq_MFT}) is a $2(V-1)$-variable hypergeometric function of Horn type \cite{horn1931hypergeometrische}. As a series of $2(V-1)$-fold summations, the summand of (\ref{eq_MFT}), called $S(\{\ell,m\})$, is a function of $2(V-1)$ summation variables $\ell_i,m_i$ ($i=2,\cdots, V$). Such a series gives rise to a hypergeometric function if it has a finite convergence region, and if the summand $S(\{\ell,m\})$ satisfies the following property: For any $i=2,\cdots,V$ and any $\ell_i,m_i\geq 0$, $S(\ell_i+1)/S(\ell_i)$ and $S(m_i+1)/S(m_i)$ are both rational functions of all $\ell_i$ and $m_i$. Clearly, we are not super well equipped to decode all properties of such hypergeometric functions. Therefore, it is important to develop different series representations. In this work, we have found series expanded in the inverse of \emph{any} vertex energy. Thus, for every $V$-site massive family tree, we have $V$ different inequivalent series expansions. Together with appropriate homogeneous parts to be discussed below, they give rise to analytical continuations of each other. This already covers many regions in the physically accessible parameter spaces. It is an interesting future problem to find more series expansions for the massive family trees.

We finish this section by making a more conjectural observation. As shown in \cite{Fan:2024iek}, any time integrals nested by a partially ordered structure can be further decomposed into a sum of standard \emph{iterated integrals} in which the time order is a total order. (That is, every mother has only one daughter.) For iterated integrals, there is a notion of grading such that we can define a transcendental weight for this class of functions \cite{Brown:2013qva}. We conjecture that it is possible to assign transcendental weights to our series representations of CISs, which count the minimally required layers of summations with the condition that the summand has a hypergeometric form. (That is, $S(\ell_i+1)/S(\ell_i)$ being rational as mentioned above.) We conjecture that the CIS of a $V$-site graph has weight $2(V-1)$, and our series representation (\ref{eq_MFT}) is optimal in the sense of counting the transcendental weight, meaning that, in general, the layer of summation cannot be further reduced without spoiling the rational form of neighboring summand ratios. 

For comparison, we can also get a series representation for an arbitrary massive tree graph with partial Mellin-Barnes representation plus the family tree decomposition as described in \cite{Xianyu:2023ytd}. However, the series obtained in this way is not optimal, because the CIS of a $V$-site graph involves $3(V-1)$ summations, among which $2(V-1)$ comes from the PMB representation of Hankel functions, and $(V-1)$ comes from the (conformal) family tree integrals. 

As we will see below, the layers of summations do not change after we include the homogeneous solutions. That is, the ``signals'' share the same transcendental weight as the ``background,'' with the only difference of being factorized (similar to the relation between $\log^2(z)$ and $\text{Li}_2(z)$). Thus, it seems that the transcendental weight can be directly defined with respect to a whole graph, not only to its CIS. However, if we go to degenerate configurations when some line and vertex energies are set equal, results from previous works \cite{Qin:2023ejc,Aoki:2024uyi} show that the weight of the function can decrease. It would be very interesting to better understand these phenomena and we leave it for a future work.

\section{Building Homogenous Solutions from Cuts}
\label{sec_hom}

In this section, we consider the (partially) homogeneous solutions to the differential equation system. Again, by a homogeneous solution, we mean part of the whole correlator that is annihilated by at least one of the $2V-2$ differential operators. Given the CIS from the last section, it is easy to obtain homogeneous solutions by taking cuts of the CIS.

\subsection{Single cut of the CIS}

We first describe how to cut a single line in the CIS to get a solution homogeneous with respect to the cut line, but inhomogeneous to all other lines. Later, we generalize the result to multiple cuts.

\begin{figure}[t]
\centering
\includegraphics[width=0.38\textwidth]{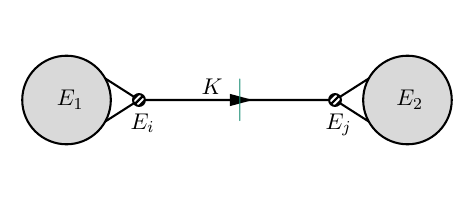}
\caption{The single cut of a tree graph over the line $K$, which connects vertices $E_i$ and $E_j$. $E_1$ ($E_2$) is the largest vertex energy of the left (right) subgraph, which may or may not coincide with $E_i$ ($E_j$). We take $E_1>E_2$.}
\label{fig_singlecut}
\end{figure}

We still work with a general graph $\G$ of $V$ vertices, as shown in Fig.\;\ref{fig_singlecut}. We fix the order for all vertex energy $E_1>E_{i_2}>\cdots E_{i_V}$, and we work with the region where all line energies $K_{i}$ $(i=2,\cdots,V)$ are smaller than all vertex energies. Then, we want to take one cut of the CIS for $\G$ with respect to a particular line $K$, which connects two vertices $E_i$ and $E_j$. Without loss of generality, we assume that $E_i$ and the maximal energy $E_1$ are on the same side with respect to the line $K$, which we call the ``left'' side, and $E_j$ is on the other side, namely the ``right'' side. (In particular, $E_i$ can be $E_1$.) Also, for convenience, let us define, for the argument of this subsection:
\begin{align}
  &r_i\equiv K/E_i,
  &&r_j\equiv K/E_j.
\end{align}
Then, by a single cut of CIS, we mean the part of the full correlator that is annihilated only by $\mathcal{D}_{r_i}$ and $\mathcal{D}_{r_j}$ but not any other differential operators in our differential equation system\footnote{Similar to the CIS, this definition specifies the single cut only up to terms homogeneous with respect to other lines. This ambiguity is removed by imposing correct analytic behavior, to be detailed below.}. 

It is easier to switch to the bulk time integral perspective at this point. Let us isolate the part of time integrals related to vertices $E_i$ and $E_j$:
\bge
\label{eq_tGbipart}
  \G=\sum_{\aa_i,\aa_j=\pm}\int\di z_i\,\ii\aa_i(-z_i)^{p_i}e^{\ii\aa_iz_i}\di z_j\ii\aa_j(-z_j)^{p_j}e^{\ii\aa_jz_j}\wt D_{\aa_i\aa_j}^{(\wt\nu)}(r_iz_i,r_jz_j) \mathcal{I}_{\aa_i}^\text{(L)}\mathcal{I}_{\aa_j}^\text{(R)},
\ede
where $\mathcal{I}_{\aa_i}^\text{(L)}$ consists of time integrals over all left-side vertices except for the factors explicitly shown in (\ref{eq_tGbipart}), with the SK index $\aa_i$ not summed. Likewise, $\mathcal{I}_{\aa_j}^\text{(R)}$ collects time integrals over all right-side vertices that are not explicitly shown in (\ref{eq_tGbipart}).

It is important to note that, since we are executing only one cut, the left and right subgraphs are themselves fully nested. (That is, they are also CISs of the two subgraphs.) Therefore, all vertices in $\mathcal{I}_{\aa_i}^\text{(L)}$ share the same SK index as the $z_i$ vertex. That is why we can use a single SK index $\aa_i$ ($\aa_j$) for all left-side (right-side) time integrals.

The CIS we have found in (\ref{eq_MFT}) corresponds to a piece analytic in $1/E_1\to 0$ (up to powers with exponent $p_{2\cdots V}$, which could be nonanalytic if the exponent is noninteger). From the bulk integral perspective, this corresponds to the contribution from terms in the inhomogeneous propagator $\wt{D}_{\pm\pm}^{(\wt\nu)}(r_iz_i,r_jz_j)$ that place $\tau_i$ earlier than $\tau_j$. Explicitly, we rewrite the inhomogeneous bulk propagator as:
\begin{align}
\label{eq_Dsplit}
  \wt{D}_{\pm\pm}^{(\wt\nu)}(r_iz_i,r_jz_j)=\Big[\wt{D}_{\pm\mp}^{(\wt\nu)}(r_iz_i,r_jz_j)-\wt{D}_{\mp\pm}^{(\wt\nu)}(r_iz_i,r_jz_j)\Big]\theta(r_jz_j-r_iz_i)+\wt{D}_{\mp\pm}^{(\wt\nu)}(r_iz_i,r_jz_j).
\end{align}
Then, the CIS is contributed by the first term on the right-hand side, and the second term, together with the contributions from the other SK branches ($\aa_i=-\aa_j$), gives rise to the cut part. As we shall see from the result of this section, the cut part is fully nonanalytic as $1/E_1\to 0$, and corresponds to the cosmological collider signals of both nonlocal and local types. This is well understood as a cutting rule for computing cosmological collider signals \cite{Tong:2021wai,Qin:2023bjk,Qin:2023nhv}. 

It is important that the cut part in (\ref{eq_Dsplit}) is asymmetric with respect to the replacement $i\leftrightarrow j$. The asymmetry arises from the fact that we have given a particular order to all vertex energies. It is clear from the argument leading to (\ref{eq_Dsplit}) that the ``direction'' of the cut is determined by comparing the maximal vertex energies of the two subgraphs $E_1$ and $E_2$ (see Fig.\;\ref{fig_singlecut}), rather than by comparing the two endpoint energies of the cut line ($E_i$ and $E_j$).  

It is now clear that our asymmetric cutting rule means that, if the largest vertex energy $E_1$ is on the left subgraph, then we do the replacement:
\bge
  \wt{D}_{\pm\pm}^{(\wt\nu)}(r_iz_i,r_jz_j)\To \wt{D}_{\mp\pm}^{(\wt\nu)}(r_iz_i,r_jz_j).
\ede
Then the integral is:  
\begin{align}
  \label{eq_CutKtG}
  \mathop\text{Cut}_{K}\big[\G\big]  =- \int\di z_i\di z_j(-z_i)^{p_i}(-z_j)^{p_j} \Big[e^{+\ii(z_i+z_j)}\mathcal{I}_{+}^\text{(L)}-e^{-\ii(z_i-z_j)}\mathcal{I}_{-}^\text{(L)}\Big]\mathcal{I}_{+}^\text{(R)}\wt D_{-+}^{(\wt\nu)}(r_iz_i,r_jz_j) +\text{c.c.} .
\end{align}
Now the $z_i$ and $z_j$ integrals are factorized, so we can compute them separately. This can be done by a direct late-time expansion of the homogeneous propagator\footnote{We emphasize that this expansion leads to a series expression for the final result of finite convergence region, and thus is \emph{not} an approximation.}:
\bge
\label{eq_DmpInMode}
  \wt{D}_{-+}^{(\wt\nu)}(r_iz_i,r_jz_j)=\si(r_iz_i)\si^*(r_jz_j),
\ede
where the mode function $\si(rz)$ for the  scalar with mass $\wt\nu$ can be written as:
\begin{align}
\label{eq_ModeExpansion}
  \si(rz)= &~\sum_{\cc=\pm}\sum_{m=0}^\infty A_{m,\cc}(r)(-z)^{2m+\ii\cc\wt\nu+3/2}\\
  \label{eq_Amc}
  A_{m,\cc}=&~\sqrt{\FR{2}{\pi}}\FR{e^{+\pi\cc\wt\nu/2}\Gamma(-m-\ii\cc_1\wt\nu)}{m!}\Big(\FR{r}{2}\Big)^{2m+\ii\cc\wt\nu+3/2}.
\end{align}
Now, for later convenience, let us define the following integrals:
\begin{align}
\label{eq_gLpmc}
  \gL_\pm^{\cc_1}=&\pm\ii\int \di z_i(-z_i)^{p_i+2m_1+\ii\cc_1\wt\nu+3/2}e^{\pm\ii z_i}\mathcal{I}_{\pm}^\text{(L)},\\
\label{eq_gRpmc}
  \gR_\pm^{\cc_2}=&\pm\ii\int \di z_j(-z_j)^{p_j+2m_2+\ii\cc_2\wt\nu+3/2}e^{\pm\ii z_i} \mathcal{I}_{\pm}^\text{(R)},
\end{align}
where $\cc_{1,2}$ are parameters labeling the shadow pair $\pm\ii\wt\nu$ associated with the internal line $K$.  
These are the time integrals of two subgraphs after the cuts, with the SK indices (subscripts) not summed. We also define the corresponding integrals with the SK indices summed: 
\begin{align}
  &\gL^\cc= \gL_+^\cc+\gL_-^\cc,
  &&\gR^\cc= \gR_+^\cc+\gR_-^\cc.
\end{align}
Then, after substituting the expansion (\ref{eq_DmpInMode}) and (\ref{eq_ModeExpansion}) into the integral (\ref{eq_CutKtG}). it is easy to see that the single cut $\mathop\text{Cut}\limits_{K}[\G]$ can be expressed as a sum of factorized terms as:
\begin{align}
\label{eq_CutKtGmid}
   \mathop\text{Cut}_{K}\big[\G\big]
  =&\sum_{\{\cc,m\}}\Big[A_{m_1,\cc_1}(r_i)A_{m_2,\cc_2}^*(r_j)\gL_+^{\cc_1}\gR_+^{-\cc_2}     
    +A_{m_1,\cc_1}^*(r_i)A_{m_2,\cc_2}(r_j)\gL_+^{-\cc_1}\gR_-^{\cc_2}  \Big] +\text{c.c.},
\end{align}
However, the integrals $\gL_{\pm}^\cc$ and $\gR_{\pm}^\cc$ in the above expression are not yet in the form of CISs for the left and right subdiagrams, since they carry unsummed SK indices. It is desirable if we can combine them into real CISs with SK indices summed. This can be partially achieved by exploiting the explicit expressions of $A_{m,\cc}$. After substituting (\ref{eq_Amc}) into (\ref{eq_CutKtGmid}) and with a little bit of algebra, we get:
\begin{align}
\label{eq_CutKtGmid2}
\mathop\text{Cut}_{K}\big[\G\big]
=&~\FR{2}{\pi}\FR{1}{m_1!m_2!}\Big(\FR{r_i}2\Big)^{2m_1+3/2}\Big(\FR{r_j}2\Big)^{2m_2+3/2}  
  \gL^+\bigg[\Gamma[-m_1-\ii\wt\nu,-m_2-\ii\wt\nu]\Big(\FR{r_ir_j}4\Big)^{\ii\wt\nu}  \gR^+ \n\\
&+\Gamma[-m_1-\ii\wt\nu,-m_2+\ii\wt\nu]\Big(\FR{r_i}{r_j}\Big)^{+\ii\wt\nu} (e^{+\pi\wt\nu}\gR_+^{-}+e^{-\pi\wt\nu}\gR_-^{-}) \bigg]+\text{c.c.}.
\end{align}
Most of the terms in this expression are already written in terms of the complete CISs $\gL^\cc$ and $\gR^\cc$, except for the last term in the square bracket, where we have a pair of $\gR_{\pm}^{-}$ and they cannot be trivially combined into a single CIS $\gR^-$. To solve this problem, we go back to the explicit expression of a general CIS discussed in the previous section. Recall that all massive family trees discussed before are constructed recursively from the 1-site graph $\mft{1}$ in (\ref{eq_1site}), which is itself a bulk time integral with the SK index summed. Thus, to get a massive family tree with an unsummed SK index, we only need to work with the 1-site integral (\ref{eq_1site}) without the summation over $\aa=\pm$. Explicitly, let us denote the 1-site integral with unsummed SK index as $\mft{1}_\pm$, then:
\begin{align}
\label{eq_1siteSK}
  &\mft{1}_\pm = e^{\mp\ii\pi p/2}\Gamma(1+p);
  &&\mft{1}=\mft{1}_+ + \mft{1}_- = 2\cos(\pi p/2)\Gamma(1+p).
\end{align}
Now, we can choose to carry out the recursive construction of any tree graph based on either $\mft{1}$ or $\mft{1}_\pm$. The net effect is that we change all $p$ in (\ref{eq_1siteSK}) by the sum of $p$'s on all the vertices of the graph which we call $p_\text{tot}$. Apart from the cosine or phase factors, all other factors in the final CIS are actually ignorant of the SK index. From this analysis, we get the following result that holds for arbitrary massive family trees:
\begin{align}
  \FR{\mft{1\cdots N}_\pm}{\mft{1\cdots N}}=\FR{e^{\mp \ii\pi p_\text{tot}/2}}{2\cos(\pi p_\text{tot}/2)}.
\end{align}
In particular, we have $ \gR_\pm^-/\gR^-=e^{\mp\ii\pi p_\text{R}/2}/[2\cos(\pi p_\text{R}/2)]$, where, again, $p_\text{R}$ is total twist (namely, the sum of $p$'s on all vertices) of $\gR^-$  \emph{after} the shift $p_j\to p_j+2m_2-\ii\wt\nu+3/2$. (See (\ref{eq_gRpmc}).) Then, we can use this result to rewrite the single cut of the graph as:
\begin{align}
 \mathop\text{Cut}_{K}\big[\G\big]
  =&\sum_{m_1,m_2=0}^\infty\FR{2}{\pi}\FR{1}{m_1!m_2!}\Big(\FR{r_i}2\Big)^{2m_1+3/2}\Big(\FR{r_j}2\Big)^{2m_2+3/2} 
 \gL^+\Bigg\{\Gamma[-m_1-\ii\wt\nu,-m_2-\ii\wt\nu]\Big(\FR{r_ir_j}4\Big)^{\ii\wt\nu}  \gR^+ \n\\
&+\Gamma[-m_1-\ii\wt\nu,-m_2+\ii\wt\nu]\Big(\FR{r_i}{r_j}\Big)^{+\ii\wt\nu}  \FR{\cos\big[\fr{\pi(p_\text{R}+2\ii\wt\nu)}2\big]}{\cos\big(\fr{\pi p_\text{R}}2\big)}\gR^- \Bigg\}+\text{c.c.}
\end{align}
In this form, we have succeeded in writing the single cut of a graph in terms of two CISs of the subgraphs. 

\paragraph{Summary and remarks} Let us summarize our result in a clearer form for easier generalization in the following. Suppose that we are considering a $V$-vertex graph $\G_V$ with $E_1$ chosen as the largest energy variable, and we take the single cut of $\G_V$ with respect to a given line $K_\al$ that connecting Vertices $i$ and $j$. Further assuming that Vertex $i$ is on the same side as the maximal-energy vertex $E_1$. Then, the single cut $\mathop\text{Cut}\limits_{K_\al} \big[\G_V\big]$ can be expressed as the sum of products of \emph{augmented} and \emph{flattened} CISs of two subdiagrams, as:
\begin{keyeqn}
\begin{align}
\label{eq_SingleCut}
  \mathop\text{Cut}_{K_\al} \big[\G_V\big] = \mft{\wh{1}\cdots i^\sharp\cdots }\Big\{\mft{ \cdots j^{\sharp}\cdots } + \mft{\cdots j^{\flat}\cdots }\Big\} + (\wt\nu_\al\to-\wt\nu_\al).
\end{align}
\end{keyeqn}
Here we added a hat to $\wh{1}$ to highlight that $E_1$ is chosen as the largest vertex energy in this expression. Also, we have defined the following \emph{tuned} massive family trees with one entry augmented ($\sharp$) or flattened ($\flat$), as:
\begin{align}
\label{eq_augm}
  \mft{\cdots i^\sharp \cdots}\equiv &\sum_{m=0}^\infty\sqrt{\FR{2}{\pi}}\FR{\Gamma(-m-\ii\wt\nu_\al)}{m!}\Big(\FR{K_\al}{2E_i}\Big)^{2m+\ii\wt\nu_\al+3/2}\mft{\cdots i \cdots}_{p_i\to p_i+2m+\ii\wt\nu_\al+3/2}~,\\ 
\label{eq_flat}
  \mft{\cdots i^\flat \cdots}\equiv &\sum_{m=0}^\infty\sqrt{\FR{2}{\pi}}\FR{\Gamma(-m+\ii\wt\nu_\al)}{m!} \Big(\FR{K_\al}{2E_i}\Big)^{2m-\ii\wt\nu_\al+3/2}\n\\
  &\times \bigg\{\FR{\cos\big[\fr{\pi(p_\text{tot}+2\ii\wt\nu_\al)}2\big]}{\cos\big(\fr{\pi p_\text{tot}}2\big)}\mft{\cdots i  \cdots}\bigg\}_{p_i\to p_i+2m-\ii\wt\nu_\al+3/2}~,
\end{align}
where $\wt\nu_\al$ is the mass of Line $K_\al$, $p_\text{tot}$ is the total twist of the subgraph \emph{before} the shift $p_i\to p_i+2m-\ii\wt\nu_\al+3/2$. Moreover, we rewrite complex conjugation in previous derivations as $(\wt\nu_\al\to-\wt\nu_\al)$, assuming that $\ii\wt\nu_\al$ is the only complex variable that needs to flip sign under complex conjugation. In particular, we assume that both subgraphs before the tuning are real. The appearance of $(\wt\nu_\al\to-\wt\nu_\al)$ is easy to understand: each bulk massive scalar gives rise to two late-time falloffs with conformal dimensions $\Delta_\pm=\fr32\pm\ii\wt\nu_\al$, which are called shadow conjugates of each other. We will borrow this terminology and call $(\wt\nu_\al\to-\wt\nu_\al)$ the shadow conjugate. 

Clearly, the cut is not symmetric with respect to the two subgraphs, due to our choice of energy orderings. We can think of the cut as being directional, flowing from the subgraph containing the largest vertex energy to the other subgraph, as shown in Fig.\;\ref{fig_singlecut}. Then, the subgraph containing the largest vertex energy is augmented, and the other subgraph is both augmented and flattened. (Note that the augmentation and flattening of the same graph are not complex conjugates of each other.) 

The two subgraphs are themselves CISs. Therefore, we can apply the result of (\ref{eq_MFT}) to directly write down their expression. In particular, the CIS of the $i$-subgraph is still expressed as a series expanded in $1/E_1$ since it contains the maximal energy site $E_1$. The other subgraph that contains Vertex $j$, however, is independent of $E_1$. Therefore, we need to find the largest vertex energy within this subgraph, called $E_2$, and expand the CIS of $j$-subgraph in terms of $1/E_2$. 

In fact, it is still true that the whole subgraph is expanded in the inverse of the locally largest vertex energy (say, $E_1$) even after being tuned (augmented/flattened). This is clearly true if the tuned vertex ($E_i$) coincides with the vertex carrying the locally maximal energy, as one can easily see from (\ref{eq_augm})-(\ref{eq_flat}). On the other hand, if $i\neq 1$, then we need to go back to the explicit expression for the CIS in (\ref{eq_MFT}), in which the vertex energy $E_i$ appears in the tuned CIS through the power factor $(E_i/E_1)^{\ell_i+(p_i+2m\pm\ii\wt\nu_\al+3/2)+1}$. Combining it with the other $E_i$-dependent energy ratio factor $(K_\al/2E_i)^{2m\pm\ii\wt\nu_\al+3/2}$ in (\ref{eq_augm})-(\ref{eq_flat}), we have:
\bge
\label{eq_EratioRecombined}
   \Big(\FR{K_\al}{2E_i}\Big)^{2m\pm\ii\wt\nu_\al+3/2}\Big(\FR{E_i}{E_1}\Big)^{\ell_i+(p_i+2m\pm\ii\wt\nu_\al+3/2)+1}=\Big(\FR{K_\al}{2E_1}\Big)^{2m\pm\ii\wt\nu_\al+3/2}\Big(\FR{E_i}{E_1}\Big)^{\ell_i+p_i +1}.
\ede 
Therefore, everything in (\ref{eq_augm})-(\ref{eq_flat}) is actually expressed as expansion in the inverse of the locally maximal vertex energy $E_1$(\footnote{By comparing  (\ref{eq_EratioRecombined}) with the expression of the CIS in (\ref{eq_MFT}), one may want to interpret the augmentation/flattening of a graph as shifting the summation variable $m_i$ instead of shifting $p_i$. This is indeed possible if $E_i$ is not the locally maximal energy $E_1$. When $E_i$ and $E_1$ coincide, however, this interpretation does not work since we do not introduce a summation variable $m$ to the largest-energy site. Thus we choose the interpretation of shifting $p_i$ for maximal generality, at the expense of obscuring the fact that all energies in (\ref{eq_augm})-(\ref{eq_flat}) are actually divided by the maximal energy $E_1$.}).

The physical meaning of the single cut is quite clear: it represents the part of the whole diagram that shows the nonanalytic energy dependences due to the on-shell resonance of the $K_\al$ line, but not to any other lines. In particular, the first term in (\ref{eq_SingleCut}) involving two augmented subgraphs, together with its complex conjugate, gives rise to the so-called ``nonlocal signal'' in the context of CC physics. The complex power $K_\al^{\pm 2\ii\wt\nu_\al}$ in the nonlocal signal shows that it is nonanalytic as the line energy $K_\al\to 0$. As such, it gives rise to a characteristic oscillatory behavior as we send $K_\al\to 0$ but keep all other energies fixed\footnote{We note that the single cut does not give all the nonlocal signals with respect to $K_\al$. The full nonlocal signal in $K_\al$ corresponds to the sum of either cutting or not cutting all other lines. (The $K_\al$ line is always cut.) }. As is known, it can be viewed as from the two-point correlator of the conformal operator arising from a late-time expansion of the massive line $K_\al$. In particular, it is symmetric with respect to the two sides, being ignorant about the relative sizes of vertex energies. Therefore, the two subgraphs in this part also appear symmetrically and are both augmented. 

On the other hand, the second term in (\ref{eq_SingleCut}) with both augmented and flattened subgraphs, together with the complex conjugate, contributes to the so-called ``local signal.'' The local signal can be thought of as from the two-point correlator of the late-time conformal operator with its shadow conjugate. As such, it is asymmetric with respect to the exchange of two subgraphs, and is sensitive to the relative sizes of vertex energies. The generation of nonlocal and local signals from cutting the diagram symmetrically or asymmetrically are also investigated in previous works \cite{Tong:2021wai,Qin:2023bjk,Qin:2023nhv,Liu:2024xyi,Ema:2024hkj}. 

\subsection{Multiple cuts}

We can easily generalize the above discussion to multiple cuts of a diagram. Again, a $V$-vertex graph $\G_V$ with $C$ cuts over Line $K_{\al_1},\cdots,K_{\al_C}$ corresponds to the part of the solution that is annihilated by $2C$ differential operators involving $K_{\al_n}$ $(n=1,\cdots,C)$, but not by any of the rest of $2V-2C-2$ operators. Also, the cut is sensitive to energy ordering. So we still work with the region where all vertex energies are arbitrarily ordered, but every line energy is required to be smaller than its two adjacent vertex energies. 

Repeating the derivation of the previous subsection, we see that, cutting an arbitrary number of lines in a diagram can be done in the following way:
\begin{enumerate}
  \item The removal of $C$ cut lines renders the original graph $\G_V$ to $C+1$ disjoint subgraphs, which we call $\mathcal{H}_1,\cdots,\mathcal{H}_{C+1}$. Since the vertex energies have been given an order from the very beginning, each subgraph has its own maximal vertex energy, which we call $E_1',\cdots,E_{C+1}'$, respectively. 
  
  \item Every cut line $K_{\al}$ connects two subgraphs, say $\mathcal{H}_i$ and $\mathcal{H}_j$. Then, we compare their own locally maximal energies $E_i'$ and $E_j'$ to determine the direction of the cut: If $E_i'>E_j'$, the cut is directed to $\mathcal{H}_j$, and otherwise to $\mathcal{H}_i$. That is, the cut always carries a direction flowing from the subgraph carrying the larger maximal vertex energy to the other one.
  
  \item The result of the $C$ cuts is then expressed as the sum and products of the CISs of $C+1$ subgraphs. Explicitly, every cut generates a pair of terms together with their shadow conjugates, in which the ``outgoing'' vertex is augmented in both terms, and the ``ingoing'' vertex is augmented and flattened in the two terms, respectively. Note that augmentation and flattening are defined with respect to a vertex \emph{and} a cut line, so it is possible that a vertex is both augmented and flattened at the same time.
\end{enumerate}

Following the same procedure of deriving the single cut, we can find a general expression for a subgraph augmented and flattened an arbitrary number of times. Suppose a fully nested subgraph (also a CIS) is augmented $A$ times, acting on the vertices $E_{1},\cdots,E_{A}$ (with twists $p_1,\cdots,p_A$), and flattened $F$ times, acting on the vertices $E_{1}',\cdots,E_{F}'$ (with twists $p_1',\cdots,p_F'$). These $A+F$ vertices do not have to be distinct. Further assume that the $A$ augmentations, $\sharp_1,\cdots,\sharp_A$, are associated with cut lines $K_1,\cdots,K_A$ with mass $\wt\nu_1,\cdots,\wt\nu_A$, and the $F$ flattenings, $\flat_1,\cdots,\flat_F$, are associated with cut lines $K'_1,\cdots,K'_F$ with mass $\wt\nu'_1,\cdots,\wt\nu'_F$. Then, the $A$-fold augmented and $F$-fold flattened CIS is given by:
\begin{keyeqn}
\begin{align}
\label{eq_MFTaf}
  &\mft{(\cdots)^{\sharp_1\cdots\sharp_A \flat_{1}\cdots \flat_{F}}}
  =\sum_{\{\ell, m\}}\Big(\FR{2}\pi\Big)^{(A+F)/2}\bigg\{\FR{\cos\big[\pi(p_\text{tot}/2+\ii\wt\nu_{1\cdots F}')\big]}{\cos(\pi p_\text{tot}/2)}\mft{\cdots}\bigg\}_{\subalign{p_{\al}&\to p_{\al}+2\ell_\al+\ii\wt\nu_\al+3/2 \\ p_{\be}'&\to p_{\be}'+2m_\be-\ii\wt\nu_\be'+3/2}}\n\\
  &~\times\prod_{\al=1}^A \FR{\Gamma(-\ell_\al-\ii\wt\nu_\al)}{\ell_\al!}\Big(\FR{K_\al}{2E_{\al}}\Big)^{2\ell_\al+\ii\wt\nu_\al+3/2}\prod_{\be=1}^F \FR{\Gamma(-m_\be+\ii\wt\nu_\be')}{m_\be!}\Big(\FR{K_\be'}{2E_{\be}'}\Big)^{2m_\be-\ii\wt\nu_\be'+3/2}.
\end{align}
\end{keyeqn}

In a graph with multiple cuts, importantly, the direction of a cut generally does not agree with either the induced order or the imposed order: The cut direction is always determined by comparing the two locally maximal energies of the two subgraphs joined by the cut line. However, it is easy to observe the following two facts: First, the direction of the single cut always coincides with the induced order of the graph, as already shown in the previous subsection; Second, the direction of the maximal cut (namely, all lines being cut) coincides with the imposed order of the graph. Thus we see that, although the CIS of a graph is ignorant about the relative sizes of all but the largest vertex energies, the full expression of the correlator is sensitive to energy ordering through the cuts. 

\paragraph{Example 1: Triple cuts} The rules for taking multiple cuts are best illustrated with examples. Here we present two examples. More examples will be given in Sec.\;\ref{sec_examples}. 
\begin{figure}[t]
\centering
\includegraphics[width=0.65\textwidth]{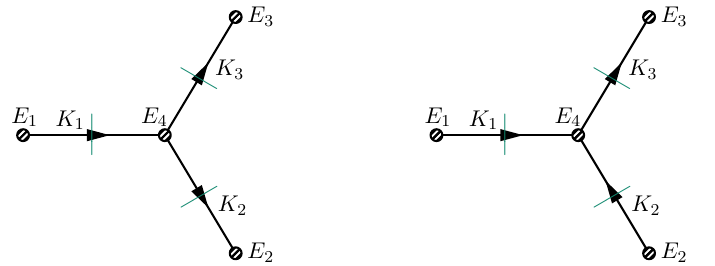}
\caption{Taking the triple cuts of a 4-vertex star graph. The vertex energies are ordered as $E_1>E_4>E_2,E_3$ in the left graph, and $E_1,E_2>E_4>E_3$ in the right graph.  }
\label{fig_3cuts}
\end{figure}

In our first example, we take the maximal cut of a 4-site star graph, as shown in Fig.\;\ref{fig_3cuts}. We take Vertex $i$ to have vertex energy $E_i$ and twist $p_i$ ($i=1,\cdots,4$), and Line $\al$ to have line energy $K_\al$ and mass $\wt\nu_\al$ ($\al=1,2,3$). To start, we need to order the four vertex energies, and we choose two orderings. In the first ordering, we take $E_1>E_4>E_2,E_3$ (left graph of Fig.\;\ref{fig_3cuts}). Then, the triple cuts can be expressed as:
\begin{align}
\label{eq_triplecut1}
  \mathop{\text{Cut}}_{K_1,K_2,K_3}\big[\G_4\big]=&~ \mft{1^{\sharp_1}}\Big(\mft{4^{\sharp_1\sharp_2\sharp_3}}+\mft{4^{\flat_1\sharp_2\sharp_3}}\Big) \Big(\mft{2^{\sharp_2}}+\mft{2^{\flat_2}}\Big)\Big(\mft{3^{\sharp_3}}+\mft{3^{\flat_3}}\Big)  
 +\text{shadows}.
\end{align}
Here the terms collected in ``shadows'' mean to include all independent shadow conjugates of the cut lines, namely, $\wt\nu_i\to -\wt\nu_i$ $(i=1,2,3)$. Explicitly, in the above example: 
\begin{align}
\label{eq_shadowconj}
  (\cdots)+\text{shadows}= \Big\{\big[(\cdots)+(\wt\nu_1\to-\wt\nu_1)\big]+(\wt\nu_2\to-\wt\nu_2)\Big\}+(\wt\nu_3\to-\wt\nu_3).
\end{align}
Also, we have augmented or flattened certain vertices multiple times. Following the general formula (\ref{eq_MFTaf}), it is straightforward to write down their expressions. For instance: 
\begin{align}
  \mft{4^{\sharp_1\sharp_2\sharp_3}}=&~\sum_{m_1,m_2,m_3=0}^\infty \Big(\FR{2}\pi\Big)^{3/2}\prod_{\al=1}^3\bigg[\FR{\Gamma(-m_\al-\ii\wt\nu_\al)}{m_\al!}\Big(\FR{K_\al}{2E_{4}}\Big)^{2m_\al+\ii\wt\nu_\al+3/2}\bigg]\n\\
  &\times \mft{4}_{p_4\to p_4+2m_{123}+\ii(\wt\nu_1+\wt\nu_2+\wt\nu_3)+9/2}\;, \\
  \mft{4^{\flat_1\sharp_2\sharp_3}}=&~\sum_{m_1,m_2,m_3=0}^\infty \Big(\FR{2}\pi\Big)^{3/2}\prod_{\al=2}^3\bigg[ \FR{\Gamma(-m_\al-\ii\wt\nu_\al)}{m_\al!}\Big(\FR{K_\al}{2E_{4}}\Big)^{2m_\al+\ii\wt\nu_\al+3/2}\bigg] \FR{\Gamma(-m_1+\ii\wt\nu_1)}{m_1!}\n\\
  &\times \Big(\FR{K_1}{2E_2}\Big)^{2m_1-\ii\wt\nu_1+3/2}\bigg\{\FR{\cos\big[\fr{\pi p_4+2\ii\wt\nu_1}2\big]}{\cos\big(\fr{\pi p_4}2\big)}\mft{4}\bigg\}_{p_4\to p_4+2m_{123}+\ii(-\wt\nu_1+\wt\nu_2+\wt\nu_3)+9/2}\;,~\text{etc.}\,.
\end{align}
Clearly, both of the triply dressed family trees can be expressed in terms of trivariate Lauricella function $\text{F}_\text{C}$. The other singly dressed family trees in (\ref{eq_triplecut1}) can all be expressed as the standard Gauss's hypergeometric function ${}_2\text{F}_1$. We will not spell them out explicitly. 
 
Now consider a different ordering of vertex energies, namely $E_1,E_2>E_4>E_3$, shown in the right graph of Fig.\;\ref{fig_3cuts}. In this case, the triple cuts can be expressed as: 
\begin{align}
  \mathop{\text{Cut}}_{K_1,K_2,K_3}\big[\G_4\big]=&~\mft{1^{\sharp_1}}\mft{2^{\sharp_2}}\Big(\mft{4^{\sharp_1\sharp_2\sharp_3}}+\mft{4^{\flat_1\sharp_2\sharp_3}}+\mft{4^{\sharp_1\flat_2\sharp_3}}+\mft{4^{\flat_1\flat_2\sharp_3}}\Big)\Big(\mft{3^{\sharp_3}}+\mft{3^{\flat_3}}\Big)\n\\
  &~+\text{shadows},
\end{align} 
where the shadow conjugates are again given by (\ref{eq_shadowconj}). Thus we see that the expression for the cuts is crucially dependent on the choice of vertex energy ordering.

\begin{figure}[t]
\centering
\includegraphics[width=0.85\textwidth]{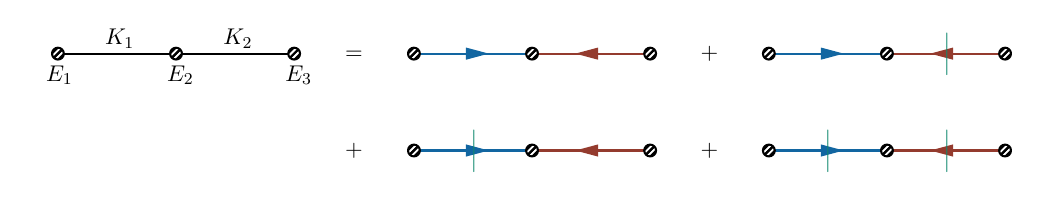}
\caption{The cuts of a 3-site chain graph with energy ordering $E_1>E_3>E_2$. The blue/red lines are correct/wrong lines. }
\label{fig_3chain_cuts}
\end{figure}

\paragraph{Example 2: inward-pointing 3-site graph} We use this example to explain a subtlety when taking cuts, as shown in Fig.\;\ref{fig_3chain_cuts}. Organizing a graph into a sum of the CIS and its cuts amounts to regrouping all the bulk propagators. The time-ordered part gives the CIS and the factorized parts give the cuts. Due to the energy orderings, both the CIS and cuts are directional. It may happen that the direction of a line changes after we cut another line. This leads to a change of directional cut, namely, how we separate out the factorized part from an inhomogeneous propagator as in (\ref{eq_Dsplit}). Below we elucidate this point with the 3-site chain and show that our prescription leads to correct answers. 

When extracting the CIS of this 3-site chain, we should follow the induced order to decide the time ordering of all internal lines. Taking all vertices to be in the $+$ branch, this means that the two bulk propagators $\wt D_{++}$ in the integrand should be written as: (See (\ref{eq_Dpmpm}) and (\ref{eq_DmpInMode}).)
\begin{align}
  \Big[\big(\si_1^*\si_2-\si_1\si_2^*\big)\theta_{21}+\si_1\si_2^*\Big]\Big[\big(\si_3^*\si_4-\si_3\si_4^*\big)\theta_{32}+\si_3\si_4^*\Big].
\end{align}
Here we use the shorthand notation for the mode functions of the internal line: $\si_1=\si(r_1z_1)$, $\si_2=\si(r_2z_2)$, $\si_3=\si(r_3z_2)$, and $\si_4=\si(r_4z_3)$. Also, $\theta_{21}\equiv\theta(r_2z_2-r_1z_1)$ and $\theta_{32}\equiv\theta(r_4z_3-r_3z_2)$.
Expanding the four terms above, we get:
\begin{align}
\label{eq_3siteMFTexpanded}
  &\underbrace{\big(\si_1^*\si_2-\si_1\si_2^*\big)\theta_{21}\big(\si_3^*\si_4-\si_3\si_4^*\big)\theta_{32}}_\text{CIS}+\underbrace{\big(\si_1^*\si_2-\si_1\si_2^*\big)\theta_{21}\si_3\si_4^*}_\text{cutting $K_2$}\n\\
  &+\si_1\si_2^*\big(\si_3^*\si_4-\si_3\si_4^*\big)\theta_{32}+\si_1\si_2^*\si_3\si_4^*.
\end{align}
Clearly, the first term gives the CIS, and the rest of the terms correspond to cuts, including two single cuts and one double cut. Naively, one would expect that the two terms with a single $\theta$ factor correspond to the two single cuts, and the last term with no $\theta$ function corresponds to the double cut. However, things are a little more complicated than that. Below let us look at those cuts more closely. 

 For the single cuts, let us first consider cutting the right line. In this case, the left subgraph is a 2-site graph whose ``local induced order'' agrees with the original induced order of the whole graph. Thus, it is straightforward to see that the single cut over the right line $K_2$ is given by the second term of the above expansion. 

However, when we look at the single cut over the left line $K_1$, we see a new phenomenon: The resulting right subgraph, which is again a 2-site chain, has a locally induced order that is opposite to the original induced order, since we have $E_3>E_2$. Consequently, according to our rule for taking cuts, the single cut over $K_1$ does not agree with the third term in the above expansion, but is given by the following term:
\bge
\label{eq_K1cut}
  \si_1\si_2^*(\si_3\si_4^*-\si_3^*\si_4)\theta_{23}.~~~~~(\text{cutting $K_1$})
\ede
This apparent mismatch is fixed by another mismatch from the double cut. According to our rule, in the double cut, we compare the maximal energies of all subgraphs to determine the directions of the cuts. In this case, it agrees with the imposed order. Accordingly, the double cut is given by:
\bge
\label{eq_doublecut}
  \si_1\si_2^*\si_3^*\si_4.~~~~~(\text{cutting $K_1$ and $K_2$})
\ede 
Then, obviously, the sum of (\ref{eq_K1cut}) and (\ref{eq_doublecut}) agree with the second line of (\ref{eq_3siteMFTexpanded}). 

Although we are showing it with a specific example, the phenomenon is general: When we take a cut of a diagram, the induced order for some subgraphs may change, which amounts to regrouping the inhomogeneous propagators. The upshot is that we should decide the direction of a cut by comparing the two locally maximal vertex energies of the two subgraphs connected by the cut line, as described before. 

\section{Complete Solutions and Selected Examples}
\label{sec_examples}

Given all the inhomogeneous and homogeneous pieces constructed in the previous two sections, we are now ready to present the full solution to an arbitrary tree graph $\G$. In this section, we will first summarize the complete solution, and then provide a few explicit examples, including the 2-site and 3-site chains, and a 4-site star graph. In all examples, we have numerically checked that our results agree perfectly with direct numerical integrations, but are way faster to evaluate than doing bulk time integral, especially when the number of vertices is large. 

\subsection{Complete solutions}

Given a (dimensionless) tree diagram $\G$ with $V$ vertices, carrying vertex energies $E_1,\cdots,E_V$ and twists $p_1,\cdots,p_V$, and $I=V-1$ internal lines, carrying line energies $K_1,\cdots,K_I$ and masses $\wt\nu_1,\cdots,\wt\nu_I$, the analytical answer for its integral (\ref{eq_dimlessIntG}) is generally a sum of products of multivariate hypergeometric functions. Working in the signal region, i.e., the region where all line energies are small compared to all vertex energies, an analytical expression for $\G$ can be expressed as the sum of all cuts of $\G$:
\begin{align}
\label{eq_GfromCuts}
  \G=\sum_{i\in 2^{\{K\}}}\mathop{\text{Cut}}_{i}\big[\G\big].
\end{align}
Here $2^{\{K\}}$ is the power set of the line-energy set $\{K_\al\}$, namely, the set of all subsets of $\{K_\al\}$ (including the empty set and $\{K_\al\}$ itself). Thus, there are a total of $2^I$ independent cuts. Among these $2^I$ terms, the uncut piece is the completely inhomogeneous solution $\text{CIS}[\G]$, and is given by (\ref{eq_MFT}). The single cut over a line $K$ is given by (\ref{eq_SingleCut}). The multiple cuts can be defined similarly. In particular, a term with $C$ cuts can be expressed as the sum of products of dressed CISs of $C+1$ subgraphs. The two types of dressing, called augmentation and flattening, are defined in (\ref{eq_augm}) and (\ref{eq_flat}). The rule for augmenting and/or flattening a CIS an arbitrary number of times is given in (\ref{eq_MFTaf}). 

With the above rule, we can directly write down the solution to an arbitrary tree graph $G$ without doing any real computation. On the other hand, the structure of an arbitrary tree graph is fully fixed by decoding the family parameter $q_i$ in the CIS, and also by all the cuts. Therefore, from the expression, it is also straightforward to recover the tree graph, including its topology and all parameters. Thus, there is a one-to-one relationship between the tree graph and the solution. In this sense, we can say that our procedure provides a WYSIWYG (what-you-see-is-what-you-get) solution to arbitrary trees.

As mentioned in Sec.\;\ref{sec_strategy}, there is a direct relation between the cut/uncut of a line and the signal/background in CC physics. From our explicit solutions, the meaning of CC signals can be better understood. There are two distinct types of signals that we can identify from a cut. 

First, the nonlocal signal, coming from the part of the cut where both subgraphs are augmented (together with shadow conjugates), corresponds to a nonanalytic power of $K_\al^{\pm 2\ii\wt\nu_\al}$ as $K_\al\to 0$ where $K_\al$ is the line energy of the cut line. Thus, a nonlocal signal is associated with a line $K_\al$, and is to be found by sending the line energy $K_\al\to 0$. 

On the other hand, there is a local signal, coming from the part of a cut where one subgraph is flattened. By examining the expressions for the cuts, we see that the local signal corresponds to a nonanalytic power of vertex energy ratios, such as $(E_1/E_2)^{\pm\ii\wt\nu_\al}$, where $E_{1,2}$ are \emph{not} vertex energies of the two endpoints of the cut line $K_\al$, but the maximally vertex energies of the two subgraphs. Therefore, a local signal is associated with a vertex $E_i$, and is to be found by sending $E_i\to \infty$. 

Given the simple rules of writing down the full solutions as in (\ref{eq_GfromCuts}), computing arbitrary massive tree graphs becomes a trivial task. Below we apply these rules to write down answers for a few simple tree diagrams, including the 2-site chain, the 3-site chain, and the 4-site star. Among them, the result for the 2-site chain is well known, while the results for the 3-site and 4-site graphs are new. 

\subsection{Two-site chain}

\begin{figure}[t]
\centering
\includegraphics[width=0.75\textwidth]{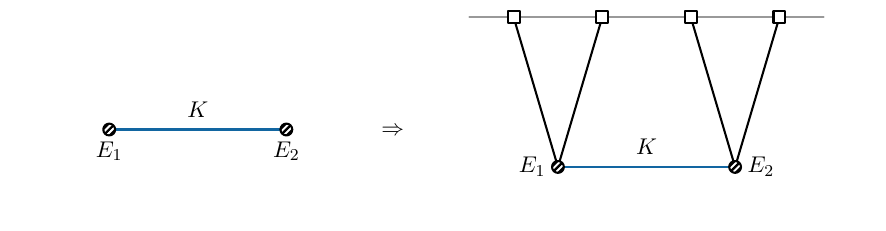}\vspace{-5mm}
\caption{The 2-site chain graph and a corresponding 4-point inflaton correlator. }
\label{fig_2chain}
\end{figure}

We first consider the simplest nontrivial example, namely a 2-site chain, which can be, but not necessarily, from a 4-point correlator $\la\varphi_{\bm k_1}\varphi_{\bm k_2}\varphi_{\bm k_3}\varphi_{\bm k_4}\ra'$ of the massless inflaton $\varphi$ with single massive exchange $\si_{\bm K}$ in the $s$-channel, namely, $\bm K=\bm k_1+\bm k_2$, shown in Fig.\;\ref{fig_2chain}. This is a well-studied case and its explicit analytical expression can be found in many works, e.g., \cite{Arkani-Hamed:2018kmz,Qin:2022fbv}. Using our notation, the dimensionless graph $\G_2$ for the 2-site chain is specified by the two vertex energies $E_1=k_{12}$ and $E_{2}=k_{34}$, together with the line energy $K\equiv |\bm K|$. 

Without loss of generality, we work in the region where $E_1>E_2$. The other region with $E_2>E_1$ can be obtained by trivially switching labels. Then, it is straightforward to write down the answer for $\G_2$ following the rules presented above:
\begin{align}
\label{eq_G2}
    \G_{2}=\text{CIS}\big[\G_{\text{2}}\big]+\mathop{\text{Cut}}_{K}\big[\G_{\text{2}}\big]=\mft{12}+\Big\{\mft{1^\sharp}\Big(\mft{2^\sharp}+\mft{2^\flat}\Big)+(\wt\nu\to-\wt\nu)\Big\}.
\end{align}
Here the CIS corresponds to the ``background,'' and is given by:
\begin{align}\label{2siteCIS}
    \mft{12}=\ &\sum_{\ell,m=0}^{\infty} 
      \FR{(-1)^{\ell}4\cos(\pi p_{12}/2)\Gamma(p_{12}+\ell+2m+5)}{\ell!\big(\fr{p_2+\ell}{2}+\fr{5}{4}\pm\fr{\ii\wt\nu}{2}\big)_{m+1}}\Big(\FR{K}{2E_1}\Big)^{2m+3}\Big(\FR{E_2}{E_1}\Big)^{p_2+\ell+1}.
\end{align}
This series can be summed into a Kampé de Fériet function \cite{Arkani-Hamed:2018kmz}, but we do not spell it out explicitly. 

The signal, on the other hand, corresponds to the single cut. Given the expression for the 1-site graph $\mft{1}$ in (\ref{eq_1site}), the tuned graphs can be found according to (\ref{eq_augm}) and (\ref{eq_flat}). In this particular example, the tuned graphs can be summed into known hypergeometric functions. It is convenient to introduce the following function:
\begin{align}
\label{eq_boldF}
  &~\mathbf{F}^{p_1}_{\wt\nu}(E_1,K)\equiv \sum_{m=0}^{\infty}\sqrt{\FR{2}{\pi}}\FR{\Gamma(-m-\ii\wt\nu)}{m!}\Big(\FR{K}{2E_1}\Big)^{2m+3/2+\ii\wt\nu}\mft{1}_{p_1\to p_1+2m+3/2+\ii\wt\nu}\n\\ 
    &=- \FR{2^{3/2+p_1}\cos\big[\fr{\pi(p_1+\ii\wt\nu+3/2)}{2}\big]}{\sin(\pi\ii\wt\nu)}\Big(\FR{K}{E_1}\Big)^{3/2+\ii\wt\nu} {}_{2}\mathcal{F}_1\left[\bgm\fr{p_1}{2}+\fr{5}{4}+\fr{\ii\wt\nu}{2},\fr{p_1}{2}+\fr{7}{4}+\fr{\ii\wt\nu}{2}\\1+\ii\wt\nu\edm \middle| \FR{K^2}{E_1^2} \right],
\end{align}
where ${}_2\mathcal{F}_1$ is the dressed Gauss's hypergeometric function and is defined in (\ref{eq_dressed2F1}), and we have defined this augmented graph into a new function $\mb{F}^{p}_{\wt\nu}$. Then, the 3 tuned 1-site graphs can be easily expressed as:
\begin{align}
  &\mft{1^\sharp}=\mathbf{F}^{p_1}_{\wt\nu}(E_1,K),
  &\mft{2^\sharp}=\mathbf{F}^{p_2}_{\wt\nu}(E_2,K),
  &&\mft{2^\flat}=\FR{\cos\big[\fr{\pi}{2}(p_2+\fr{3}{2}+\ii\wt\nu)\big]}{\cos\big[\fr{\pi}{2}(p_2+\fr{3}{2}-\ii\wt\nu)\big]}\mathbf{F}^{p_2}_{-\wt\nu}(E_2,K). 
\end{align}
Now, it is straightforward to check that our result (\ref{eq_G2}) agrees with known results in the literature. 
 
\subsection{Three-site chain}

\begin{figure}[t]
\centering
\includegraphics[width=0.85\textwidth]{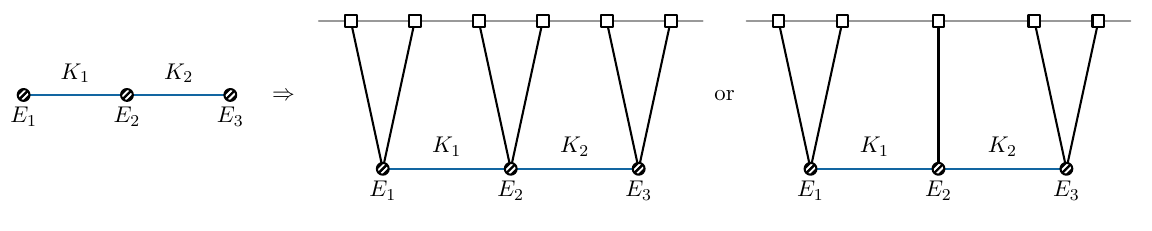}\vspace{-5mm}
\caption{The 3-site chain graph and corresponding 6-point or 5-point inflaton correlators. }
\label{fig_3chain}
\end{figure}

Next, we consider the 3-site chain with two massive exchanges. This example has been computed analytically in \cite{Xianyu:2023ytd} with PMB and FTD for a specific energy order. Related to this, a degenerate case with two massive exchanges was computed in \cite{Aoki:2024uyi}. Here we present the full result for general nondegenerate kinematics in the signal region, including all possible vertex energy orderings. A 3-site chain is specified by three vertex energies $E_1,E_2,E_3$, and two line energies $K_1,K_2$. It can be from a 5-point or 6-point inflaton correlator with two massive exchanges, as shown in Fig.\;\ref{fig_3chain}. 

According to our rule, the answer for the 3-site chain is:
\begin{align}
\label{eq_G3}
  \G_3= \text{CIS}\,\big[\G_3\big]+\mathop{\text{Cut}}_{K_1}\big[\G_3\big]+\mathop{\text{Cut}}_{K_2}\big[\G_3\big]+\mathop{\text{Cut}}_{K_1,K_2}\big[\G_3\big].
\end{align}
However, to write down the explicit expressions for each term, we need to specify the energy order. There are three independent choices, 1) $E_2>E_1,E_{3}$; 2) $E_1>E_2>E_3$; 3) $E_1>E_3>E_2$. Below we consider these cases in turn.

\paragraph{Case 1: $\bm{E_2>E_1,E_3}$} Using our terminology for massive family trees, both lines in this case are correct lines, so the graph is already a family tree by itself, and $E_2$ is the mother of $E_1$ and $E_3$. Thus, the general expression (\ref{eq_G3}) can be unpacked in the following way:
\begin{align}
\label{eq_3chainCase1}
\G_3
  =&~\mft{2(1)(3)}+\Big\{\Big(\mft{1^{\sharp_1}}+\mft{1^{\flat_1}}\Big)\mft{2^{\sharp_1} 3}+(\wt\nu_1\to-\wt\nu_1)\Big\}+\Big\{\mft{2^{\sharp_2}1}\Big(\mft{3^{\sharp_2}}+\mft{3^{\flat_2}}\Big)+(\wt\nu_2\to-\wt\nu_2)\Big\}\n\\
  &+\Big\{\Big[\Big(\mft{1^{\sharp_1}}+\mft{1^{\flat_1}}\Big)\mft{2^{\sharp_1\sharp_2}}\Big(\mft{3^{\sharp_2}}+\mft{3^{\flat_2}}\Big)+(\wt\nu_1\to-\wt\nu_1)\Big]+(\wt\nu_3\to-\wt\nu_3)\Big\}.
\end{align}
Using (\ref{eq_MFT}), we have the following expression for the CIS:
\begin{align}
    \mft{2(1)(3)}=\ &\sum_{\ell_1,\ell_2,m_1,m_2=0}^{\infty} 
  \FR{8 \cos\big(\fr{\pi p_{123}}2\big)(-1)^{\ell_{12}}\Gamma(p_{123}+\ell_{12}+2m_{12}+9)}{\ell_1!\ell_2!\big(\fr{p_1+\ell_1}{2}+\fr{5}{4}\pm\fr{\ii\wt\nu_1}{2}\big)_{m_1+1}\big(\fr{p_3+\ell_2}{2}+\fr{5}{4}\pm\fr{\ii\wt\nu_2}{2}\big)_{m_2+1}}\n\\
  &\times\Big(\FR{K_1}{2E_2}\Big)^{2m_1+3}\Big(\FR{E_1}{E_2}\Big)^{p_1+\ell_1+1} 
   \Big(\FR{K_2}{2E_2}\Big)^{2m_2+3}\Big(\FR{E_3}{E_2}\Big)^{p_3+\ell_2+1}.
\end{align}
Then we have a collection of tuned CISs. First, the singly tuned 1-site CISs can be expressed in terms of the $\mb{F}^{p}_{\wt\nu}$ function defined in (\ref{eq_boldF}), as:
\begin{align}
\label{eq_1sharp}
  &\mft{1^{\sharp_1}}=\mb{F}_{\wt\nu_1}^{p_1}(E_1,K_1),
  &&\mft{1^{\flat_1}}=\FR{\cos\big[\fr{\pi(p_1+3/2+\ii\wt\nu_1)}{2}\big]}{\cos\big[\fr{\pi(p_1+3/2-\ii\wt\nu_1)}{2}\big]}\mb{F}_{-\wt\nu_1}^{p_1}(E_1,K_1),\\
\label{eq_3sharp}
  &\mft{3^{\sharp_2}}=\mb{F}_{\wt\nu_2}^{p_3}(E_3,K_2),
  &&\mft{3^{\flat_2}}=\FR{\cos\big[\fr{\pi(p_3+3/2+\ii\wt\nu_2)}{2}\big]}{\cos\big[\fr{\pi(p_3+3/2-\ii\wt\nu_2)}{2}\big]}\mb{F}_{-\wt\nu_2}^{p_3}(E_3,K_2).
\end{align}
Second, the doubly augmented 1-site CIS $\mft{2^{\sharp_1\sharp_2}}$ can be expressed in terms of an Appell function:
\bge
  \label{eq_2doublesharp}
  \mft{2^{\sharp_1\sharp_2}}=\mathbf{G}^{p_2}_{\wt\nu_1,\wt\nu_2}(E_2,K_1,K_2),
\ede
where we define a new function $\mathbf{G}$ for later convenience:
\begin{align}
\label{eq_boldG}
  &\mathbf{G}^{p_2}_{\wt\nu_1,\wt\nu_2}(E_2,K_1,K_2)\equiv  \sum_{m_1,m_2=0}^{\infty}\FR{2}{\pi}\bigg[\prod_{\al=1}^{2}\FR{\Gamma[-m_\al-\ii\wt\nu_\al]}{m_\al!}\Big(\FR{K_\al}{2E_2}\Big)^{2m_\al+3/2+\ii\wt\nu_\al}\bigg]\mft{2}_{p_2\to p_2+2m_{12}+3+\ii\wt\nu_{12}}\n\\
   &= \FR{2^{p_2+2}\sqrt{\pi}\sin\big[\fr{\pi(p_2+\ii\wt\nu_{12})}{2}\big]}{\sin(\pi\ii\wt\nu_1)\sin(\pi\ii\wt\nu_2)}\Big(\FR{K_1}{E_2}\Big)^{3/2+\ii\wt\nu_1}\Big(\FR{K_2}{E_2}\Big)^{3/2+\ii\wt\nu_2} 
  \mathcal{F}_{4}\left[\bgm\fr{p_2+4+\ii\wt\nu_{12}}{2},\fr{p_2+5+ \ii\wt\nu_{12}}{2}\\1+\ii\wt\nu_1,1+\ii\wt\nu_2\edm\middle|\FR{K_1^2}{E_2^2},\FR{K_2^2}{E_2^2}\right].
\end{align}
Here $\mathcal{F}_4$ denotes the dressed Appell $\text{F}_4$ function, whose definition can be found in (\ref{eq_AppellF4}).

Finally, there are two singly augmented 2-site CISs. Among them, $\mft{2^{\sharp_1}3}$ can be expressed as:
\begin{align}
\label{eq_2sharp3}
  \mft{2^{\sharp_1}3}
  =&\sum_{\ell,m_1,m_2=0}^{\infty}\FR{(-1)^{\ell+m_1}4\sqrt{2/\pi}\cos\big[\fr{\pi(p_{23}+3/2+\ii\wt\nu_1)}{2}\big]\Gamma(p_{23}+\ell+2m_{12}+{13}/{2}+\ii\wt\nu_1)}{\ell!\big(\fr{p_3+\ell}{2}+\fr{5}{4}\pm\fr{\ii\wt\nu_2}{2}\big)_{m_2+1}}\n\\
  &\times\FR{\Gamma(-m_1-\ii\wt\nu_1)}{m_1!}\Big(\FR{K_1}{2E_2}\Big)^{2m_1+3/2+\ii\wt\nu_1}\Big(\FR{K_2}{2E_2}\Big)^{2m_2+3}\Big(\FR{E_3}{E_2}\Big)^{p_3+\ell+1} ,
\end{align}
%
and $\mft{2^{\sharp_2}1}$ is easily obtained from $\mft{2^{\sharp_1}3}$ by the replacement $(E_3,p_3)\to (E_1,p_1)$ and $(K_1,\wt\nu_1)\leftrightarrow (K_2,\wt\nu_2)$.

\paragraph{Case 2: $\bm{E_1>E_2>E_3}$} The 3-site graph with this energy order is also a massive family tree automatically, with three generations being $E_1$, $E_2$, and $E_3$. Thus, we can unpack the general expression (\ref{eq_G3}) as:
\begin{align}
\label{eq_3chainCase2}
  \G_3=&~\mft{123}+\mft{1^{\sharp_1}}\Big(\mft{2^{\sharp_1}3}+\mft{2^{\flat_1}3}\Big)+\mft{12^{\sharp_2}}\Big(\mft{3^{\sharp_2}}+\mft{3^{\flat_2}}\Big)\n\\
  &+\mft{1^{\sharp_1}}\Big(\mft{2^{\sharp_1\sharp_2}}+\mft{2^{\flat_1\sharp_2}}\Big)\Big(\mft{3^{\sharp_2}}+\mft{3^{\flat_2}}\Big)+\text{shadows}.
\end{align}
From now on we will not spell out the shadow conjugates. The CIS is given by:
\begin{align}\label{eq_CIS123}
    \mft{123}= &\sum_{\ell_1,\ell_2,m_1,m_2=0}^{\infty}  \FR{8\cos\big(\fr{\pi p_{123}}2\big)(-1)^{\ell_{12}}\Gamma(p_{123}+\ell_{12}+2m_{12}+9)}{\ell_1!\ell_2!\big(\frac{p_{23}+\ell_{12}+2m_2}{2}+\frac{13}{4}\pm\frac{\ii\wt\nu_1}{2}\big)_{m_1+1}\big(\fr{p_3+\ell_2}{2}+\fr{5}{4}\pm\fr{\ii\wt\nu_2}{2}\big)_{m_2+1}}\n\\
    &\times\Big(\FR{K_1}{2E_1}\Big)^{2m_1+3}\Big(\FR{E_2}{E_1}\Big)^{p_2+\ell_1+1} 
 \Big(\FR{K_2}{2E_1}\Big)^{2m_2+3}\Big(\FR{E_3}{E_1}\Big)^{p_3+\ell_2+1}  .
\end{align}
For the tuned CISs, the singly and doubly tuned 1-site CISs are the same as (\ref{eq_1sharp}), (\ref{eq_3sharp}), and (\ref{eq_2doublesharp}). The new one is:
\begin{align}
\label{eq_2flatsharp}
  \mft{2^{\flat_1\sharp_2}}=\FR{\cos\big[\fr{\pi(p_2+3+\ii\wt\nu_{12})}{2}\big]}{\cos\big[\fr{\pi(p_2+3-\ii\wt\nu_{1}+\ii\wt\nu_2)}{2}\big]}\mathbf{G}^{p_2}_{-\wt\nu_1,\wt\nu_2}(E_2,K_1,K_2),
\end{align}
where $\mathbf{G}$ is defined in (\ref{eq_boldG}).

Finally, there are three singly tuned 2-site CISs. Among them, $\mft{2^{\sharp_1}3}$ has been given in (\ref{eq_2sharp3}), and the two new ones are:
\begin{align}\label{eq_12sharp}
  \mft{12^{\sharp_2}}=&\sum_{\ell,m_1,m_2=0}^{\infty}\FR{(-1)^{\ell+m_1}4\sqrt{2/\pi}\cos\big[\fr{\pi(p_{12}+3/2+\ii\wt\nu_2)}{2}\big]\Gamma(p_{12}+\ell+2m_{12}+{13}/{2}+\ii\wt\nu_2)}{\ell!\big(\fr{p_2+\ell+2m_1}{2}+2+\fr{\ii\wt\nu_2}{2}\pm\fr{\ii\wt\nu_{1}}{2}\big)_{m_2+1}}\n\\
  &\times\FR{\Gamma(-m_1-\ii\wt\nu_2)}{m_1!}\Big(\FR{K_2}{2E_1}\Big)^{2m_1+3/2+\ii\wt\nu_2}\Big(\FR{K_1}{2E_1}\Big)^{2m_2+3}\Big(\FR{E_2}{E_1}\Big)^{p_2+\ell+1}, \end{align}
\begin{align} 
  \mft{2^{\flat_1}3}=&\sum_{\ell,m_1,m_2=0}^{\infty}\FR{(-1)^{\ell+m_1}4\sqrt{2/\pi}\cos\big[\fr{\pi(p_{23}+3/2+\ii\wt\nu_1)}{2}\big]\Gamma(p_{23}+\ell+2m_{12}+{13}/{2}-\ii\wt\nu_1)}{\ell!\big(\fr{p_3+\ell}{2}+\fr{5}{4}\pm\fr{\ii\wt\nu_{2}}{2}\big)_{m_2+1}}\n\\
  &\times\FR{\Gamma(-m_1+\ii\wt\nu_1)}{m_1!}\Big(\FR{K_1}{2E_2}\Big)^{2m_1+3/2-\ii\wt\nu_1}\Big(\FR{K_2}{2E_2}\Big)^{2m_2+3}\Big(\FR{E_3}{E_2}\Big)^{p_3+\ell+1}.
\end{align}

\paragraph{Case 3: $\bm{E_1>E_3>E_2}$} This is an interesting case, as the energy order does not fit directly into a family tree structure, as already discussed at the end of Sec.\;\ref{sec_hom}. According to this discussion, we should express (\ref{eq_G3}) as:
\begin{align}
\label{eq_3chainCase3}
  \G_3=&~\mft{123}+\mft{1^{\sharp_1}}\Big(\mft{32^{\sharp_1}}+\mft{32^{\flat_1}}\Big)+\mft{12^{\sharp_2}}\Big(\mft{3^{\sharp_2}}+\mft{3^{\flat_2}} \Big)\n\\
  &+\mft{1^{\sharp_1}}\Big(\mft{2^{\sharp_1\sharp_2}}+\mft{2^{\flat_1\sharp_2}}+\mft{2^{\sharp_1\flat_2}}+\mft{2^{\flat_1\flat_2}}\Big)\mft{3^{\sharp_2}}+\text{shadows}.
\end{align}
The CIS $\mft{123}$ is identical to Case 2, as given in (\ref{eq_CIS123}). The singly tuned and doubly tuned 1-site CISs are the same as (\ref{eq_1sharp}), (\ref{eq_3sharp}), (\ref{eq_2doublesharp}), and (\ref{eq_2flatsharp}). The new doubly tuned 1-site CISs are:
\begin{align}
  \mft{2^{\sharp_1\flat_2}}=&~\FR{\cos\big[\fr{\pi(p_2+3+\ii\wt\nu_{12})}{2}\big]}{\cos\big[\fr{\pi(p_2+3+\ii\wt\nu_1-\ii\wt\nu_2)}{2}\big]}\mathbf{G}_{\wt\nu_1,-\wt\nu_2}^{p_2}(E_2,K_1,K_2), \\
  \mft{2^{\flat_1\flat_2}}=&~\FR{\cos\big[\fr{\pi(p_2+3+\ii\wt\nu_{12})}{2}\big]}{\cos\big[\fr{\pi(p_2+3-\ii\wt\nu_{12})}{2}\big]}\mathbf{G}_{-\wt\nu_1,-\wt\nu_2}^{p_2}(E_2,K_1,K_2).
\end{align}
There are three singly tuned 2-site CISs, among which $\mft{12^{\sharp_2}}$ has been given in (\ref{eq_12sharp}), and the two new ones are:
\begin{align}
  \mft{32^{\sharp_1}}=&\sum_{\ell,m_1,m_2=0}^{\infty}\FR{(-1)^{\ell+m_1}4\sqrt{2/\pi}\cos\big[\fr{\pi(p_{23}+3/2+\ii\wt\nu_1)}{2}\big]\Gamma(p_{23}+\ell+2m_{12}+13/2+\ii\wt\nu_1)}{\ell!\big(\fr{p_2+\ell+2m_1}{2}+2+\fr{\ii\wt\nu_1}{2}\pm\fr{\ii\wt\nu_2}{2}\big)_{m_2+1}}\n\\
  &\times\FR{\Gamma(-m_1-\ii\wt\nu_1)}{m_1!}\Big(\FR{K_1}{2E_3}\Big)^{2m_1+3/2+\ii\wt\nu_1}\Big(\FR{K_2}{2E_3}\Big)^{2m_2+3}\Big(\FR{E_2}{E_3}\Big)^{p_2+\ell+1},
\end{align}
\begin{align}
  \mft{32^{\flat_1}}=&\sum_{\ell,m_1,m_2=0}^{\infty}\FR{(-1)^{\ell+m_1}4\sqrt{2/\pi}\cos\big[\fr{\pi(p_{23}+3/2+\ii\wt\nu_1)}{2}\big]\Gamma(p_{23}+\ell+2m_{12}+13/2-\ii\wt\nu_1)}{\ell!\big(\fr{p_2+\ell+2m_1}{2}+2-\fr{\ii\wt\nu_1}{2}\pm\fr{\ii\wt\nu_2}{2}\big)_{m_2+1}}\n\\
  &\times\FR{\Gamma(-m_1+\ii\wt\nu_1)}{m_1!}\Big(\FR{K_1}{2E_3}\Big)^{2m_1+3/2-\ii\wt\nu_1}\Big(\FR{K_2}{2E_3}\Big)^{2m_2+3}\Big(\FR{E_2}{E_3}\Big)^{p_2+\ell+1}.
\end{align}

At this point, we have completed the presentation of full analytical results for the 3-chain graph in three different kinematic regions. We note again that the three seemingly different expressions, (\ref{eq_3chainCase1}), (\ref{eq_3chainCase2}), and (\ref{eq_3chainCase3}), are merely different series expansions of the same object in (\ref{eq_G3}). Clearly, we can think of them as analytical continuations of each other beyond their own convergence domain. 

\subsection{Four-site star}
\label{sec_4site}

\begin{figure}[t]
\centering
\includegraphics[width=0.75\textwidth]{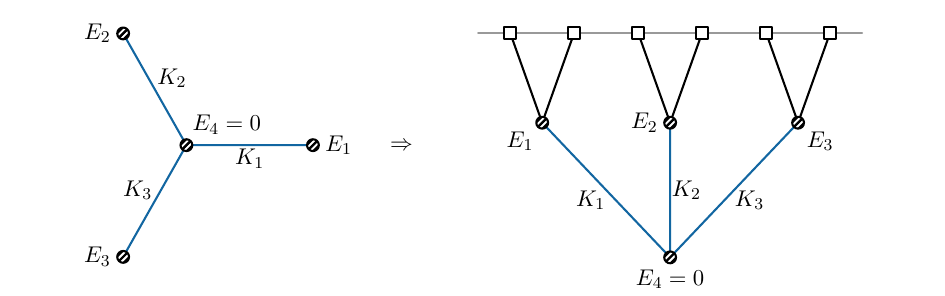} 
\caption{The 4-site star graph and a corresponding 6-point inflaton correlator. }
\label{fig_4star}
\end{figure}

As the final example, we consider a 4-star graph with triple massive exchanges. Without trivially repeating the standard construction, let us consider a special case where the central vertex is not connected to any external lines, which corresponds to a 6-point inflaton correlator, as shown in Fig.\;\ref{fig_4star}. We use this example to demonstrate how to take the limit of a vanishing vertex energy, which is $E_4\to 0$ in this case. Note that this case is somewhat beyond the signal region we originally defined, which requires all vertex energies larger than all line energies. Here, one vertex energy, $E_4$, is smaller than all line energies. We show that only a slight change of our general rule is needed to get results in this case. 

Given $E_4\to 0$, our original definition of the dimensionless graph (\ref{eq_DimlessGraph}) becomes singular for generic $p_4$. Thus, we slightly change the dimensionless graph in this case to be:
\begin{align}
  \G_4'=\Big(\FR{K_1}{2E_4}\Big)^{p_4+1} \G_4,
\end{align}
where $\G_4$ is the original dimensionless graph constructed from (\ref{eq_DimlessGraph}), and a factor of 2 is included for the notational simplicity of final expressions.  Then, we can decompose $\G_4'$ as usual: 
\begin{align}
  \G_4'=&~\text{CIS}\,\big[\G_4'\big]+{ \sum_{\al=1}^3\mathop{\text{Cut}}_{K_\al}\big[\G_4'\big]}+ { \sum_{\al\neq\be}\mathop{\text{Cut}}_{K_\al,K_\be}\big[\G_4'\big]}+\mathop{\text{Cut}}_{K_1,K_2,K_3}\big[\G_4'\big]. 
\end{align}
To write down explicit expressions, we need to pick up a particular energy order. Here we choose $E_1>E_2>E_3$. All other cases can be obtained by trivial permutations. Then, we have:
{\allowdisplaybreaks
\begin{align}
  \text{CIS}\,\big[\G_4'\big]=&~\mft{1\sla 4(2)(3)},\\
  { \sum_{\al=1}^3\mathop{\text{Cut}}_{K_\al}\big[\G_4'\big]}
  =&~\mft{1^{\sharp_1}}\Big(\mft{2\sla 4^{\sharp_1}3}+\mft{2\sla 4^{\flat_1}3}\Big)+ \mft{1\sla 4^{\sharp_2}3}\Big(\mft{2^{\sharp_2}}+\mft{2^{\flat_2}}\Big)\n\\
  &~+\mft{1\sla 4^{\sharp_3}2}\Big(\mft{3^{\sharp_3}}+\mft{3^{\flat_3}}\Big) + \text{shadows} , \\
  { \sum_{\al\neq\be}\mathop{\text{Cut}}_{K_\al,K_\be}\big[\G_4'\big]}
  =&~\mft{1^{\sharp_1}}\mft{2^{\sharp_2}}\Big(\mft{3\sla 4^{\sharp_1\sharp_2}}+\mft{3\sla 4^{\flat_1\sharp_2}}+\mft{3\sla 4^{\sharp_1\flat_2}}+\mft{3\sla 4^{\flat_1\flat_2}}\Big)\n\\
  &+\mft{1^{\sharp_1}}\Big(\mft{2\sla 4^{\sharp_1\sharp_3}}+\mft{2\sla 4^{\flat_1\sharp_3}}\Big)\Big(\mft{3^{\sharp_3}}+\mft{3^{\flat_3}}\Big)\n\\
  &+\mft{1\sla 4^{\sharp_2\sharp_3}}\Big(\mft{2^{\sharp_2}}+\mft{2^{\flat_2}}\Big)\Big(\mft{3^{\sharp_3}}+\mft{3^{\flat_3}}\Big)+ \text{shadows} , \\
  \label{eq_G43cuts}
  \mathop{\text{Cut}}_{K_1,K_2,K_3}\big[\G_4'\big]
  =&~\mft{1^{\sharp_1}}\mft{2^{\sharp_2}}\mft{3^{\sharp_3}}\Big(\mft{\sla 4^{\sharp_1\sharp_2\sharp_3}}+\mft{\sla 4^{\flat_1\sharp_2\sharp_3}}+\mft{\sla 4^{\sharp_1\flat_2\sharp_3}}+\mft{\sla 4^{\sharp_1\sharp_2\flat_3}}\n\\
  &+\mft{\sla 4^{\flat_1\flat_2\sharp_3}}+\mft{\sla 4^{\flat_1\sharp_2\flat_3}}+\mft{\sla 4^{\sharp_1\flat_2\flat_3}}+\mft{\sla 4^{\flat_1\flat_2\flat_3}}\Big)+ \text{shadows} .
\end{align}}%
In these expressions, we have slashed the entry $\sla 4$ to highlight that $E_4\to 0$ but $p_4$ is kept finite.  

Now it is straightforward to find various CISs and tuned CISs. The CIS for the full graph $\G_4'$ is:
\begin{align}
  \mft{1\sla 4(2)(3)} 
  =&\sum_{\ell_2,\ell_3,m_1,m_2,m_3=0}^{\infty}\frac{16\cos(\frac{\pi p_{1234}}{2}) \Gamma(p_{1234}+\ell_{23}+2m_{123}+13)}
  {\big(\frac{p_{234}+\ell_{23}+2m_{23}}{2}+\frac{21}{4}\pm\frac{\ii\wt\nu_1}{2}\big)_{m_1+1} }\Big(\FR{K_1}{2E_1}\Big)^{2m_1+p_4+4}\n\\
  &\times \prod_{\al=2}^3\FR{(-1)^{\ell_\al}}{\ell_\al!\big(\frac{p_\al+\ell_\al}{2}+\frac{5}{4}\pm\frac{\ii\wt\nu_\al}{2}\big)_{m_\al+1}}\Big(\FR{K_\al}{2E_1}\Big)^{2m_\al+3}\Big(\FR{E_\al}{E_1}\Big)^{p_\al+\ell_\al+1}.
\end{align}
Here we have one less summation than an ordinary 4-site family tree $\G_4$, where we would have a series of the form $(E_4/E_1)^{\ell_1+p_4+1}$. In $\G_4'$, the overall factor $(E_4/E_1)^{p_4+1}$ is replaced by $(K_1/2E_1)^{p_4+1}$, and the series $(E_4/E_1)^{\ell_1}$ vanishes for all $\ell_1>0$. 

Next, we list the tuned CISs. First, there are 5 singly tuned 1-site CISs. Among them, $\mft{1^{\sharp_1}}$ is given in (\ref{eq_1sharp}). The other four are:
\begin{align}
  \label{eq_2sharp_in4site}&\mft{2^{\sharp_2}}=\mathbf{F}^{p_2}_{\wt\nu_2}(E_2,K_2),&&\mft{2^{\flat_2}}=\FR{\cos\big[\fr{\pi(p_2+{3}/{2}+\ii\wt\nu_2)}{2}\big]}{\cos\big[\fr{\pi(p_2+{3}/{2}-\ii\wt\nu_2)}{2}\big]}\mathbf{F}^{p_2}_{-\wt\nu_2}(E_2,K_2),\\
  \label{eq_3sharp_in4site}&\mft{3^{\sharp_3}}=\mathbf{F}^{p_3}_{\wt\nu_3}(E_3,K_3),&&\mft{3^{\flat_3}}=\FR{\cos\big[\fr{\pi(p_3+{3}/{2}+\ii\wt\nu_3)}{2}\big]}{\cos\big[\fr{\pi(p_3+{3}/{2}-\ii\wt\nu_3)}{2}\big]}\mathbf{F}^{p_3}_{-\wt\nu_3}(E_3,K_3).
\end{align}

Second, there are 7 doubly tuned 2-site CISs, which can be conveniently expressed in terms of the following function:
\begin{align}
  &\mathbf{H}^{p_3,p_4}_{\wt\nu_1,\wt\nu_2,\wt\nu_3}(E_3,K_1,K_2,K_3)=\sum_{m_1,m_2,m_3=0}^{\infty}\FR{8(-1)^{m_{23}}\cos\big[\fr{\pi (p_{34}+3+\ii\wt\nu_{12})}{2}\big] \Gamma[p_{34}+2m_{123}+8+\ii\wt\nu_{12}]}{\pi \big(\fr{p_4+2m_{23}}{2}+\fr{11}{4}+\fr{\ii\wt\nu_{12}}{2}\pm\fr{\ii\wt\nu_3}{2}\big)_{m_1+1}}\n\\
  &\times\FR{\Gamma[-m_2-\ii\wt\nu_1,-m_3-\ii\wt\nu_2]}{m_2!m_3!} \Big(\FR{K_1}{2E_3}\Big)^{2m_2+p_4+5/2+\ii\wt\nu_1}\Big(\FR{K_2}{2E_3}\Big)^{2m_3+3/2+\ii\wt\nu_2}\Big(\FR{K_3}{2E_3}\Big)^{2m_1+3}.
\end{align}
Then: 
{\allowdisplaybreaks\begin{align}
  \mft{34^{\sharp_1\sharp_2}}=&~\mathbf{H}^{p_3,p_4}_{\wt\nu_1,\wt\nu_2,\wt\nu_3}(E_3,K_1,K_2,K_3),\\
  \mft{34^{\flat_1\sharp_2}}=&~\FR{\cos\big[\fr{\pi(p_{34}+3+\ii\wt\nu_{12})}{2}\big]}{\cos\big[\fr{\pi(p_{34}+3-\ii\wt\nu_1+\ii\wt\nu_2)}{2}\big]}\mathbf{H}^{p_3,p_4}_{-\wt\nu_1,\wt\nu_2,\wt\nu_3}(E_3,K_1,K_2,K_3),\\
  \mft{34^{\sharp_1\flat_2}}=&~\FR{\cos\big[\fr{\pi(p_{34}+3+\ii\wt\nu_{12})}{2}\big]}{\cos\big[\fr{\pi(p_{34}+3+\ii\wt\nu_1-\ii\wt\nu_2)}{2}\big]}\mathbf{H}^{p_3,p_4}_{\wt\nu_1,-\wt\nu_2,\wt\nu_3}(E_3,K_1,K_2,K_3),\\
  \mft{34^{\flat_1\flat_2}}=&~\FR{\cos\big[\fr{\pi(p_{34}+3-\ii\wt\nu_{12})}{2}\big]}{\cos\big[\fr{\pi(p_{34}+3-\ii\wt\nu_{12})}{2}\big]}\mathbf{H}^{p_3,p_4}_{-\wt\nu_1,-\wt\nu_2,\wt\nu_3}(E_3,K_1,K_2,K_3),\\
  \mft{24^{\sharp_1\sharp_3}}=&~\mathbf{H}^{p_2,p_4}_{\wt\nu_1,\wt\nu_3,\wt\nu_2}(E_2,K_1,K_3,K_2),\\
  \mft{24^{\flat_1\sharp_3}}=&~\FR{\cos\big[\fr{\pi(p_{24}+3+\ii\wt\nu_{13})}{2}\big]}{\cos\big[\fr{\pi(p_{24}+3-\ii\wt\nu_1+\ii\wt\nu_3)}{2}\big]}\mathbf{H}^{p_2,p_4}_{-\wt\nu_1,\wt\nu_3,\wt\nu_2}(E_2,K_1,K_3,K_2),\\
  \mft{14^{\sharp_2\sharp_3}}=&~\Big(\FR{K_1}{K_2}\Big)^{p_4+1}\mathbf{H}^{p_1,p_4}_{\wt\nu_2,\wt\nu_3,\wt\nu_1}(E_1,K_2,K_3,K_1).
\end{align}}%

Third, there are 4 singly tuned 3-site CISs, which can be conveniently expressed in terms of the following $\mathbf{I}$ function:
\begin{align}
  &~\mathbf{I}^{p_2,p_4,p_3}_{\wt\nu_1,\wt\nu_2,\wt\nu_3}(E_2,E_3,K_1,K_2,K_3)\n\\
  \equiv&\sum_{\ell_2,m_1,m_2,m_3=0}^{\infty}\FR{8\cos\big[\fr{\pi(p_{234}+3/2+\ii\wt\nu_1)}2\big](-1)^{\ell_{2}+m_3}\Gamma(p_{234}+\ell_{2}+2m_{123}+21/2+\ii\wt\nu_1)}{\sqrt{\pi/2}\,\ell_2!\big(\frac{p_{34}+\ell_{2}+2m_{23}}{2}+4+\fr{\ii\wt\nu_1}{2}\pm\frac{\ii\wt\nu_2}{2}\big)_{m_1+1}\big(\fr{p_3+\ell_2}{2}+\fr{5}{4}\pm\fr{\ii\wt\nu_3}{2}\big)_{m_2+1}}\n\\
  &\times\FR{\Gamma(-m_3-\ii\wt\nu_1)}{m_3!}\Big(\FR{K_1}{2E_2}\Big)^{p_4+2m_3+5/2+\ii\wt\nu_1}\Big(\FR{K_2}{2E_2}\Big)^{2m_1+3}\Big(\FR{K_3}{2E_2}\Big)^{2m_2+3}\Big(\FR{E_3}{E_2}\Big)^{p_3+\ell_2+1}.
\end{align}
Then, we have:
\begin{align}
\mft{24^{\sharp_1}3}=&~\mathbf{I}^{p_2,p_4,p_3}_{\wt\nu_1,\wt\nu_2,\wt\nu_3}(E_2,E_3,K_1,K_2,K_3)\\
  \mft{24^{\flat_1}3}
  =&~\FR{\cos\big[\fr{\pi(p_{234}+3/2+\ii\wt\nu_1)}2\big]}{\cos\big[\fr{\pi(p_{234}+3/2-\ii\wt\nu_1)}2\big]}\mathbf{I}^{p_2,p_4,p_3}_{-\wt\nu_1,\wt\nu_2,\wt\nu_3}(E_2,E_3,K_1,K_2,K_3), \\
  \mft{14^{\sharp_3}2} 
  =&~\Big(\FR{K_1}{K_3}\Big)^{p_4+1}\mathbf{I}^{p_1,p_4,p_2}_{\wt\nu_3,\wt\nu_1,\wt\nu_2}(E_1,E_2,K_3,K_1,K_2), \\
  \mft{14^{\sharp_2}3} 
  =&~\Big(\FR{K_1}{K_2}\Big)^{p_4+1}\mathbf{I}^{p_1,p_4,p_3}_{\wt\nu_2,\wt\nu_1,\wt\nu_3}(E_1,E_3,K_2,K_1,K_3). 
\end{align}

Finally, there are apparently 8 triply tuned 1-site CISs in (\ref{eq_G43cuts}), but with the main entry slashed. We cannot apply the general formula for the augmentation or flattening in this case since we cannot expand the expression in small $1/E_4$. However, the 1-site graph is simple enough and is nothing but the completely homogeneous solution to our differential equations, as discussed in (\ref{sec_lauricella}). From (\ref{eq_LauricellaMB}), we see that the $N$-tuned 1-site graph is nothing but a Lauricella function. In our case, when setting $E_4\to 0$, the time integral (\ref{eq_CHS}) yields a $\de$-function for Mellin variables. Then, completing the Mellin integral, it is straightforward to see that the triply tuned graphs are all expressible in terms of Appell functions, very similar to the well-known triple-$K$ integral in the context of momentum-space CFT. Then, in the region of $E_1>E_2,E_3$, we have:
\begin{align}
  &\mft{4^{\sharp_1\sharp_2\sharp_3}}=0,
  &&\mft{4^{\flat_1\sharp_2\flat_3}}=2\sin(\ii\pi\wt\nu_{1})\mathbf{J}^{p_4}_{-\wt\nu_1,\wt\nu_2,\wt\nu_3},\\
  &\mft{4^{\sharp_1\flat_2\sharp_3}}=2\sin(\ii\pi\wt\nu_{2})\mathbf{J}^{p_4}_{\wt\nu_1,-\wt\nu_2,\wt\nu_3},
  &&\mft{4^{\sharp_1\sharp_2\flat_3}}=2\sin(\ii\pi\wt\nu_{3})\mathbf{J}^{p_4}_{\wt\nu_1,\wt\nu_2,-\wt\nu_3},\\
  &\mft{4^{\flat_1\flat_2\sharp_3}}=2\sin(\ii\pi\wt\nu_{12})\mathbf{J}^{p_4}_{-\wt\nu_1,-\wt\nu_2,\wt\nu_3},
  &&\mft{4^{\flat_1\sharp_2\flat_3}}=2\sin(\ii\pi\wt\nu_{13})\mathbf{J}^{p_4}_{-\wt\nu_1,\wt\nu_2,-\wt\nu_3},\\
  &\mft{4^{\sharp_1\flat_2\flat_3}}=2\sin(\ii\pi\wt\nu_{23})\mathbf{J}^{p_4}_{\wt\nu_1,-\wt\nu_2,-\wt\nu_3},
  &&\mft{4^{\flat_1\flat_2\flat_3}}=2\sin(\ii\pi\wt\nu_{123})\mathbf{J}^{p_4}_{-\wt\nu_1,-\wt\nu_2,-\wt\nu_3},
\end{align}
where $\mb{J}^{p_4}_{\wt\nu_1,\wt\nu_2,\wt\nu_3}$ is a triple-$K$ integral and can be expressed in terms of an Appell function: 
\begin{align}
  \mb{J}^{p_4}_{\wt\nu_1,\wt\nu_2,\wt\nu_3}=&~\FR{\sqrt{2\pi}\sin\big[\fr{\pi(p_4-1/2-\ii\wt\nu_1+\ii\wt\nu_{23})}{2}\big]}{\sin(\ii\pi\wt\nu_1)\sin(\ii\pi\wt\nu_2)\sin(\ii\pi\wt\nu_3)}\Big(\FR{K_2}{K_1}\Big)^{3/2+\ii\wt\nu_2}\Big(\FR{K_3}{K_1}\Big)^{3/2+\ii\wt\nu_3}\n\\
  &\times\mathcal{F}_4\bigg[\bgm\fr{p_4+11/2+\ii\wt\nu_{123}}{2},\fr{p_4+11/2-\ii\wt\nu_1+\ii\wt\nu_{23}}{2}\\1+\ii\wt\nu_2,1+\ii\wt\nu_3\edm\bigg|\FR{K_2^2}{K_1^2},\FR{K_3^2}{K_1^2}\bigg].
\end{align} 
This completes the presentation of the 4-site star graph $\G_4'$.

\section{Discussions and Outlooks}
\label{sec_conclusion}

The program of cosmological collider (CC) physics has been actively investigated in recent years, with many interesting models found and many signal targets identified for future cosmological observations. Confronting CC models with real data calls for efficient and accurate computation of inflationary correlators. Meanwhile, cosmological correlators possess unique analytical structure which encodes rich information about QFTs in cosmological backgrounds, which deserve more systematic studies. Clearly, the analytical approach to cosmological correlators plays an essential role in both aspects.

The studies of cosmological correlators have been progressing fast in recent years. Many analytical calculations that were considered too difficult to be done a few years ago have now been made possible. Currently, the analytical computation of massive inflationary correlators at the tree level is largely a solved problem, and complete or partial results on massive loop correlators were obtained as well.

In this work, we carried out a systematic study of tree-level inflationary correlators with an arbitrary number of massive exchanges. We found a complete system of differential equations for these correlators and solved them in terms of multivariate hypergeometric functions. There is a natural structure emerged from this solution, with the completely inhomogeneous solution (CIS) expressed as a massive family tree, and the homogeneous solutions as the cuts of the CIS. With these results, we can now directly write down complete analytical results for arbitrary massive tree correlators in generic kinematics without doing any real computations, much like what we can do for tree amplitudes in flat spacetime. Also, compared with results from PMB and FTD, which are the only known previous methods that can analytically compute arbitrary massive trees, our results here have a more economic summation structure.

In earlier studies, the computation of nested bulk time integrals was often cited as a difficulty. Now, with family tree and massive family tree decompositions, the bulk time integrals are no longer obstacles. We choose to work with bulk time integrals directly, which can be important and useful for several reasons. First, the inflaton is known to be weakly coupled (small non-Gaussianity), and typical CC models with massive fields are perturbative in the bulk. Thus, it is more natural to work directly in the bulk. Second, for phenomenological studies of CC physics, most particle models with beyond-standard-model new physics have been built in the bulk spacetime, and it is important to find maps between bulk fields/interactions in these models and the signal shapes in inflationary correlators. Third, in realistic CC models, dS boosts and even the dilation are often broken. Thus it is important to develop techniques that do not rely on these symmetries. In these cases, it is often more convenient to work directly with bulk time integrals than pursuing a boundary correspondence.

On the other hand, from the explicit results obtained in this work, we can understand why the tree-level amplitudes are necessarily much more complicated than their flat-space counterparts. These inflationary correlators involve multi-layer nested time integrals of Hankel functions, which are often multivariate hypergeometric functions of high transcendental weights. Having high weights means that these functions can in no way be reduced to simpler and more familiar functions for generic parameters. This is in contrast to other amplitude stories such as the Parke-Taylor formula \cite{Parke:1986gb}, where the intermediate steps can be lengthy but the final result is simple. Therefore, whatever method we choose, we have to deal with those hypergeometric functions so long as we take the 3-momentum representation. 

The hypergeometric functions/series identified in this work can be covered by known function classes such as Horn type \cite{horn1931hypergeometrische} or GKZ type \cite{gel1989hypergeometric}, but our functions are a very special subclass: They are all massive family trees, with possible augmentation and flattening, and are fully determined by the twists and masses. Thus, it would be useful to develop specialized methods and techniques to understand this particular subclass of multivariate hypergeometric functions. In this work, we have made the first step, identified these functions and achieved expanding these functions in the signal regions, namely, when all vertex energies are larger than the line energies. It would be very useful to develop more expansion schemes, ideally around all possible singular regions. This type of expansions will be very useful for us to understand the analytical properties, including the singularity structures and functional relations, of these hypergeometric functions.

Our work has made a step towards more systematic studies of general massive inflation correlators. With the new results obtained here, many directions are now open for further explorations. We end this work by listing a few of them. 

First, we have only considered generic kinematics in this work, meaning that all kinematic variables are not constrained. However, in realistic models, it also happens that some kinematic variables are constrained, which we call degenerate configurations. For instance, in the case of a two-point mixing between a massless external line and a massive internal line, we have a constraint $E=K$ relating the corresponding vertex and line energies. It would be very interesting to push our results to various types of degenerate configurations. 

Second, it was known that the differential equation for single exchange can be easily derived for fields with more complicated dispersion relations, such as non-unit sound speed, the helical chemical potential, or even when there is nontrivial scale dependence beyond pure dS \cite{Qin:2022fbv,Jazayeri:2022kjy,Pimentel:2022fsc,Qin:2023ejc,Aoki:2023wdc}. For general graphs beyond the single exchange, can we find the complete differential equation systems and even solve them for these cases? We plan to address these questions in future works. 

Third, it is known that the differential equations for tree correlators can be easily generalized to loop integrands in the case of conformal scalar amplitudes \cite{He:2024olr,Baumann:2024mvm}. The massive amplitudes should be similar. Can we use these differential equations to better understand the analytical structure of loop graphs?  

Finally but importantly, the differential equation system presented here could be a useful starting point for developing new numerical strategies for computing and studying massive inflationary correlators. Can we use these differential equations to find more efficient methods for computing these graphs, especially when the vertex number is large? We leave all these interesting questions for future studies.

\paragraph{Acknowledgments} We thank Song He, Xuhang Jiang, and Denis Werth for useful discussions. Results of this work have been presented by ZX in the workshop ``Looping the primordial universe'' at CERN, and we thank the organizers and participants of the workshop for useful discussions and feedback. This work is supported by NSFC under Grants No.\ 12275146 and No.\ 12247103, the National Key R\&D Program of China (2021YFC2203100), and the Dushi Program of Tsinghua University. 

\newpage
\begin{appendix}

\section{Notations}
\label{app_notation}

\subsection{List of symbols}

We list some frequently used symbols in Table \ref{tab_notations}, together with the numbers of equations where these notations are defined or first appear. 
\begin{table}[tbph] 
\centering
\caption{List of selected symbolic notations }
\vspace{2mm}
\begin{tabular}{lll}
\toprule[1.5pt]
Notation 
&\multicolumn{1}{c}{Description} & Equation \\ \hline 
    $E_i$ & Vertex energy & Below (\ref{eq_Ctau})\\
    $K_\al$ & Line energy & Below (\ref{eq_Ctau}) \\
    $r$ & Energy ratio; In particular, $r_{(\al i)}\equiv K_{\al}/E_i$ & (\ref{eq_r_ali})\\
    $C_{\mathsf{a}}(k;\tau)$ & Bulk-to-boundary propagator of a conformal scalar & (\ref{eq_CSProp})\\
    $\sigma(rz)$ & Mode function of a massive scalar & (\ref{eq_ModeExpansion})\\
    $D_{\mathsf{ab}}^{(\wt\nu)}(K;\tau_1,\tau_2)$ & Bulk propagator of a massive scalar & (\ref{eq_Dmp})-(\ref{eq_Dpmpm})\\
    $\wt{D}^{(\wt\nu)}_{\aa\bb}(K_{\alpha}\tau_i,K_{\alpha}\tau_j)$ & Dimensionless bulk propagator& (\ref{eq_wtD}) \\
    $G^{(\wt\nu)}(K;\tau_1,\tau_2)$ & Bulk propagator for wavefunction coefficients & (\ref{eq_G_wavefunc})\\
    $\wh{\mathcal{G}}(\bm k_1,\cdots,\bm k_N)$ & Tree graph for an inflation correlator & (\ref{eq_GraphInt})\\
    $\mathcal{G}(\{E\},\{K\})$ & Dimensionless tree graph & (\ref{eq_DimlessGraph})\\
    ${\Psi}(\bm k_1,\cdots,\bm k_N)$ & Tree graph for a wavefunction coefficient &(\ref{eq_wavefunction})\\
    $\mathsf{C}_{\al}[\G]$ & Contraction of the graph $\G$ & Above (\ref{eq_DEtG})\\
    $\mathcal{D}_{(\al i)}$ & Differential operator with respect to $r_{(\al i)}$ & (\ref{eq_Dri})\\
    $\vartheta_{(\al i)}$ & Euler operator & Above (\ref{eq_varthetai}) \\
    $\vartheta_{\{i\}}$ & Sum of Euler operators & (\ref{eq_varthetai}) \\
    $\mathcal{N}(i)$ & Neighbor set of $i$ (All lines attached to Vertex $i$)  & Below (\ref{eq_varthetai})  \\
    $\text{CIS}\;[\G]$ & Completely inhomogeneous solution of $\G$ & Below (\ref{eq_Gpattern}); (\ref{eq_MFT}) \\
    $\mathop{\text{Inh}}\limits_{K}\big[\G\big]$ & Inhomogeneous solution of $\G$ with respect to $K$ & (\ref{eq_tGinhsmallr1}) \\
    $\mft{\cdots}$ & Massive family tree & (\ref{eq_1site}); App.\;\ref{app_MFT} \\
    $\mathscr{P}(\cdots)$ & Partial order of a given tree diagram & (\ref{eq_PartOrderTree}) \\
    $\mathop\text{Cut}\limits_{K}\big[\G\big]$ & Cut of $\G$ with respect to $K$ & (\ref{eq_CutKtG}) \\
    $\mft{\cdots i^\sharp \cdots}$ & Augmented CIS & (\ref{eq_augm}) \\
    $\mft{\cdots i^\flat \cdots}$ & Flattened CIS & (\ref{eq_flat})  \\
\bottomrule[1.5pt] 
\end{tabular}
\label{tab_notations}
\end{table} 

\subsection{Notations for massive family trees}
\label{app_MFT} 

In the main text, we use a notation $\mft{\cdots}$ to represent general massive family trees, which correspond to CISs of general massive tree graphs. The use of this notation is in parallel with the family tree notation $\big[\cdots\big]$ introduced in \cite{Fan:2024iek}, but also with some differences. Here we briefly introduce this notation. 

Similar to a general dimensionless tree graph (\ref{eq_dimlessIntG}), an $N$-site \emph{massive family tree} with the partial order $\mathscr{P}$ is defined by the following integral: 
\begin{align} 
\mft{\mathscr{P}(1\cdots N)} \equiv \sum_{\aa=\pm} \int_{-\infty}^0\prod_{i=1}^N\Big[\di z_i\, (\ii\aa)(-z_i)^{p_i}e^{\ii \aa z_i}\Big]\prod_{\al=1}^I R_{\aa}^{(\wt\nu_\al)}(r_{(\al i)} z_{i},r_{(\al j)}z_{j}),
\end{align}
where $z_i\equiv E_i\tau_i$ is the dimensionless time variable of Site $i$, $r_{(\al i)}\equiv K_\al/E_i$, $R_{\pm}^{(\wt\nu_\al)}(r_iz_i,r_jz_j)$ is the inhomogeneous and retarded bulk propagator for a scalar of mass $\wt\nu_\al$:
\begin{align}
  R_{\pm}^{(\wt\nu)}(r_iz_i,r_jz_j)=\Big[\wt{D}_{\pm\mp}^{(\wt\nu)}(r_iz_i,r_jz_j)-\wt{D}_{\mp\pm}^{(\wt\nu)}(r_iz_i,r_jz_j)\Big]\theta(r_jz_j-r_iz_i),
\end{align}
and $\wt{D}_{\aa\bb}^{(\wt\nu)}$ is given in (\ref{eq_Dmp})-(\ref{eq_Dpmpm}). Clearly, the retarded propagator $R_{\pm}^{(\wt\nu)}(r_iz_i,r_jz_j)$ is endowed with a time direction flowing from Site $i$ to Site $j$. For convenience, we call $i$ the mother of $j$ and thus $j$ the daughter of $i$. Then the partial order $\mathscr{P}(1\cdots N)$ is defined with respect to this time flow, which, as explained in the main text, corresponds to the induced order of the graph, and the ancestor of the whole family tree is the maximal energy site. 

We see that a massive family tree is fully specified by a partially ordered tree structure, together with a set of parameters $(E_i,p_i)$ $(i=1,\cdots,N)$, and $(K_i,\wt\nu_i)$ $(i=2,\cdots,N)$, and the label of line variables $(K_i,\wt\nu_i)$ is identified with the label of the vertex to which the line is directed. With this understood, we only need to spell out the labels of all vertices when specifying a massive family tree. This observation motivates us to use a shorthand notation for massive family trees, similar to the notation for family trees in \cite{Fan:2024iek}. Here we quote the rule for this notation: 
\begin{enumerate}
  \item Every massive family tree integral is represented by a string of numbers within a pair of blackboard-bold square brackets $\mft{\cdots}$. (We use $\mft{\cdots}$ to distinguish the massive family tree with the original family tree notation $\big[\cdots\big]$ introduced in \cite{Fan:2024iek}.) 
  \item The leftmost number denotes the earliest member. If a mother has only one daughter, write the daughter's number directly to the right of the mother's number. If a mother has more than one daughter, use a pair of parentheses to enclose the \emph{entire} subfamily of each of these daughters, and write all these subfamilies (in parentheses) to the right of the mother number. The order of subfamilies at the same level is arbitrary. Here, a subfamily of a site means herself and all her descendants.\end{enumerate}
Here are a few examples which illustrate how the rule works:
\begin{align}
  \mft{1234}=&\sum_{\aa=\pm} \int_{-\infty}^0\prod_{i=1}^4\Big[\di z_i\, (\ii\aa)(-z_i)^{p_i}e^{\ii \aa z_i}\Big]R_{12}R_{23}R_{34},\\
  \mft{12(3)(4)}=&\sum_{\aa=\pm} \int_{-\infty}^0\prod_{i=1}^4\Big[\di z_i\, (\ii\aa)(-z_i)^{p_i}e^{\ii \aa z_i}\Big]R_{12}R_{23}R_{24},\\
  \mft{1(23)(4)}=&\sum_{\aa=\pm} \int_{-\infty}^0\prod_{i=1}^4\Big[\di z_i\, (\ii\aa)(-z_i)^{p_i}e^{\ii \aa z_i}\Big] R_{12}R_{14}R_{23},
\end{align}
where we write $R_{\aa}^{(\wt\nu_{\al})}(r_{(\al i)} z_{i},r_{(\al j)}z_{j})=R_{ij}$.

Although the notation above is designed to express a partially ordered tree structure, it is also straightforward to generalize it to an arbitrary directional tree without partial orders. As described in Sec.\;\ref{sec_ordering}, in a general directional tree graph, we can still introduce an induced partial order, with respect to which all lines are either ``correct'' or ``wrong.'' Then, we can simply add a back-pointing arrow to mark out all the wrong lines. For instance:
\begin{align}
  \mft{12\overleftarrow{3}4}=&\sum_{\aa=\pm} \int_{-\infty}^0\prod_{i=1}^4\Big[\di z_i\, (\ii\aa)(-z_i)^{p_i}e^{\ii \aa z_i}\Big]R_{12}R_{32}R_{34}.
\end{align} 
Clearly, once we allow for wrong lines in a family tree, the notation above is no longer unique. There could be many different ways to express the same directional tree graph, which is however not a problem for us.

\section{Special Functions}
\label{app_functions}

All massive family trees and their augmentation/flattening are multivariate hypergeometric functions. However, only a handful of simple cases are named, and even fewer are well studied. In this appendix, we collect the definitions of a few named hypergeometric functions appearing in the main text. 

\paragraph{Gamma functions} As a preparation, we introduce a shorthand notation for products of Euler $\Gamma$ functions:
\begin{align}
  \Gamma[a_1,\cdots,a_n]\equiv \Gamma(a_1)\cdots \Gamma(a_n),
\end{align}
It is also convenient to abbreviate ratios of $\Gamma$ products as:
\begin{align}
    \Gamma\bgb a_1,\cdots,a_m\\b_1,\cdots,b_n\edb\equiv\FR{\Gamma[a_1,\cdots,a_m]}{\Gamma[b_1,\cdots,b_n]}.
\end{align}
We also draw heavy use of the Pochhammer symbol $(a)_n$, defined as:
\begin{align}
  (a)_n\equiv\FR{\Gamma(a+n)}{\Gamma(n)}.
\end{align}

\paragraph{Hypergeometric functions} The most well-known hypergeometric function is probably Gauss's hypergeometric function, which is defined by the following series when it is convergent and by its analytical continuation outside the convergence region:
\begin{align}
  {}_2\text{F}_1\bigg[\bgm a,b\\c\edm\bigg|z\bigg]\equiv\sum_{n=0}^{\infty}\FR{(a)_n(b)_n}{(c)_n}\FR{z^n}{n!}.
\end{align}
In the main text, we use the dressed version of Gauss's hypergeometric function, which is defined as
\begin{align}
\label{eq_dressed2F1}
  {}_2\mathcal{F}_1\bigg[\bgm a,b\\c\edm\bigg|z\bigg]\equiv{\Gamma\bigg[\bgm a,b\\c\edm\bigg]}{}_2\text{F}_1\bigg[\bgm a,b\\c\edm\bigg|z\bigg]=\sum_{n=0}^{\infty}\Gamma\bigg[\bgm a+n,b+n\\c+n\edm\bigg]\FR{z^n}{n!}.
\end{align}
The most studied bivariate hypergeometric functions are Appell series \cite{Slater:1966}. In the paper, we use the (dressed) Appell $\text{F}_4$ function, which is defined by the following series and its analytical continuation: 
\begin{align}
\label{eq_AppellF4}
  \mathcal{F}_4\bigg[\bgm a,b\\c_1,c_2\edm\bigg|x,y\bigg]\equiv\sum_{m,n=0}^{\infty}\Gamma\bigg[\bgm a+m+n,b+m+n\\c_1+m,c_2+n\edm\bigg]\FR{x^my^n}{m!n!}.
\end{align}
At the level of $n$-variate hypergeometric functions, known results are rather sparse. Among them a relatively well-studied class is Lauricella's functions \cite{Matsumoto_2020}. In this paper, we used the Lauricella's $\text{F}_\text{C}$ function, defined as:
\begin{align}
  \text{F}_\text{C}\left[\bgm a,b\\ c_1,\cdots,c_N\edm\middle|z_1,\cdots,z_N\right]\equiv 
  \sum_{n_1,\cdots,n_N=0}^\infty(a)_{n_{1\cdots N}}(b)_{n_{1\cdots N}}\prod_{i=1}^N\FR{1}{(c_i)_{n_i}}\FR{z_i^{n_i}}{n_i!}.
\end{align}

\end{appendix}

\newpage
\bibliography{CosmoCollider} 

\providecommand{\href}[2]{#2}\begingroup\raggedright\begin{thebibliography}{100}

\bibitem{Chen:2009we}
X.~Chen and Y.~Wang, ``{Large non-Gaussianities with Intermediate Shapes from
  Quasi-Single Field Inflation},''
  \href{http://dx.doi.org/10.1103/PhysRevD.81.063511}{{\em Phys. Rev. D}
  {\bfseries 81} (2010) 063511},
  \href{http://arxiv.org/abs/0909.0496}{{\ttfamily arXiv:0909.0496
  [astro-ph.CO]}}.

\bibitem{Chen:2009zp}
X.~Chen and Y.~Wang, ``{Quasi-Single Field Inflation and Non-Gaussianities},''
  \href{http://dx.doi.org/10.1088/1475-7516/2010/04/027}{{\em JCAP} {\bfseries
  1004} (2010) 027},
\href{http://arxiv.org/abs/0911.3380}{{\ttfamily arXiv:0911.3380 [hep-th]}}.

\bibitem{Arkani-Hamed:2015bza}
N.~Arkani-Hamed and J.~Maldacena, ``{Cosmological Collider Physics},''
\href{http://arxiv.org/abs/1503.08043}{{\ttfamily arXiv:1503.08043 [hep-th]}}.

\bibitem{Chen:2016nrs}
X.~Chen, Y.~Wang, and Z.-Z. Xianyu, ``{Loop Corrections to Standard Model
  Fields in Inflation},'' \href{http://dx.doi.org/10.1007/JHEP08(2016)051}{{\em
  JHEP} {\bfseries 08} (2016) 051},
\href{http://arxiv.org/abs/1604.07841}{{\ttfamily arXiv:1604.07841 [hep-th]}}.

\bibitem{Chen:2016uwp}
X.~Chen, Y.~Wang, and Z.-Z. Xianyu, ``{Standard Model Background of the
  Cosmological Collider},''
  \href{http://dx.doi.org/10.1103/PhysRevLett.118.261302}{{\em Phys. Rev.
  Lett.} {\bfseries 118} no.~26, (2017) 261302},
\href{http://arxiv.org/abs/1610.06597}{{\ttfamily arXiv:1610.06597 [hep-th]}}.

\bibitem{Chen:2016hrz}
X.~Chen, Y.~Wang, and Z.-Z. Xianyu, ``{Standard Model Mass Spectrum in
  Inflationary Universe},''
  \href{http://dx.doi.org/10.1007/JHEP04(2017)058}{{\em JHEP} {\bfseries 04}
  (2017) 058},
\href{http://arxiv.org/abs/1612.08122}{{\ttfamily arXiv:1612.08122 [hep-th]}}.

\bibitem{Lee:2016vti}
H.~Lee, D.~Baumann, and G.~L. Pimentel, ``{Non-Gaussianity as a Particle
  Detector},'' \href{http://dx.doi.org/10.1007/JHEP12(2016)040}{{\em JHEP}
  {\bfseries 12} (2016) 040},
\href{http://arxiv.org/abs/1607.03735}{{\ttfamily arXiv:1607.03735 [hep-th]}}.

\bibitem{An:2017hlx}
H.~An, M.~McAneny, A.~K. Ridgway, and M.~B. Wise, ``{Quasi Single Field
  Inflation in the non-perturbative regime},''
  \href{http://dx.doi.org/10.1007/JHEP06(2018)105}{{\em JHEP} {\bfseries 06}
  (2018) 105},
\href{http://arxiv.org/abs/1706.09971}{{\ttfamily arXiv:1706.09971 [hep-ph]}}.

\bibitem{An:2017rwo}
H.~An, M.~McAneny, A.~K. Ridgway, and M.~B. Wise, ``{Non-Gaussian Enhancements
  of Galactic Halo Correlations in Quasi-Single Field Inflation},''
  \href{http://dx.doi.org/10.1103/PhysRevD.97.123528}{{\em Phys. Rev. D}
  {\bfseries 97} no.~12, (2018) 123528},
  \href{http://arxiv.org/abs/1711.02667}{{\ttfamily arXiv:1711.02667
  [hep-ph]}}.

\bibitem{Iyer:2017qzw}
A.~V. Iyer, S.~Pi, Y.~Wang, Z.~Wang, and S.~Zhou, ``{Strongly Coupled
  Quasi-Single Field Inflation},''
  \href{http://dx.doi.org/10.1088/1475-7516/2018/01/041}{{\em JCAP} {\bfseries
  1801} no.~01, (2018) 041},
\href{http://arxiv.org/abs/1710.03054}{{\ttfamily arXiv:1710.03054 [hep-th]}}.

\bibitem{Kumar:2017ecc}
S.~Kumar and R.~Sundrum, ``{Heavy-Lifting of Gauge Theories By Cosmic
  Inflation},'' \href{http://dx.doi.org/10.1007/JHEP05(2018)011}{{\em JHEP}
  {\bfseries 05} (2018) 011},
\href{http://arxiv.org/abs/1711.03988}{{\ttfamily arXiv:1711.03988 [hep-ph]}}.

\bibitem{Tong:2018tqf}
X.~Tong, Y.~Wang, and S.~Zhou, ``{Unsuppressed primordial standard clocks in
  warm quasi-single field inflation},''
  \href{http://dx.doi.org/10.1088/1475-7516/2018/06/013}{{\em JCAP} {\bfseries
  1806} no.~06, (2018) 013},
\href{http://arxiv.org/abs/1801.05688}{{\ttfamily arXiv:1801.05688 [hep-th]}}.

\bibitem{Chen:2018sce}
X.~Chen, W.~Z. Chua, Y.~Guo, Y.~Wang, Z.-Z. Xianyu, and T.~Xie, ``{Quantum
  Standard Clocks in the Primordial Trispectrum},''
  \href{http://dx.doi.org/10.1088/1475-7516/2018/05/049}{{\em JCAP} {\bfseries
  1805} no.~05, (2018) 049},
\href{http://arxiv.org/abs/1803.04412}{{\ttfamily arXiv:1803.04412 [hep-th]}}.

\bibitem{Chen:2018xck}
X.~Chen, Y.~Wang, and Z.-Z. Xianyu, ``{Neutrino Signatures in Primordial
  Non-Gaussianities},'' \href{http://dx.doi.org/10.1007/JHEP09(2018)022}{{\em
  JHEP} {\bfseries 09} (2018) 022},
\href{http://arxiv.org/abs/1805.02656}{{\ttfamily arXiv:1805.02656 [hep-ph]}}.

\bibitem{Chua:2018dqh}
W.~Z. Chua, Q.~Ding, Y.~Wang, and S.~Zhou, ``{Imprints of Schwinger Effect on
  Primordial Spectra},'' \href{http://dx.doi.org/10.1007/JHEP04(2019)066}{{\em
  JHEP} {\bfseries 04} (2019) 066},
\href{http://arxiv.org/abs/1810.09815}{{\ttfamily arXiv:1810.09815 [hep-th]}}.

\bibitem{Wu:2018lmx}
Y.-P. Wu, ``{Higgs as heavy-lifted physics during inflation},''
  \href{http://dx.doi.org/10.1007/JHEP04(2019)125}{{\em JHEP} {\bfseries 04}
  (2019) 125},
\href{http://arxiv.org/abs/1812.10654}{{\ttfamily arXiv:1812.10654 [hep-ph]}}.

\bibitem{Saito:2018omt}
R.~Saito and T.~Kubota, ``{Heavy Particle Signatures in Cosmological
  Correlation Functions with Tensor Modes},''
  \href{http://dx.doi.org/10.1088/1475-7516/2018/06/009}{{\em JCAP} {\bfseries
  06} (2018) 009}, \href{http://arxiv.org/abs/1804.06974}{{\ttfamily
  arXiv:1804.06974 [hep-th]}}.

\bibitem{Li:2019ves}
L.~Li, T.~Nakama, C.~M. Sou, Y.~Wang, and S.~Zhou, ``{Gravitational Production
  of Superheavy Dark Matter and Associated Cosmological Signatures},''
  \href{http://dx.doi.org/10.1007/JHEP07(2019)067}{{\em JHEP} {\bfseries 07}
  (2019) 067}, \href{http://arxiv.org/abs/1903.08842}{{\ttfamily
  arXiv:1903.08842 [astro-ph.CO]}}.

\bibitem{Lu:2019tjj}
S.~Lu, Y.~Wang, and Z.-Z. Xianyu, ``{A Cosmological Higgs Collider},''
  \href{http://dx.doi.org/10.1007/JHEP02(2020)011}{{\em JHEP} {\bfseries 02}
  (2020) 011}, \href{http://arxiv.org/abs/1907.07390}{{\ttfamily
  arXiv:1907.07390 [hep-th]}}.

\bibitem{Liu:2019fag}
T.~Liu, X.~Tong, Y.~Wang, and Z.-Z. Xianyu, ``{Probing P and CP Violations on
  the Cosmological Collider},''
  \href{http://dx.doi.org/10.1007/JHEP04(2020)189}{{\em JHEP} {\bfseries 04}
  (2020) 189}, \href{http://arxiv.org/abs/1909.01819}{{\ttfamily
  arXiv:1909.01819 [hep-ph]}}.

\bibitem{Hook:2019zxa}
A.~Hook, J.~Huang, and D.~Racco, ``{Searches for other vacua. Part II. A new
  Higgstory at the cosmological collider},''
  \href{http://dx.doi.org/10.1007/JHEP01(2020)105}{{\em JHEP} {\bfseries 01}
  (2020) 105}, \href{http://arxiv.org/abs/1907.10624}{{\ttfamily
  arXiv:1907.10624 [hep-ph]}}.

\bibitem{Hook:2019vcn}
A.~Hook, J.~Huang, and D.~Racco, ``{Minimal signatures of the Standard Model in
  non-Gaussianities},''
  \href{http://dx.doi.org/10.1103/PhysRevD.101.023519}{{\em Phys. Rev. D}
  {\bfseries 101} no.~2, (2020) 023519},
  \href{http://arxiv.org/abs/1908.00019}{{\ttfamily arXiv:1908.00019
  [hep-ph]}}.

\bibitem{Kumar:2018jxz}
S.~Kumar and R.~Sundrum, ``{Seeing Higher-Dimensional Grand Unification In
  Primordial Non-Gaussianities},''
  \href{http://dx.doi.org/10.1007/JHEP04(2019)120}{{\em JHEP} {\bfseries 04}
  (2019) 120},
\href{http://arxiv.org/abs/1811.11200}{{\ttfamily arXiv:1811.11200 [hep-ph]}}.

\bibitem{Kumar:2019ebj}
S.~Kumar and R.~Sundrum, ``{Cosmological Collider Physics and the Curvaton},''
  \href{http://dx.doi.org/10.1007/JHEP04(2020)077}{{\em JHEP} {\bfseries 04}
  (2020) 077}, \href{http://arxiv.org/abs/1908.11378}{{\ttfamily
  arXiv:1908.11378 [hep-ph]}}.

\bibitem{Alexander:2019vtb}
S.~Alexander, S.~J. Gates, L.~Jenks, K.~Koutrolikos, and E.~McDonough,
  ``{Higher Spin Supersymmetry at the Cosmological Collider: Sculpting SUSY
  Rilles in the CMB},'' \href{http://dx.doi.org/10.1007/JHEP10(2019)156}{{\em
  JHEP} {\bfseries 10} (2019) 156},
  \href{http://arxiv.org/abs/1907.05829}{{\ttfamily arXiv:1907.05829
  [hep-th]}}.

\bibitem{Wang:2019gbi}
L.-T. Wang and Z.-Z. Xianyu, ``{In Search of Large Signals at the Cosmological
  Collider},'' \href{http://dx.doi.org/10.1007/JHEP02(2020)044}{{\em JHEP}
  {\bfseries 02} (2020) 044},
\href{http://arxiv.org/abs/1910.12876}{{\ttfamily arXiv:1910.12876 [hep-ph]}}.

\bibitem{Wang:2019gok}
D.-G. Wang, ``{On the inflationary massive field with a curved field
  manifold},'' \href{http://dx.doi.org/10.1088/1475-7516/2020/01/046}{{\em
  JCAP} {\bfseries 01} (2020) 046},
  \href{http://arxiv.org/abs/1911.04459}{{\ttfamily arXiv:1911.04459
  [astro-ph.CO]}}.

\bibitem{Wang:2020uic}
Y.~Wang and Y.~Zhu, ``{Cosmological Collider Signatures of Massive Vectors from
  Non-Gaussian Gravitational Waves},''
  \href{http://dx.doi.org/10.1088/1475-7516/2020/04/049}{{\em JCAP} {\bfseries
  04} (2020) 049}, \href{http://arxiv.org/abs/2001.03879}{{\ttfamily
  arXiv:2001.03879 [astro-ph.CO]}}.

\bibitem{Li:2020xwr}
L.~Li, S.~Lu, Y.~Wang, and S.~Zhou, ``{Cosmological Signatures of Superheavy
  Dark Matter},'' \href{http://dx.doi.org/10.1007/JHEP07(2020)231}{{\em JHEP}
  {\bfseries 07} (2020) 231}, \href{http://arxiv.org/abs/2002.01131}{{\ttfamily
  arXiv:2002.01131 [hep-ph]}}.

\bibitem{Wang:2020ioa}
L.-T. Wang and Z.-Z. Xianyu, ``{Gauge Boson Signals at the Cosmological
  Collider},'' \href{http://dx.doi.org/10.1007/JHEP11(2020)082}{{\em JHEP}
  {\bfseries 11} (2020) 082}, \href{http://arxiv.org/abs/2004.02887}{{\ttfamily
  arXiv:2004.02887 [hep-ph]}}.

\bibitem{Fan:2020xgh}
J.~Fan and Z.-Z. Xianyu, ``{A Cosmic Microscope for the Preheating Era},''
  \href{http://dx.doi.org/10.1007/JHEP01(2021)021}{{\em JHEP} {\bfseries 01}
  (2021) 021}, \href{http://arxiv.org/abs/2005.12278}{{\ttfamily
  arXiv:2005.12278 [hep-ph]}}.

\bibitem{Aoki:2020zbj}
S.~Aoki and M.~Yamaguchi, ``{Disentangling mass spectra of multiple fields in
  cosmological collider},''
  \href{http://dx.doi.org/10.1007/JHEP04(2021)127}{{\em JHEP} {\bfseries 04}
  (2021) 127}, \href{http://arxiv.org/abs/2012.13667}{{\ttfamily
  arXiv:2012.13667 [hep-th]}}.

\bibitem{Bodas:2020yho}
A.~Bodas, S.~Kumar, and R.~Sundrum, ``{The Scalar Chemical Potential in
  Cosmological Collider Physics},''
  \href{http://dx.doi.org/10.1007/JHEP02(2021)079}{{\em JHEP} {\bfseries 02}
  (2021) 079}, \href{http://arxiv.org/abs/2010.04727}{{\ttfamily
  arXiv:2010.04727 [hep-ph]}}.

\bibitem{Maru:2021ezc}
N.~Maru and A.~Okawa, ``{Non-Gaussianity from $X, Y$ gauge bosons in
  Cosmological Collider Physics},''
  \href{http://arxiv.org/abs/2101.10634}{{\ttfamily arXiv:2101.10634
  [hep-ph]}}.

\bibitem{Lu:2021gso}
S.~Lu, ``{Axion isocurvature collider},''
  \href{http://dx.doi.org/10.1007/JHEP04(2022)157}{{\em JHEP} {\bfseries 04}
  (2022) 157}, \href{http://arxiv.org/abs/2103.05958}{{\ttfamily
  arXiv:2103.05958 [hep-th]}}.

\bibitem{Sou:2021juh}
C.~M. Sou, X.~Tong, and Y.~Wang, ``{Chemical-potential-assisted particle
  production in FRW spacetimes},''
  \href{http://dx.doi.org/10.1007/JHEP06(2021)129}{{\em JHEP} {\bfseries 06}
  (2021) 129}, \href{http://arxiv.org/abs/2104.08772}{{\ttfamily
  arXiv:2104.08772 [hep-th]}}.

\bibitem{Lu:2021wxu}
Q.~Lu, M.~Reece, and Z.-Z. Xianyu, ``{Missing scalars at the cosmological
  collider},'' \href{http://dx.doi.org/10.1007/JHEP12(2021)098}{{\em JHEP}
  {\bfseries 12} (2021) 098}, \href{http://arxiv.org/abs/2108.11385}{{\ttfamily
  arXiv:2108.11385 [hep-ph]}}.

\bibitem{Pinol:2021aun}
L.~Pinol, S.~Aoki, S.~Renaux-Petel, and M.~Yamaguchi, ``{Inflationary flavor
  oscillations and the cosmic spectroscopy},''
  \href{http://dx.doi.org/10.1103/PhysRevD.107.L021301}{{\em Phys. Rev. D}
  {\bfseries 107} no.~2, (2023) L021301},
  \href{http://arxiv.org/abs/2112.05710}{{\ttfamily arXiv:2112.05710
  [hep-th]}}.

\bibitem{Cui:2021iie}
Y.~Cui and Z.-Z. Xianyu, ``{Probing Leptogenesis with the Cosmological
  Collider},'' \href{http://dx.doi.org/10.1103/PhysRevLett.129.111301}{{\em
  Phys. Rev. Lett.} {\bfseries 129} no.~11, (2022) 111301},
  \href{http://arxiv.org/abs/2112.10793}{{\ttfamily arXiv:2112.10793
  [hep-ph]}}.

\bibitem{Tong:2022cdz}
X.~Tong and Z.-Z. Xianyu, ``{Large spin-2 signals at the cosmological
  collider},'' \href{http://dx.doi.org/10.1007/JHEP10(2022)194}{{\em JHEP}
  {\bfseries 10} (2022) 194}, \href{http://arxiv.org/abs/2203.06349}{{\ttfamily
  arXiv:2203.06349 [hep-ph]}}.

\bibitem{Reece:2022soh}
M.~Reece, L.-T. Wang, and Z.-Z. Xianyu, ``{Large-field inflation and the
  cosmological collider},''
  \href{http://dx.doi.org/10.1103/PhysRevD.107.L101304}{{\em Phys. Rev. D}
  {\bfseries 107} no.~10, (2023) L101304},
  \href{http://arxiv.org/abs/2204.11869}{{\ttfamily arXiv:2204.11869
  [hep-ph]}}.

\bibitem{Chen:2022vzh}
X.~Chen, R.~Ebadi, and S.~Kumar, ``{Classical cosmological collider physics and
  primordial features},''
  \href{http://dx.doi.org/10.1088/1475-7516/2022/08/083}{{\em JCAP} {\bfseries
  08} (2022) 083}, \href{http://arxiv.org/abs/2205.01107}{{\ttfamily
  arXiv:2205.01107 [hep-ph]}}.

\bibitem{Niu:2022quw}
X.~Niu, M.~H. Rahat, K.~Srinivasan, and W.~Xue, ``{Gravitational wave probes of
  massive gauge bosons at the cosmological collider},''
  \href{http://dx.doi.org/10.1088/1475-7516/2023/02/013}{{\em JCAP} {\bfseries
  02} (2023) 013}, \href{http://arxiv.org/abs/2211.14331}{{\ttfamily
  arXiv:2211.14331 [hep-ph]}}.

\bibitem{Niu:2022fki}
X.~Niu, M.~H. Rahat, K.~Srinivasan, and W.~Xue, ``{Parity-odd and even
  trispectrum from axion inflation},''
  \href{http://dx.doi.org/10.1088/1475-7516/2023/05/018}{{\em JCAP} {\bfseries
  05} (2023) 018}, \href{http://arxiv.org/abs/2211.14324}{{\ttfamily
  arXiv:2211.14324 [hep-ph]}}.

\bibitem{Chen:2023txq}
X.~Chen, J.~Fan, and L.~Li, ``{New inflationary probes of axion dark matter},''
  \href{http://dx.doi.org/10.1007/JHEP12(2023)197}{{\em JHEP} {\bfseries 12}
  (2023) 197}, \href{http://arxiv.org/abs/2303.03406}{{\ttfamily
  arXiv:2303.03406 [hep-ph]}}.

\bibitem{Chakraborty:2023qbp}
P.~Chakraborty and J.~Stout, ``{Light scalars at the cosmological collider},''
  \href{http://dx.doi.org/10.1007/JHEP02(2024)021}{{\em JHEP} {\bfseries 02}
  (2024) 021}, \href{http://arxiv.org/abs/2310.01494}{{\ttfamily
  arXiv:2310.01494 [hep-th]}}.

\bibitem{Tong:2023krn}
X.~Tong, Y.~Wang, C.~Zhang, and Y.~Zhu, ``{BCS in the sky: signatures of
  inflationary fermion condensation},''
  \href{http://dx.doi.org/10.1088/1475-7516/2024/04/022}{{\em JCAP} {\bfseries
  04} (2024) 022}, \href{http://arxiv.org/abs/2304.09428}{{\ttfamily
  arXiv:2304.09428 [hep-th]}}.

\bibitem{Jazayeri:2023xcj}
S.~Jazayeri, S.~Renaux-Petel, and D.~Werth, ``{Shapes of the cosmological
  low-speed collider},''
  \href{http://dx.doi.org/10.1088/1475-7516/2023/12/035}{{\em JCAP} {\bfseries
  12} (2023) 035}, \href{http://arxiv.org/abs/2307.01751}{{\ttfamily
  arXiv:2307.01751 [hep-th]}}.

\bibitem{Jazayeri:2023kji}
S.~Jazayeri, S.~Renaux-Petel, X.~Tong, D.~Werth, and Y.~Zhu, ``{Parity
  violation from emergent nonlocality during inflation},''
  \href{http://dx.doi.org/10.1103/PhysRevD.108.123523}{{\em Phys. Rev. D}
  {\bfseries 108} no.~12, (2023) 123523},
  \href{http://arxiv.org/abs/2308.11315}{{\ttfamily arXiv:2308.11315
  [hep-th]}}.

\bibitem{Aoki:2023tjm}
S.~Aoki, ``{Continuous spectrum on cosmological collider},''
  \href{http://dx.doi.org/10.1088/1475-7516/2023/04/002}{{\em JCAP} {\bfseries
  04} (2023) 002}, \href{http://arxiv.org/abs/2301.07920}{{\ttfamily
  arXiv:2301.07920 [hep-th]}}.

\bibitem{McCulloch:2024hiz}
C.~McCulloch, E.~Pajer, and X.~Tong, ``{A cosmological tachyon collider:
  enhancing the long-short scale coupling},''
  \href{http://dx.doi.org/10.1007/JHEP05(2024)262}{{\em JHEP} {\bfseries 05}
  (2024) 262}, \href{http://arxiv.org/abs/2401.11009}{{\ttfamily
  arXiv:2401.11009 [hep-th]}}.

\bibitem{Craig:2024qgy}
N.~Craig, S.~Kumar, and A.~McCune, ``{An Effective Cosmological Collider},''
  \href{http://arxiv.org/abs/2401.10976}{{\ttfamily arXiv:2401.10976
  [hep-ph]}}.

\bibitem{Baumann:2022jpr}
D.~Baumann, D.~Green, A.~Joyce, E.~Pajer, G.~L. Pimentel, C.~Sleight, and
  M.~Taronna, ``{Snowmass White Paper: The Cosmological Bootstrap},'' in {\em
  {2022 Snowmass Summer Study}}.
\newblock 3, 2022.
\newblock \href{http://arxiv.org/abs/2203.08121}{{\ttfamily arXiv:2203.08121
  [hep-th]}}.

\bibitem{Arkani-Hamed:2018kmz}
N.~Arkani-Hamed, D.~Baumann, H.~Lee, and G.~L. Pimentel, ``{The Cosmological
  Bootstrap: Inflationary Correlators from Symmetries and Singularities},''
  \href{http://dx.doi.org/10.1007/JHEP04(2020)105}{{\em JHEP} {\bfseries 04}
  (2020) 105}, \href{http://arxiv.org/abs/1811.00024}{{\ttfamily
  arXiv:1811.00024 [hep-th]}}.

\bibitem{Baumann:2019oyu}
D.~Baumann, C.~Duaso~Pueyo, A.~Joyce, H.~Lee, and G.~L. Pimentel, ``{The
  cosmological bootstrap: weight-shifting operators and scalar seeds},''
  \href{http://dx.doi.org/10.1007/JHEP12(2020)204}{{\em JHEP} {\bfseries 12}
  (2020) 204}, \href{http://arxiv.org/abs/1910.14051}{{\ttfamily
  arXiv:1910.14051 [hep-th]}}.

\bibitem{Pimentel:2022fsc}
G.~L. Pimentel and D.-G. Wang, ``{Boostless cosmological collider bootstrap},''
  \href{http://dx.doi.org/10.1007/JHEP10(2022)177}{{\em JHEP} {\bfseries 10}
  (2022) 177}, \href{http://arxiv.org/abs/2205.00013}{{\ttfamily
  arXiv:2205.00013 [hep-th]}}.

\bibitem{Jazayeri:2022kjy}
S.~Jazayeri and S.~Renaux-Petel, ``{Cosmological bootstrap in slow motion},''
  \href{http://dx.doi.org/10.1007/JHEP12(2022)137}{{\em JHEP} {\bfseries 12}
  (2022) 137}, \href{http://arxiv.org/abs/2205.10340}{{\ttfamily
  arXiv:2205.10340 [hep-th]}}.

\bibitem{Qin:2022fbv}
Z.~Qin and Z.-Z. Xianyu, ``{Helical inflation correlators: partial
  Mellin-Barnes and bootstrap equations},''
  \href{http://dx.doi.org/10.1007/JHEP04(2023)059}{{\em JHEP} {\bfseries 04}
  (2023) 059}, \href{http://arxiv.org/abs/2208.13790}{{\ttfamily
  arXiv:2208.13790 [hep-th]}}.

\bibitem{Aoki:2023wdc}
S.~Aoki, T.~Noumi, F.~Sano, and M.~Yamaguchi, ``{Analytic formulae for
  inflationary correlators with dynamical mass},''
  \href{http://dx.doi.org/10.1007/JHEP03(2024)073}{{\em JHEP} {\bfseries 03}
  (2024) 073}, \href{http://arxiv.org/abs/2312.09642}{{\ttfamily
  arXiv:2312.09642 [hep-th]}}.

\bibitem{Qin:2023ejc}
Z.~Qin and Z.-Z. Xianyu, ``{Closed-form formulae for inflation correlators},''
  \href{http://dx.doi.org/10.1007/JHEP07(2023)001}{{\em JHEP} {\bfseries 07}
  (2023) 001}, \href{http://arxiv.org/abs/2301.07047}{{\ttfamily
  arXiv:2301.07047 [hep-th]}}.

\bibitem{Aoki:2024uyi}
S.~Aoki, L.~Pinol, F.~Sano, M.~Yamaguchi, and Y.~Zhu, ``{Cosmological
  Correlators with Double Massive Exchanges: Bootstrap Equation and
  Phenomenology},'' \href{http://arxiv.org/abs/2404.09547}{{\ttfamily
  arXiv:2404.09547 [hep-th]}}.

\bibitem{Hillman:2021bnk}
A.~Hillman and E.~Pajer, ``{A differential representation of cosmological
  wavefunctions},'' \href{http://dx.doi.org/10.1007/JHEP04(2022)012}{{\em JHEP}
  {\bfseries 04} (2022) 012}, \href{http://arxiv.org/abs/2112.01619}{{\ttfamily
  arXiv:2112.01619 [hep-th]}}.

\bibitem{Chen:2023iix}
J.~Chen and B.~Feng, ``{Towards Systematic Evaluation of de Sitter Correlators
  via Generalized Integration-By-Parts Relations},''
  \href{http://arxiv.org/abs/2401.00129}{{\ttfamily arXiv:2401.00129
  [hep-th]}}.

\bibitem{Chen:2024glu}
J.~Chen, B.~Feng, and Y.-X. Tao, ``{Multivariate hypergeometric solutions of
  cosmological (dS) correlators by $\text{d} \log$-form differential
  equations},'' \href{http://arxiv.org/abs/2411.03088}{{\ttfamily
  arXiv:2411.03088 [hep-th]}}.

\bibitem{Arkani-Hamed:2017fdk}
N.~Arkani-Hamed, P.~Benincasa, and A.~Postnikov, ``{Cosmological Polytopes and
  the Wavefunction of the Universe},''
  \href{http://arxiv.org/abs/1709.02813}{{\ttfamily arXiv:1709.02813
  [hep-th]}}.

\bibitem{Arkani-Hamed:2023bsv}
N.~Arkani-Hamed, D.~Baumann, A.~Hillman, A.~Joyce, H.~Lee, and G.~L. Pimentel,
  ``{Kinematic Flow and the Emergence of Time},''
  \href{http://arxiv.org/abs/2312.05300}{{\ttfamily arXiv:2312.05300
  [hep-th]}}.

\bibitem{Arkani-Hamed:2023kig}
N.~Arkani-Hamed, D.~Baumann, A.~Hillman, A.~Joyce, H.~Lee, and G.~L. Pimentel,
  ``{Differential Equations for Cosmological Correlators},''
  \href{http://arxiv.org/abs/2312.05303}{{\ttfamily arXiv:2312.05303
  [hep-th]}}.

\bibitem{Fan:2024iek}
B.~Fan and Z.-Z. Xianyu, ``{Cosmological Amplitudes in Power-Law FRW
  Universe},'' \href{http://arxiv.org/abs/2403.07050}{{\ttfamily
  arXiv:2403.07050 [hep-th]}}.

\bibitem{He:2024olr}
S.~He, X.~Jiang, J.~Liu, Q.~Yang, and Y.-Q. Zhang, ``{Differential equations
  and recursive solutions for cosmological amplitudes},''
  \href{http://arxiv.org/abs/2407.17715}{{\ttfamily arXiv:2407.17715
  [hep-th]}}.

\bibitem{Hillman:2019wgh}
A.~Hillman, ``{Symbol Recursion for the dS Wave Function},''
  \href{http://arxiv.org/abs/1912.09450}{{\ttfamily arXiv:1912.09450
  [hep-th]}}.

\bibitem{De:2023xue}
S.~De and A.~Pokraka, ``{Cosmology meets cohomology},''
  \href{http://dx.doi.org/10.1007/JHEP03(2024)156}{{\em JHEP} {\bfseries 03}
  (2024) 156}, \href{http://arxiv.org/abs/2308.03753}{{\ttfamily
  arXiv:2308.03753 [hep-th]}}.

\bibitem{Benincasa:2024leu}
P.~Benincasa and G.~Dian, ``{The Geometry of Cosmological Correlators},''
  \href{http://arxiv.org/abs/2401.05207}{{\ttfamily arXiv:2401.05207
  [hep-th]}}.

\bibitem{Benincasa:2024lxe}
P.~Benincasa and F.~Vaz\~ao, ``{The Asymptotic Structure of Cosmological
  Integrals},'' \href{http://arxiv.org/abs/2402.06558}{{\ttfamily
  arXiv:2402.06558 [hep-th]}}.

\bibitem{Baumann:2024mvm}
D.~Baumann, H.~Goodhew, and H.~Lee, ``{Kinematic Flow for Cosmological Loop
  Integrands},'' \href{http://arxiv.org/abs/2410.17994}{{\ttfamily
  arXiv:2410.17994 [hep-th]}}.

\bibitem{Wang:2021qez}
L.-T. Wang, Z.-Z. Xianyu, and Y.-M. Zhong, ``{Precision calculation of
  inflation correlators at one loop},''
  \href{http://dx.doi.org/10.1007/JHEP02(2022)085}{{\em JHEP} {\bfseries 02}
  (2022) 085}, \href{http://arxiv.org/abs/2109.14635}{{\ttfamily
  arXiv:2109.14635 [hep-ph]}}.

\bibitem{Qin:2022lva}
Z.~Qin and Z.-Z. Xianyu, ``{Phase information in cosmological collider
  signals},'' \href{http://dx.doi.org/10.1007/JHEP10(2022)192}{{\em JHEP}
  {\bfseries 10} (2022) 192}, \href{http://arxiv.org/abs/2205.01692}{{\ttfamily
  arXiv:2205.01692 [hep-th]}}.

\bibitem{Xianyu:2022jwk}
Z.-Z. Xianyu and H.~Zhang, ``{Bootstrapping one-loop inflation correlators with
  the spectral decomposition},''
  \href{http://dx.doi.org/10.1007/JHEP04(2023)103}{{\em JHEP} {\bfseries 04}
  (2023) 103}, \href{http://arxiv.org/abs/2211.03810}{{\ttfamily
  arXiv:2211.03810 [hep-th]}}.

\bibitem{Qin:2023nhv}
Z.~Qin and Z.-Z. Xianyu, ``{Nonanalyticity and on-shell factorization of
  inflation correlators at all loop orders},''
  \href{http://dx.doi.org/10.1007/JHEP01(2024)168}{{\em JHEP} {\bfseries 01}
  (2024) 168}, \href{http://arxiv.org/abs/2308.14802}{{\ttfamily
  arXiv:2308.14802 [hep-th]}}.

\bibitem{Qin:2023bjk}
Z.~Qin and Z.-Z. Xianyu, ``{Inflation correlators at the one-loop order:
  nonanalyticity, factorization, cutting rule, and OPE},''
  \href{http://dx.doi.org/10.1007/JHEP09(2023)116}{{\em JHEP} {\bfseries 09}
  (2023) 116}, \href{http://arxiv.org/abs/2304.13295}{{\ttfamily
  arXiv:2304.13295 [hep-th]}}.

\bibitem{Xianyu:2023ytd}
Z.-Z. Xianyu and J.~Zang, ``{Inflation correlators with multiple massive
  exchanges},'' \href{http://dx.doi.org/10.1007/JHEP03(2024)070}{{\em JHEP}
  {\bfseries 03} (2024) 070}, \href{http://arxiv.org/abs/2309.10849}{{\ttfamily
  arXiv:2309.10849 [hep-th]}}.

\bibitem{Liu:2024xyi}
H.~Liu, Z.~Qin, and Z.-Z. Xianyu, ``{Dispersive Bootstrap of Massive Inflation
  Correlators},'' \href{http://arxiv.org/abs/2407.12299}{{\ttfamily
  arXiv:2407.12299 [hep-th]}}.

\bibitem{Werth:2024mjg}
D.~Werth, ``{Spectral Representation of Cosmological Correlators},''
  \href{http://arxiv.org/abs/2409.02072}{{\ttfamily arXiv:2409.02072
  [hep-th]}}.

\bibitem{Qin:2024gtr}
Z.~Qin, ``{Cosmological Correlators at the Loop Level},''
  \href{http://arxiv.org/abs/2411.13636}{{\ttfamily arXiv:2411.13636
  [hep-th]}}.

\bibitem{Sleight:2019mgd}
C.~Sleight, ``{A Mellin Space Approach to Cosmological Correlators},''
  \href{http://dx.doi.org/10.1007/JHEP01(2020)090}{{\em JHEP} {\bfseries 01}
  (2020) 090}, \href{http://arxiv.org/abs/1906.12302}{{\ttfamily
  arXiv:1906.12302 [hep-th]}}.

\bibitem{Sleight:2019hfp}
C.~Sleight and M.~Taronna, ``{Bootstrapping Inflationary Correlators in Mellin
  Space},'' \href{http://dx.doi.org/10.1007/JHEP02(2020)098}{{\em JHEP}
  {\bfseries 02} (2020) 098}, \href{http://arxiv.org/abs/1907.01143}{{\ttfamily
  arXiv:1907.01143 [hep-th]}}.

\bibitem{Sleight:2020obc}
C.~Sleight and M.~Taronna, ``{From AdS to dS exchanges: Spectral
  representation, Mellin amplitudes, and crossing},''
  \href{http://dx.doi.org/10.1103/PhysRevD.104.L081902}{{\em Phys. Rev. D}
  {\bfseries 104} no.~8, (2021) L081902},
  \href{http://arxiv.org/abs/2007.09993}{{\ttfamily arXiv:2007.09993
  [hep-th]}}.

\bibitem{Sleight:2021plv}
C.~Sleight and M.~Taronna, ``{From dS to AdS and back},''
  \href{http://dx.doi.org/10.1007/JHEP12(2021)074}{{\em JHEP} {\bfseries 12}
  (2021) 074}, \href{http://arxiv.org/abs/2109.02725}{{\ttfamily
  arXiv:2109.02725 [hep-th]}}.

\bibitem{Werth:2023pfl}
D.~Werth, L.~Pinol, and S.~Renaux-Petel, ``{Cosmological Flow of Primordial
  Correlators},'' \href{http://arxiv.org/abs/2302.00655}{{\ttfamily
  arXiv:2302.00655 [hep-th]}}.

\bibitem{Pinol:2023oux}
L.~Pinol, S.~Renaux-Petel, and D.~Werth, ``{The Cosmological Flow: A Systematic
  Approach to Primordial Correlators},''
  \href{http://arxiv.org/abs/2312.06559}{{\ttfamily arXiv:2312.06559
  [astro-ph.CO]}}.

\bibitem{Werth:2024aui}
D.~Werth, L.~Pinol, and S.~Renaux-Petel, ``{CosmoFlow: Python Package for
  Cosmological Correlators},''
  \href{http://arxiv.org/abs/2402.03693}{{\ttfamily arXiv:2402.03693
  [astro-ph.CO]}}.

\bibitem{Matsumoto_2020}
K.~Matsumoto, {\em Appell and Lauricella Hypergeometric Functions},
  pp.~79--100.
\newblock Cambridge University Press, 2020.

\bibitem{Chen:2017ryl}
X.~Chen, Y.~Wang, and Z.-Z. Xianyu, ``{Schwinger-Keldysh Diagrammatics for
  Primordial Perturbations},''
  \href{http://dx.doi.org/10.1088/1475-7516/2017/12/006}{{\em JCAP} {\bfseries
  1712} no.~12, (2017) 006},
\href{http://arxiv.org/abs/1703.10166}{{\ttfamily arXiv:1703.10166 [hep-th]}}.

\bibitem{Goodhew:2020hob}
H.~Goodhew, S.~Jazayeri, and E.~Pajer, ``{The Cosmological Optical Theorem},''
  \href{http://dx.doi.org/10.1088/1475-7516/2021/04/021}{{\em JCAP} {\bfseries
  04} (2021) 021}, \href{http://arxiv.org/abs/2009.02898}{{\ttfamily
  arXiv:2009.02898 [hep-th]}}.

\bibitem{Slater:1966}
L.~J. Slater, {\em Generalized Hypergeometric Functions}.
\newblock Cambridge University Press, 1966.

\bibitem{horn1931hypergeometrische}
J.~Horn, ``Hypergeometrische funktionen zweier ver{\"a}nderlichen,'' {\em
  Mathematische Annalen} {\bfseries 105} no.~1, (1931) 381--407.

\bibitem{Brown:2013qva}
F.~Brown, \href{http://dx.doi.org/10.1017/CBO9781139208642.006}{``{Iterated
  integrals in quantum field theory},''} in {\em {6th Summer School on
  Geometric and Topological Methods for Quantum Field Theory}}, pp.~188--240.
\newblock 2013.

\bibitem{Tong:2021wai}
X.~Tong, Y.~Wang, and Y.~Zhu, ``{Cutting rule for cosmological collider
  signals: a bulk evolution perspective},''
  \href{http://dx.doi.org/10.1007/JHEP03(2022)181}{{\em JHEP} {\bfseries 03}
  (2022) 181}, \href{http://arxiv.org/abs/2112.03448}{{\ttfamily
  arXiv:2112.03448 [hep-th]}}.

\bibitem{Ema:2024hkj}
Y.~Ema and K.~Mukaida, ``{Cutting rule for in-in correlators and cosmological
  collider},'' \href{http://arxiv.org/abs/2409.07521}{{\ttfamily
  arXiv:2409.07521 [hep-th]}}.

\bibitem{Parke:1986gb}
S.~J. Parke and T.~R. Taylor, ``{An Amplitude for $n$ Gluon Scattering},''
  \href{http://dx.doi.org/10.1103/PhysRevLett.56.2459}{{\em Phys. Rev. Lett.}
  {\bfseries 56} (1986) 2459}.

\bibitem{gel1989hypergeometric}
I.~M. Gel'fand, A.~V. Zelevinskii, and M.~M. Kapranov, ``Hypergeometric
  functions and toral manifolds,'' {\em Funktsional'nyi Analiz i ego
  Prilozheniya} {\bfseries 23} no.~2, (1989) 12--26.

\end{thebibliography}\endgroup
\bibliographystyle{utphys}

\end{document}